\documentclass[apj]{emulateapj}
\usepackage{apjfonts}
\usepackage{amsbsy}
\usepackage{longtable}
\newcommand{\mgii}{\ion{Mg}{2}}

\newcommand{\etal}{\ensuremath{\mbox{et~al.}}}
\newcommand{\hmsol}{\mbox{$h_{71}^{-1}\,{\rm M}_\odot$}}

\shorttitle{Galaxy Clusters in the Line of Sight to Background Quasars - III}
\shortauthors{Andrews \etal}

\begin{document}

\slugcomment{June 2013}

\title{Galaxy Clusters in the Line of Sight to Background Quasars - III Multi-Object Spectroscopy$^1$}

\author{H. Andrews\altaffilmark{2}, L. F.\ Barrientos\altaffilmark{2}, 
S. L\'opez\altaffilmark{3}, 
P.\ Lira\altaffilmark{3}, 
N. Padilla\altaffilmark{2}, 
D. G.\ Gilbank\altaffilmark{7}, 
I.\ Lacerna\altaffilmark{2}, 
M. J.\ Maureira\altaffilmark{3}, 
E.\ Ellingson\altaffilmark{4}, 
M. D.\ Gladders\altaffilmark{5,6} \&
H. K. C.\ Yee\altaffilmark{8}.}

\altaffiltext{1}{
Based on observations obtained at the Gemini Observatory, which is operated by the
Association of Universities for Research in Astronomy, Inc., under a cooperative agreement
with the NSF on behalf of the Gemini partnership: the National Science Foundation (United
States), the Science and Technology Facilities Council (United Kingdom), the
National Research Council (Canada), CONICYT (Chile), the Australian Research Council
(Australia), Minist\'erio da Ci\^{e}ncia e Tecnologia (Brazil) 
and Ministerio de Ciencia, Tecnolog\'ia e Innovaci\'on Productiva  (Argentina)
}
\altaffiltext{2}{Departamento de Astronom\'ia y Astrof\'isica, Pontificia Universidad Cat\'olica de Chile, Avenida Vicu\~na Mackenna 4860,
    Santiago, Chile. {\tt barrientos@astro.puc.cl}}
\altaffiltext{3}{Departamento de Astronom\'ia, Universidad de Chile, Casilla 36-D, Santiago, Chile.}
\altaffiltext{4}{Center for Astrophysics and Space Astronomy, University of Colorado at Boulder, Campus Box 389, Boulder, CO 80309-0389, USA.}
\altaffiltext{5}{Kavli Institute for Cosmological Physics, University of Chicago, 5640 South Ellis Avenue, Chicago, IL 60637, USA}
\altaffiltext{6}{Department of Astronomy and Astrophysics, University of Chicago, 5640 South Ellis Avenue, Chicago, IL 60637, USA}
\altaffiltext{7}{South African Astronomical Observatory, PO Box 9, Observatory, 7935, South Africa.}
\altaffiltext{8}{Department of Astronomy and Astrophysics, University of Toronto, 60 St. George St., Toronto, Ont., Canada M5S 3H8.}

\begin{abstract}

We present Gemini/GMOS-S multi-object spectroscopy of 31 galaxy cluster candidates at redshifts between 0.2 and 1.0 and centered on QSO sight-lines taken from \citet{l2008}. The targets were selected based on the presence of a intervening \mgii\ absorption system at a similar redshift to that of a galaxy cluster candidate lying at a projected distance $<$ 2 $h_{71}^{-1}$Mpc from the QSO sight-line (a ``photometric-hit''). The absorption systems span rest-frame equivalent widths between 0.015 and 2.028 \AA. Our aim was 3-fold: 1) identify the absorbing galaxies and determine their impact parameters, 2) confirm the galaxy cluster candidates in the vicinity of each quasar sightline, and 3) determine whether the absorbing galaxies reside in galaxy clusters.  In this way we are able to characterize the absorption systems associated with cluster members. Our main findings are: 1) the identification of 10 out of 24 absorbing galaxies with redshifts between $0.2509 \leq z_{gal} \leq 1.0955$, up to an impact parameter of  $142 h_{71}^{-1}$ kpc and a maximum velocity difference of  $280$ km/s. 2) The spectroscopic confirmation of 20 out of 31 cluster/group candidates, with most of the confirmed clusters/groups at $z < 0.7$. This relatively low efficiency results from the fact that we centered our observations on the QSO location, and thus occasionally some of the cluster centers were outside the instrument FOV. 3) Following from the results above, the spectroscopic confirmation of 10 out of 14 photometric hits within $\sim$ 650 km/s from galaxy clusters/groups, in addition to 2 new ones related to galaxy group environments. These numbers imply  efficiencies of 71\% in finding such systems with MOS spectroscopy. This is a remarkable result since we defined a photometric hit as those cluster-absorber pairs having a redshift difference $\Delta z = 0.1$.

The general population of our confirmed absorbing galaxies have luminosities $L_{B} \sim L_{B}^{\ast}$ and mean rest-frame colors ($R_{c} - z'$) typical of S$_{cd}$ galaxies. From this sample, absorbing cluster-galaxies hosting weak absorbers are consistent with lower star formation activity than the rest, which produce strong absorption and agree with typical \mgii\ absorbing galaxies found in the literature. Our spectroscopic confirmations lend support to the selection of photometric hits made in \citet{l2008}. 
\end{abstract}

\keywords{cosmology: observations---intergalactic medium---quasars: absorption lines}

\section{INTRODUCTION}

\bigskip

Galaxies hosting \mgii\ absorption systems seen in the spectra of background QSOs (hereafter \mgii\ \textit{absorbing galaxies}) appear to be a quite heterogeneous sample at $z \lesssim$ 1. They span a rather broad range of spectral types and brightness, but concentrate towards high luminosities ($L \sim L_{B}^{\ast}$) with colors and spectral features typical of S$_{b}$ or S$_{c}$ spiral field galaxies \citep[e.g., ][]{zm2007}, lying at impact parameters $\sim$ 40--100 $h_{71}^{-1}$kpc from the quasar sight-line \citep{sd1994, la1996, csk2005, kc2005, kc2007, ch2010}.

Models assuming that absorbing gas resides within galactic haloes are in general agreement with observable data available for \mgii\ absorbers \citep{sk2002, lz2001}. However, the very origin of these absorption systems (e.g., whether the absorptions are specifically produced in the disk or halo, by SN\,II ejecta, gas accretion, stripped gas, stellar outflows, etc.) has not yet been well established.

Correlations between \mgii\ absorption strength and galactic halo masses \citep{bm2006, gc2009} or galaxy type \citep{zm2007, ru2010, me2011} have been thoroughly studied, and suggest a link between \mgii\ absorber intensities and the star formation in their host galaxies. However, such studies have mostly considered strong \mgii\ absorbers ($W_{0}^{2796} >$ 1.0 \AA). Thus, we are still lacking larger galaxy surveys that include weak \mgii\ absorption systems (i.e., $W_{0}^{2796} <$ 0.3 \AA).

Furthermore, despite some \mgii\ systems having been found to populate galaxy group or cluster environments \citep{be1992, sd1992, bb1995, cc1998, ch2010, gauthier2013}, no studies have searched for galaxy group environments in a systematic fashion.

The Quasars behind Clusters project (QbC) \citep[hereafter Paper I]{l2008} is the first \mgii\ survey designed to specifically target galaxy cluster/group environments,  \mgii\ systems are sought in lines-of-sight (LOS) towards quasars known to intersect galaxy cluster/group candidates at photometric redshifts $z_{clus}^{phot} \sim$ 0.2--0.9 drawn from the Red-Sequence Cluster Survey 1 \citep[RCS1,][]{gy2000}. Using a sample of $\sim$ 400 QSO-cluster pairs at cluster-centric projected distances $d <$ 2 $h_{71}^{-1}$Mpc, \citet{l2008} detected differences between the number density of absorbers most probably located in galaxy cluster/group environments and those related to the field. While the strongest \mgii\ systems ($W_{0}^{2796} >$ 2 \AA) were found to be up to 10 times more abundant in clusters than those produced in the field, weak \mgii\ systems ($W_{0}^{2796} <$ 0.3 \AA) did not show a similar excess. The proposed explanation for this signal was that weak systems should occur in galactic haloes that have been truncated due to environmental effects, i.e. galaxy harassment or ram pressure stripping. This interpretation was later used by \citet[PaperII]{pad2009} to put constraints on the sizes of baryonic haloes around cluster galaxies.

The association between \mgii\ absorptions and galaxy cluster/group candidates in Paper I is subject to uncertainties in the photometric redshifts of cluster/group candidates, from which the cluster redshift path of the survey $\Delta z_{cluster}$ was derived. Therefore it becomes necessary to reduce the photometric redshift uncertainties of the clusters, in order to establish better the relation between absorbers and their environment.

The work presented here is based on Gemini/GMOS - S MOS observations of 31 cluster candidates drawn from Paper I. Our aim is 3-fold: 1) find the \mgii\ absorbing galaxies and calculate their LOS impact parameters, 2) verify the overabundance of galaxies in the vicinity of each quasar sight-line due to the presence of clusters of galaxies and 3) determine whether the absorbing galaxies reside in galaxy clusters or not.

This paper is organized as follows: in section $\S$2 we describe our data and the details of the spectroscopic observations and reduction steps for a sample of 23 \mgii\ systems. Results are given in $\S$3, specifically the detection of absorbing galaxies and a general view of their properties is presented in $\S$3.1; the spectroscopic confirmation of cluster/group candidates is detailed in $\S$3.2 (with the caveat of having a QSO in the middle of the field limits the ability of confirming clusters); the confirmation of spectroscopic hits is presented in $\S$3.3. A discussion of our results and their implications is given in $\S$4; our  concluding remarks are outlined in $\S$5. The cosmological parameters adopted in this study are $\Omega_{m} =$ 0.27, $\Omega_{\Lambda} =$ 0.73 and $H_{0} =$ 71 $h_{71}$ km s${-1}$ Mpc$^{-1}$.

\bigskip

\section{DATA}


In Paper I, the matching of a \mgii\ system to the presence of a galaxy cluster candidate, was termed a \textit{photometric hit}. Explicitly, for a \mgii\ absorbing system at $z_{abs}$, a photometric hit was defined as $z_{abs} \in [z_{min}, z_{max}]$ where $z_{min} = z_{clus}^{phot} - \delta z_{clus}^{phot}$ and $z_{max} = z_{clus}^{phot} + \delta z_{clus}^{phot}$, $z_{clus}^{phot}$ the photometric redshift of a RCS1 cluster/group candidate with a redshift uncertainty  $\delta z_{clus}^{phot}$. We set $\delta z_{clus}^{phot}= 0.1$, except where $z_{clus}^{phot} - \delta z_{clus}^{phot} < z_{EW}$. $z_{EW}$ is the minimum redshift at which a system with $W_{0}^{2796} =$ 0.05 \AA\ could be detected as a 3$\sigma$ detection; in these cases $z_{min}$ was set to $z_{EW}$.

The spectroscopic confirmation of RCS1 cluster/group candidates leads to a new definition of a \textit{spectroscopic hit}. By definition, a spectroscopic hit corresponds to a \mgii\ absorption (assumed to originate in a galactic halo) located in the environment of a spectroscopically confirmed galaxy cluster.

\bigskip

\subsection{Region selection}

Our data consist of multi-object spectroscopy covering 9 fields centered on different quasar sight-lines, each one presenting one or more photometric hits.

These LOS were drawn from a list of photometric hits presented in Paper I, detected in  high resolution spectroscopic data using MIKE echelle spectrograph mounted on the 6.5m Magellan telescope at Las Campanas Observatory. In this sample, \mgii\ absorption systems were detected at a $>$ 3$\sigma$ detection level in both doublet lines and have an uncertainty $\delta z_{abs}$ $\sim$ 10$^{-4}$ (see Paper I for further details).  Spanning a wide range of absorption redshifts (0.2507 $\leq$ $z_{abs}$ $\leq$ 1.0951) and rest-frame equivalent widths (0.015 $\leq$ $W_{0}^{2796}$ $\leq$ 2.028 \AA), they are representative of the high resolution sample in Paper I.

The sample of galaxy cluster/group candidates lying at cluster-centric projected distances $d <$ 2 $h_{71}^{-1}$Mpc from the quasar sight-lines consists of 31 candidates with photometric redshifts between 0.173 $\leq$ $z_{clus}^{phot}$ $\leq$ 1.032 and richness parameters $B_{gc} <$ 1037 (with a mean value $\overline{B_{gc}}$ $=$ 346; more details in $\S$3.2). These fields were available for observations during the second semester of 2008. 

Table \ref{absorption_systems} shows the \mgii\ absorption systems and the number of RCS1 cluster/group candidates lying at a cluster-centric impact parameter $d <$ 2 $h_{71}^{-1}$Mpc from each LOS studied in this work. The LOS to the quasar HE2149$-$2745A, also included in the analysis presented in Paper I, was taken from the literature and is outside the field covered by the RCS1 and SDSS (\citealt[hereafter W06]{wm2006}; \citealt[hereafter M06]{mw2006}). Absorption line redshifts were redefined with respect to paper I to match the strongest MgII velocity component.

Our analysis makes use of photometric data from both RCS1 galaxy cluster and object catalogs, where the latter provided ($R_{c} - z'$) and $z'$ magnitudes for all extended sources in our fields down to limiting magnitudes $R_{c} =$ 24.1 and $z' =$ 23.1 \citep{gy2000, y1991}. 

In the following, all apparent magnitudes are given in the AB system and are corrected for galactic extinction according to the dust maps of \citet{sf1998}. 

\bigskip

\subsection{MOS Observations}

Our spectroscopic data were obtained with the Gemini Multi-Object Spectrograph (GMOS) at Gemini South Telescope in Cerro Pach\'on, Chile. The GMOS field-of-view (5.5$\arcmin \times$ 5.5$\arcmin$) is wide enough to probe Mpc-scale projected distances at various redshifts. This is necessary since we consider cluster/group candidates at cluster-centric impact parameters $d <$ 2 $h_{71}^{-1}$Mpc from the LOS at photometric redshifts $z_{clus}^{phot} \sim$ 0.2--1 (which translate to angular separations of $\sim$ 10 $\arcmin$-- 4 $\arcmin$, respectively). The spectroscopic data were acquired in queue mode between July 2008 and December 2008 (program ID: GS-2008A-Q-10; PI: S. L\'opez).  In order to obtain good S/N ratio of the spectra and maximize spectroscopic completeness, two masks per field were designed: for faint targets ($R_{c} \gtrsim$ 21) an exposure time of 2 $\times$ 3\,000 s was chosen, whereas for brighter ones ($R_{c} <$ 21) the exposure time was 2 $\times$ 1\,800 s.

Since the goals of the MOS observations were to identify \mgii\ absorbing galaxies and confirm galaxy clusters, masks were centered on QSO LOS and position angles chosen to maximize the number of cluster candidates (i.e. their cluster centers) within the field-of-view. 

The 400 lines/mm grating (R400$_{-}$G5325) was chosen and two central wavelengths were used (670 nm and 695 nm) ---one for each of the two sets of exposures--- in order to combine two spectra of the same object, avoiding the loss of information falling into the gaps between the three CCDs of Gemini mosaic detector. The slit width for all targets was set to 1$\arcsec$, while the slit length varied from 7$\arcsec$ to 10$\arcsec$ to ensure sufficient sky counts for good sky subtraction.

Target selection for each mask focussed mainly on sources near the LOS of the background quasars, with the purpose of detecting absorbing galaxies. More specifically, galaxies at an impact parameter $\rho <$ 150 $h_{71}^{-1}$kpc from the LOS (measured at the absorption redshift $z_{abs}$) were categorized as  first priority targets in the selection algorithm. This limit permits  comparison with impact parameters found in studies of the halo cross sections of \mgii\ absorbing galaxies in the field (\citealt[ref. therein]{csk2005}; \citealt{ch2010}). Within this study we emphasize that $\rho$ [$h_{71}^{-1}$kpc] refers to galaxy impact parameters to the LOS, whilst $d$ [$h_{71}^{-1}$Mpc] refers to cluster-centric impact parameters to the LOS.

Based on the $R_{c}$ and $z'$ color magnitude diagram for galaxy-type objects around the cluster/group candidate coordinates \citep{gy2005}, in addition to already known color magnitude relations derived from composite clusters of galaxies \citep{gb2008}, we selected objects of second priority in order to detect potential brightest cluster members. Second priority objects with an impact parameter $\rho <$ 150 $h_{71}^{-1}$kpc from the LOS (at $z_{abs}$) were re-categorized as first priority objects.

Finally,  third priority targets were chosen by performing a visual inspection of objects not selected as potential cluster members. The bias in this category towards bright/bluer galaxies permits the selection of galaxies in the outer regions of the clusters potentially exhibiting bluer colors.

The total number of selected targets was 440. Table \ref{spectroscopy_targets} lists the number of spectroscopic sources selected in each field ($N_{\rm tot}$), and those at an impact parameter $\rho <$ 150 $h_{71}^{-1}$kpc from the respective LOS ($N_{\rho\,<\,\rm 150\,\,\,h_{71}^{-1}kpc}$) with a magnitude $R_{c} \leq R_{\rm faint}$, where $R_{\rm faint}$ is the magnitude of the faintest object inside this region.  

\bigskip

\subsection{MOS data reduction}

Data reduction was performed using the GEMINI IRAF package version 1.9, following standard IRAF v2.14 reduction procedures. The dispersion solution was found using 40 to 45 spectral lines of the CuAr arc lamp distributed among the whole wavelength range. A  4th or 5th order Chebyshev polynomial was fitted to the data and the resulting r.m.s. of the fits ranged between 0.15 to 0.22 \AA. The resulting wavelength interval starts at $\sim$ 4000--4500\AA\ for some spectra, ending at $\sim$ 8500--9000\AA\ in others; the exact wavelength range depends on the position of the slit in the pre-image. The final data had a dispersion of $\sim$ 1.365 \AA/pixel with a resolution element of FWHM $\sim$ 7 \AA\ ($R \sim$ 1000) equivalent to $\sim$ 350 km/s at $\lambda =$ 6000\AA.

The final signal-to-noise ratio $S/N$ of our spectra ranged between 5--20 per pixel at 6000\AA. Flux spectra were not calibrated as we were only interested in obtaining galaxy redshifts. We were able to use $\sim$ 88\% of the reduced spectra, the rest suffered from low $S/N$, artifacts or fringing.

We measured redshifts by fitting gaussian profiles to the spectral features (using the task rvidlines in the NOAO.RV IRAF package), and through visual inspection of the 2D spectra whenever necessary. Air-to-vacuum and heliocentric corrections were applied to each spectrum.

For early type galaxies, redshifts were measured primarily with the lines Ca\,II\,H, Ca\,II\,K and the G band absorption lines. For late type galaxies, the H$_{\beta}$, [O\,III]\,4959\AA\ and/or [O\,III]\,5007\AA\ were used. Whenever possible,  [N\,II]\,6548\AA, H$_{\alpha}$ 6563\AA, [N\,II]\,6583\AA, [S\,II]\,6716\AA\ and/or [S\,II]\,6731\AA\ emission features were also considered. Nearly 55\% of the galaxies show spectral features typical of early type galaxies, of which $\sim$ 63\% also show  [O\,II]\,3727\AA\ emission. Galaxies showing late type features comprise 45\% of the entire sample.

The mean redshift uncertainty of our galaxy catalog was $\sim$ 0.0005, equivalent to $\lesssim$ 100 km/s at $z =$ 0.55. All redshifts were classified  as type 1, type 2 or type 3 according to the reliability of our measurements, with type 1 the most reliable, type 3 the least reliable. Galaxies with type 1 redshifts have spectra with two or more clear spectral features that could be modeled using Gaussian profiles. Redshifts of type 2 were measured using only one or two detectable spectral lines alongside the possible presence of weaker ones, in such a way that those redshifts are still dependable. Those redshifts obtained by using only one (predominantly [O\,II]\,3727\AA) or two spectral lines of poor signal-to-noise were classified as type 3. In total, 55\% of our galaxies were assigned type 1 redshifts, 23\% have type 2 redshifts, and the remaining 22\% have type 3 redshifts.

In order to increase the total number of galaxies with spectroscopic redshifts, a search for Luminous Red Galaxies and spectroscopic targets from the DR7 SDSS was performed, increasing our spectroscopic sample by 47, adding an average of 1 and 3 galaxies per field respectively (without considering HE2149$-$2745A which is outside the SDSS area). Of these, approximately 10\% had spectroscopy from both our Gemini data and the SDSS database yielding redshift agreements of $\sim$ 10$^{-4}$ even for our less reliable class 3 redshifts. We also searched for objects in the NASA/IPAC Extragalactic Database adding 420 more redshifts to our data. The total number of galaxies with redshifts from all available sources is 43 from SDSS, 420 from NED, and 383 from our Gemini survey.

In even numbered Figures between \ref{camp022300} and \ref{camp231958}, we show a snapshot of each 5.5$\arcmin \times$ 5.5$\arcmin$ pre-image field observed with GMOS. Targets have been marked with a circle and are identified with a number. Moreover a small snapshot of each field is shown in the lower parts of these figures, indicating the center positions of the galaxy cluster/group candidates lying at $d <$ 2 $h_{71}^{-1}$Mpc from the quasar sight-lines. Their center positions are shown in circles indicating a physical radius of 0.5 $h_{71}^{-1}$Mpc from the cluster center positions. The redshift histograms (odd numbered Figures \ref{his022300}--\ref{his231958}) show all available redshifts for each field (see more in $\S$3.2). The vertical lines in these plots indicate the absorption redshifts $z_{abs}$ and the photometric redshifts of galaxy cluster/group candidates present in the fields. Tables \ref{tab022300}--\ref{tab231958} specify the number and redshift of each target, the redshift reliability classifier, and magnitudes.

\bigskip

\section{RESULTS}

\medskip

\subsection{Mg\,II Absorbing Galaxies}

We define an absorbing galaxy as the closest galaxy to the LOS, observed with GMOS as having a redshift within typical galactic stellar velocity dispersions ($<$ 300 km/s) from a \mgii\ absorption system  \citep[see e.g.,][]{sd1994, {lb1997}}. Previous surveys have found absorption galaxies at a few times $10$ kpc. Our selection of targets for spectroscopic observations primarily focused on galaxies residing at impact parameters $\rho <$ 150 $h_{71}^{-1}$kpc from the LOS and having $R_{c}$-band magnitudes brighter than $\sim$ 23.5 in GMOS pre-images. 

\begin{figure*}[ht!]
\begin{center}
\includegraphics[scale=0.55]{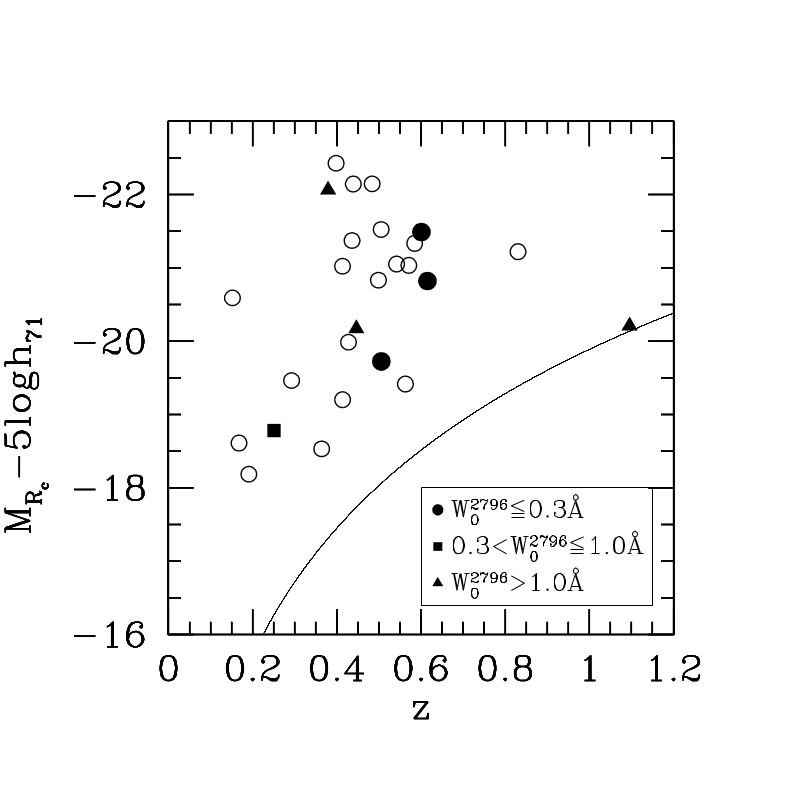}
\caption{$R_{c}$-band absolute magnitude distribution 
with redshift. Galaxies responsible for weak \mgii\ 
absorptions ($W_{0}^{2796} \leq$ 0.3 \AA) are shown as 
filled circles; those producing absorptions of 
intermediate strength (0.3 $< W_{0}^{2796} \leq$ 1.0 \AA) 
are shown as filled squares; and the ones related 
to stronger absorptions with $W_{0}^{2796} >$ 1.0 \AA\ 
are shown as filled triangles. Interlopers 
found in our fields are shown in open circles. These are galaxies observed near the LOS that 
do not produce any \mgii\ absorption detected in the spectra 
of the quasars. Absorbing 
galaxies in the field centered on HE2149$-$2745A do 
not appear in this plot since there is no available 
SDSS nor RCS1 photometry for them. The solid line 
shows a magnitude limit of $R_{c} =$ 23.5.} 
\label{absolute_magnitudes}
\end{center} 
\end{figure*}

The exposure time (see $\S$2.2) and bright wings of the quasar spatial profile restricted the search for \mgii\ absorbing galaxies to objects typically brighter than $M_{R_{c}} \sim -$20 (at $z \sim$ 0.6) and at impact parameters $\rho \gtrsim$ 2$\arcsec$ equivalent to $\sim$ 8 $h_{71}^{-1}$kpc at $z =$ 0.25 and 16 $h_{71}^{-1}$kpc from the LOS at $z =$ 1. 

Table \ref{absorbing_galaxies} summarizes the \mgii\ absorbing galaxies detected in our sample. The table lists the absorption systems of each LOS; their rest-frame equivalent widths $W_{0}^{2796}$ [\AA] and errors $\sigma_{W_{0}^{2796}}$ [\AA], the redshift at which the \mgii\ absorbing galaxy was found $z_{gal}$, its impact parameter to the LOS $\rho$ [$h_{71}^{-1}$kpc], if the absorption was considered a photometric hit in this work, the spectral features used to determine its redshift; and the absolute magnitude $M_{R_{c}}$ corrected for k-dimming according to \citet{fs1995}. These $k$-corrections have been computed by matching the observed galaxy colors and  those for a wide range of galaxies SED at the measured redshift. Absorbing galaxies in the field centered on HE2149$-$2745A lack RCS1 and SDSS photometry, and as such no absolute magnitudes can be determined for them. 

\begin{figure*}[ht!]
\begin{center}
\includegraphics[scale=0.45]{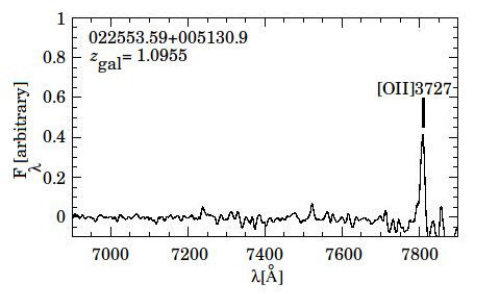}
 \includegraphics[scale=0.34]{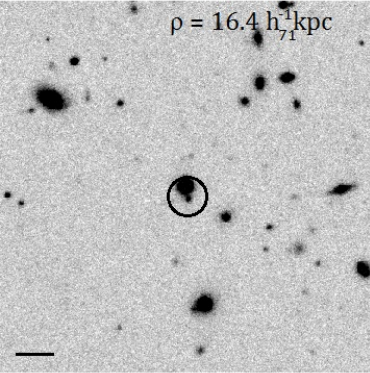}
\vspace{0.25cm}
\includegraphics[scale=0.45]{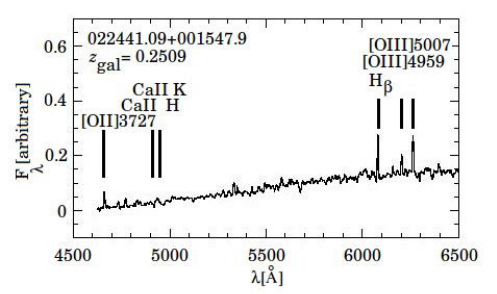}
 \includegraphics[scale=0.34]{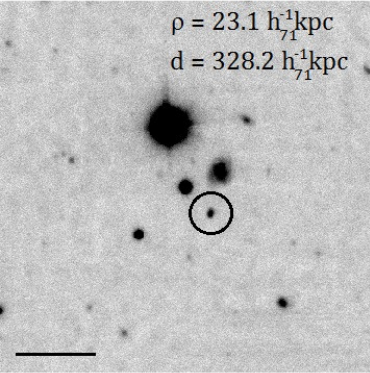}
\vspace{0.25cm}
\includegraphics[scale=0.45]{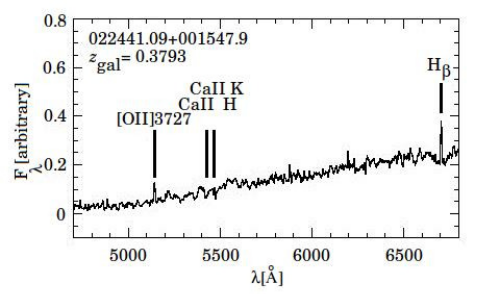}
\includegraphics[scale=0.34]{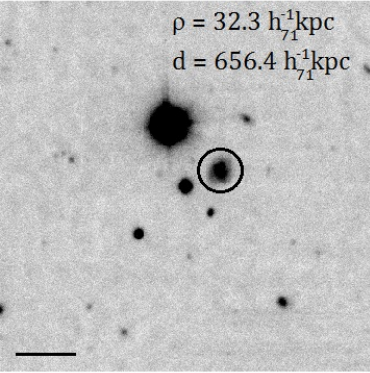}
\includegraphics[scale=0.45]{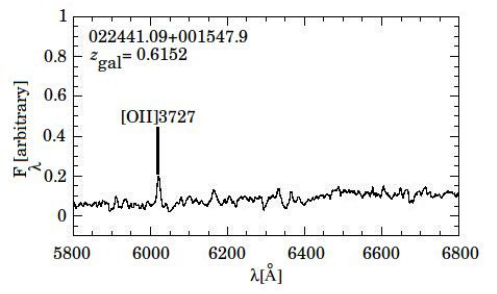}
\includegraphics[scale=0.34]{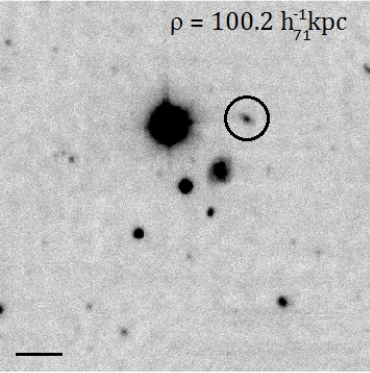}
\vspace{0.25cm}
\includegraphics[scale=0.45]{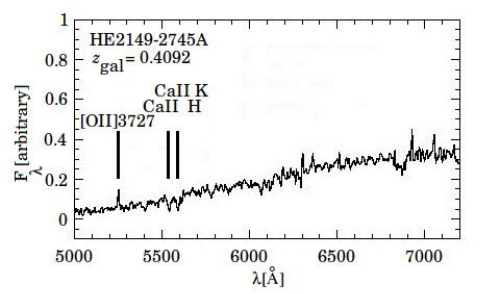}
\includegraphics[scale=0.34]{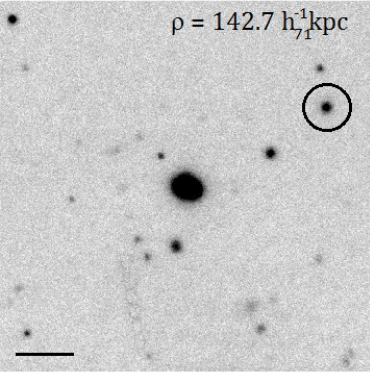}
\includegraphics[scale=0.45]{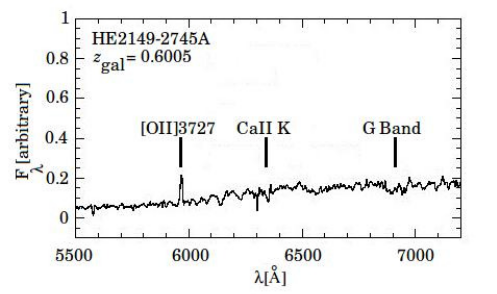}
\includegraphics[scale=0.34]{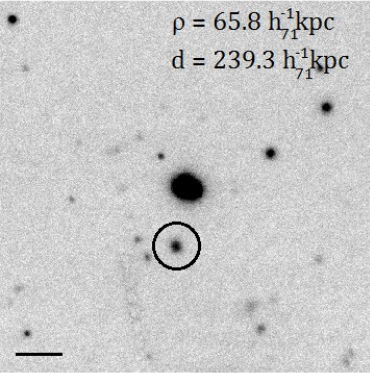}
\vspace{0.25cm}
\includegraphics[scale=0.45]{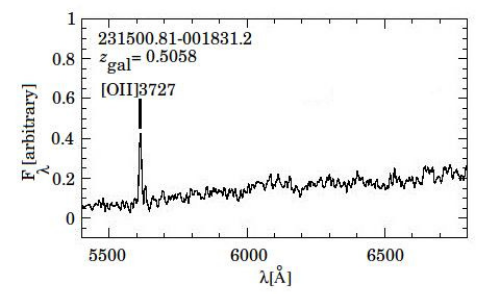}
\includegraphics[scale=0.34]{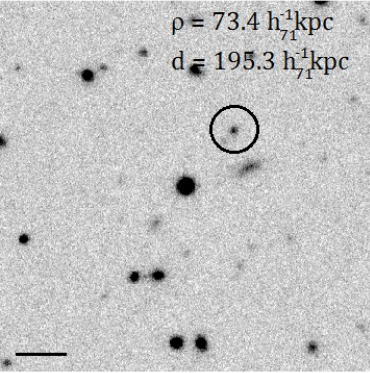}
\includegraphics[scale=0.45]{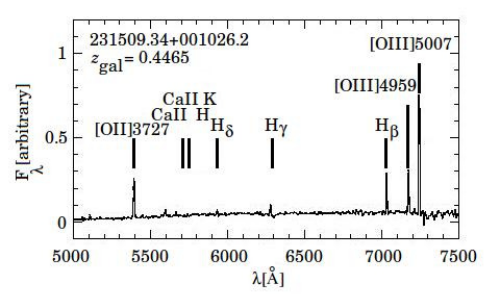}
\includegraphics[scale=0.34]{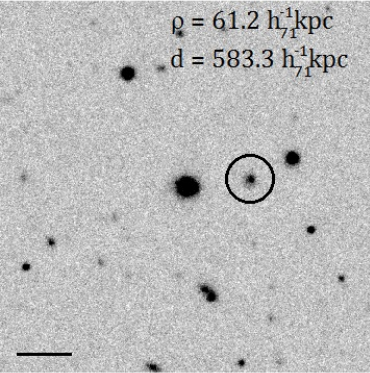}
\vspace{0.25cm}
\includegraphics[scale=0.45]{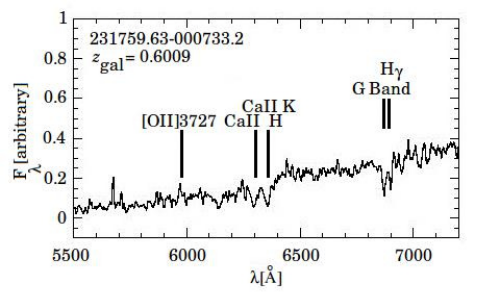}
\includegraphics[scale=0.34]{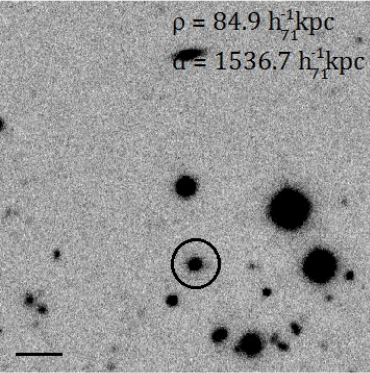}\\
\caption{Spectra of 9 absorbing galaxies confirmed in our sample 
together with a snapshot of the field centered on the respective quasar in 
a window 60$\arcsec$ wide. 
All 
spectra have been smoothed with a boxcar of 5 pixels. In each image, the black line 
represents a physical distance of 50 $h_{71}^{-1}$kpc at $z_{gal}$. Each absorbing 
galaxy is enclosed in a circle, and their impact parameters to 
the LOS $\rho$ [$h_{71}^{-1}$kpc] 
are given in the upper right corner of the image. In those fields 
where the absorption is classified 
as a photometric hit by Paper I, the projected distance 
between the cluster/group candidate center and the LOS 
$d$ [$h_{71}^{-1}$kpc] is also specified. In the case of the LOS to the quasar 
231500.81$-$001831.2, only the impact parameter to 
the closest cluster/group candidate is shown.}
\label{spectra}

\end{center} 
\hspace{0.3cm}
\end{figure*}

Out of a total of 24 absorption systems, \mgii\ absorbing galaxies were identified in 10 cases: 9 with our  spectroscopic search and 1 from the literature, leading to a success rate of 41.7\%. These galaxies have redshifts between 0.2509 $\leq z_{gal} \leq$ 1.0955 ($\overline{z} =$ 0.55). Detection of galaxies at $z >$ 1 was unlikely because the typical [O\,II]\,3\,727 \AA\ emission feature reaches $\lambda >$ 7\,500 \AA, a region of the spectra contaminated by fringing. The one high-$z$ galaxy we detected, a very bright source with $M_{R_{c}} = -$20.21 (see Figures \ref{spectra} and \ref{camp022553}), shows a clear [O\,II]\,3727\AA\ emission feature in its spectrum. A characterization of these absorbing galaxies is detailed in section 3.4.

The median velocity difference $\delta v_{gal} \equiv c(z_{gal} - z_{abs})/( 1 + z )$ is $\sim -$50 km/s spanning a range between $\sim -$280 and 57 km/s, and being consistent with galactic kinematics \citep[e.g.,][]{sk2002}. Despite these small velocity differences indicating genuine matches, we cannot exclude absorption from galaxies below our detection limit. 

The low success rate results from observational design. Slit lengths are an important constraint when trying to observe all objects within a certain region using multi-object spectrographs. This, combined with low S/N sources and fringing, did not allow us to obtain redshifts for all objects within 150 $h_{71}^{-1}$kpc from the LOS. Consequently, the missing absorbing galaxies may: lie at lower impact parameters to the LOS, be hidden behind bright/large foreground objects within the region (see for example Figures \ref{camp022441} and \ref{camp022839}), or are too faint for reliable detection. Further and deeper spectroscopic observations must be planned to identify more absorbing systems and improve the number of matches to galaxy counterparts.

Figure \ref{absolute_magnitudes} presents the $R_{c}$-band absolute magnitude versus redshift for our absorbing galaxies and interlopers, i.e., galaxies near the LOS that do not produce any \mgii\ absorption detectable in the quasar spectra. The survey is complete down to $M^*$ ($z=0$) for galaxies to $z < 0.7$.

Our absorbing galaxies have absolute magnitudes spanning a range between $M_{R_{c}} -$ 5$\log h_{71} = -$18.78 ($z_{gal} =$ 0.2509, $W_{0}^{2796} =$ 0.732 \AA) to $-$22.07 ($z_{gal} =$ 0.3793, $W_{0}^{2796} =$ 1.181 \AA), with a median value of $M_{R_{c}} = -$20.21 comparable to $M^{\ast}$ of present day galaxies \citep{b2001}. Moreover, interlopers show similar magnitude ranges and are also comparable to $M^{\ast}$ galaxies. Thus no difference in brightness can be distinguished between \mgii\ absorbing galaxies and interlopers. The median impact parameter of our ten absorbing galaxies is 63.5 $h_{71}^{-1}$kpc. The highest redshift \mgii\ absorbing galaxy in our sample ($z_{gal} =$ 1.0955, $W_{0}^{2796} =$ 1.685 \AA) has an impact parameter at the minimum distance probed by our spectroscopic campaign (16.4 $h_{71}^{-1}$kpc), while the largest impact parameter, 142.7 $h_{71}^{-1}$kpc ($z_{gal} =$ 0.4092, $W_{0}^{2796} =$ 0.228 \AA), is approximately 43 km/s away from the absorption redshift. 

Spectra of the absorbing galaxies are shown in Figure \ref{spectra}. We do not show the galaxy at $z_{gal} =$ 0.6030 responsible for the weak absorption of $W_{0}^{2796} =$ 0.015 \AA\ seen in the spectrum of the gravitationally lensed quasar HE2149$-$2745A \citep{wk1996}. This galaxy, taken from the literature, appears to be the lensing galaxy of the system as published in \citet{ec2007}. Despite no \mgii\ absorption was reported in that case, Paper I did detect an absorption with $W_{0}^{2796} =$ 0.015 \AA (the weakest in our sample) by analyzing a high resolution spectrum of the QSO. 

Three out of the ten galaxies show only emission lines, and more than half have Ca\,II\,K, Ca\,II\,H and G band absorption transitions among their spectral features. With the exception of the lensing galaxy, a common feature within our absorbing galaxies is the presence of the [O\,II]\,3727\AA\ emission line, denoting some level of recent star formation activity (see Table \ref{absorbing_galaxies}). Complementary to this, the mean ($R_{c} - z'$) color of our galaxies is 0.57, typical of S$_{bc}$-S$_{cd}$ galaxies at $\overline{z} =$ 0.55 \citep{fs1995}. 

\bigskip

\subsection{Galaxy Clusters}

The second goal of this work is to reduce the photometric redshift uncertainty of RCS1 cluster group/candidates ($\delta z_{clus}^{phot} =$ 0.1) by confirming them spectroscopically. This will allow us to properly establish the possible connection between \mgii\ absorbers and galaxy overdensities. 

As mentioned before, the sample contains 31 galaxy cluster/group candidates from which 30 are from the RCS1 and 1 is taken from the literature. The RCS1 galaxy cluster/group candidates were detected with a significance $>$ 3$\sigma_{RCS}$ and are mostly poor: 27 have richness parameters $B_{gc}$ $<$ 800 and only 3 have 800 $< B_{gc} <$ 1100 \citep{yl1999}, corresponding to clusters with Abell 0--1 richness classes \citep{yl1999}.

The confirmation algorithm we adopted is divided in three steps: 1) detect overdensities in redshift space, 2) associate galaxy cluster candidates (i.e., those without spectroscopic redshifts) with observed redshift overdensities, and 3) define cluster membership to estimate cluster redshifts $z_{clus}^{spec}$ and rest-frame velocity dispersions $\sigma_{v}$ [km/s].

We detect redshift overdensities by looking at redshift histograms for each field. Odd numbered Figures \ref{his022300}--\ref{his231958} show redshift histograms, using all redshift classes, with a bin size of 0.005. At $z =$ 0.173 and $z =$ 1.032 (the minimum and maximum photometric redshifts of our cluster/group candidate sample) this bin translates to velocity bin widths of $\sim$ 1\,280 and 740 km/s respectively. 

To define the extent of the redshift overdensities, we focus on $\pm$ 1\,000 km/s, centered on each redshift peak identified in the redshift histograms, with a rest-frame bin width of 250 km/s. We concentrated specifically on peaks near the photometric redshifts of RCS1 cluster/group candidates. Since our survey strategy was designed to spectroscopically confirm an average of $\sim$ 3 clusters per field, and ---as a result of the spectroscopic mask design procedure--- we observed $\sim$ 50 objects per field, then a more conservative approach had to be adopted when stating the considerable number of members to obtain a reliable estimate of $z_{clus}^{spec}$. Consequently, peaks of more than 15 members were considered sufficient to identify a cluster of galaxies. 


\begin{figure*}[ht!]
\begin{center}
\includegraphics[scale=0.55]{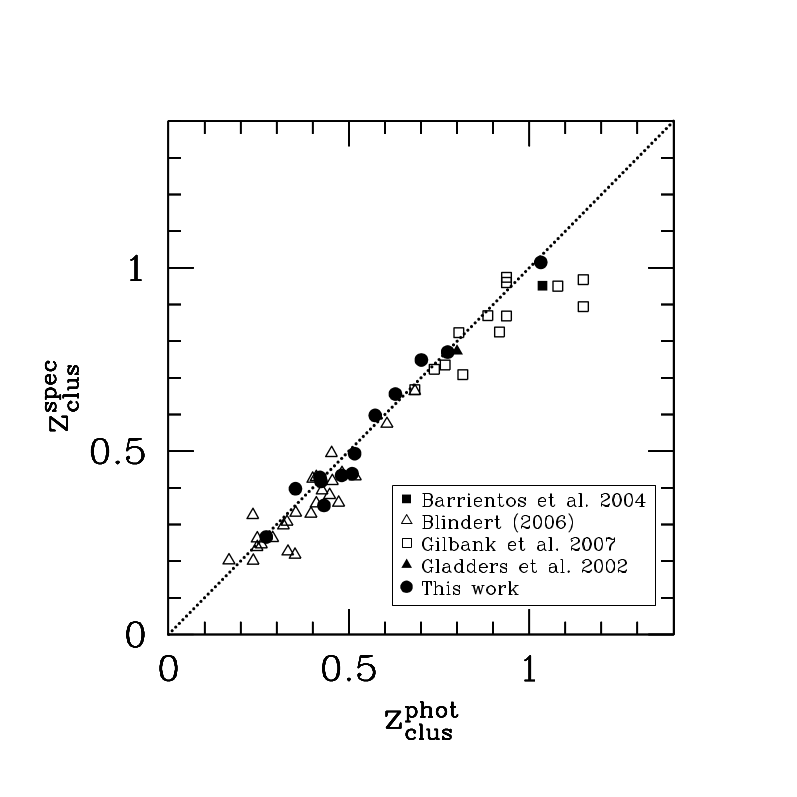}
\caption{Comparison of RCS1 cluster photometric redshifts and the 
spectroscopic redshifts we obtained in 
this work (filled circles), together with the data of \citet{bg2004} 
(filled squares), \citet{b2006} 
(open triangles), \citet{gy2007} 
(open squares) and \citet{gy2002} 
(filled triangles). The dotted line represents the one-to-one 
relation. Our absolute redshift differences span the range $\delta z \leq$ 0.08, with 
a mean 
value of $\overline{\delta z} =$ 0.03.}
\label{zcomp}
\end{center} 
\end{figure*}

The association of cluster/group candidates with observed redshift overdensities was established once the redshift of central galaxies was obtained, i.e. $i)$ early type galaxies located at projected distances $d_{clus} <$ 500 $h_{71}^{-1}$kpc from the cluster/group candidate center coordinates given by the RCS1, $ii)$ with similar redshifts ($| \delta v | <$ 1\,000 km/s) to the galaxy cluster/group candidate photometric redshifts and within their photometric redshift uncertainty ($\delta z_{clus}^{phot} =$ 0.1), and $iii)$ showing colors and magnitudes following a red sequence in ($R_{c} - z'$) vs $z'$, built by including all extended sources at $d_{clus} <$ 500 $h_{71}^{-1}$kpc from the cluster center as defined by the RCS1 photometric object catalog \citep{gy2000}. To obtain these color magnitude relations, we performed rough color cuts in the color magnitude diagrams, we fitted a linear regression ($R_{c} - z'$) $= a_{0}z' + a_{1}$, and we obtained the standard uncertainties of the constants $\sigma_{a_{0}}$ and $\sigma_{a_{1}}$.

The conditions described above are satisfied in most cases where RCS1 cluster/group centers reside within the pre-image field-of-view. However, as $\sim$ 32\% of our cluster/group candidate centers fall outside, a different approach had to be applied in order to establish the association between the RCS1 candidates and the redshift overdensities. For these cases, at least two of the following conditions had to be fulfilled: $i)$ galaxies residing at a cluster-centric projected distance $d_{clus} <$ 2 $h_{71}^{-1}$Mpc, $ii)$ having similar redshifts ($| \delta v | <$ 1\,000 km/s) near the galaxy cluster/group candidate photometric redshifts and within their photometric redshift uncertainty ($\delta z_{clus}^{phot} =$ 0.1), and/or $iii)$ falling at less than $\pm$ 3$\sigma_{a_{1}}$ from the color magnitude relation that includes all photometric extended sources at $d_{clus} <$ 500 $h_{71}^{-1}$kpc from the cluster/group candidate center, taking into account, at large cluster-centric distances, that galaxies tend to be morphologically different and have bluer colors \citep{d1980}. In the last criterion, to mitigate against fore- and background contamination, we additionally require that first two conditions are also satisfied.

One special case is the field centered on HE2149$-$2745A. Here, the overdensity found at $z \sim$ 0.6 (see histogram in Figure \ref{hishe2149}) is consistent with the galaxy cluster detection of M06 and W06, who detected a red-sequence at a photometric redshift $z_{clus}^{phot} =$ 0.590 and that was then spectroscopically confirmed at $z_{clus}^{spec} =$ 0.6030. Additionally, we find two more overdensities in the field at $z \sim$ 0.2768 and $z \sim$ 0.7395. Whilst the former has already been identified by M06, the latter is reported here for the first time due to the depth of our observations. 

To assign cluster members, we followed the procedure of \citet{f1996} and \citet{b2006}, where galaxies at $| \delta v | \leq$ 4000 km/s from each redshift peak (for all redshift flags) are subjected to an interloper rejection scheme using both galaxy angular position and cluster-centric radial velocity. More specifically, this technique utilizes overlapping and shifting cluster-centric distance bins of size $r_{gap}$ (or larger) so that each bin contains at least $n_{bin}$ galaxies. For each bin, a velocity fixed-gap rejection scheme is applied to discard galaxies at $\geq v_{gap}$ km/s from their neighbors. Here, the values of $n_{bin}$, $r_{gap}$ and $v_{gap}$ were chosen to be 10, 0.5 $h_{71}^{-1}$Mpc and 1000 km/s respectively, motivated by the small number of input redshifts per cluster and small range of cluster richness. 

Cluster redshifts, $z_{clus}^{spec}$, and velocity dispersions $\sigma_{v}$ [km/s] were determined with biweight estimators of location and scale from \citet{bf1990}. Uncertainties in $\sigma_{v}$ were obtained using the jackknife method, and were corrected to the rest-frame. 

Table \ref{galaxy_clusters} provides a summary of the results concerning the confirmation of RCS1 cluster/group candidates in our sample. All 31 cluster/group candidates lying at an impact parameter $d <$ 2 $h_{71}^{-1}$Mpc from each LOS are listed, as well as their photometric redshifts $z_{clus}^{phot}$, spectroscopic redshifts $z_{clus}^{spec}$ and rest-frame velocity dispersion estimates $\sigma_{v}$ [km/s] (both measured irrespective of redshift reliability flag). The final column includes comments about each galaxy cluster confirmation. 

From a total of 31 cluster/group candidates 20 were spectroscopically confirmed spanning a redshift range 0.2659 $\leq z_{clus}^{spec} \leq$ 1.0152. Of these, 10 have a significant number of members (N $\geq$ 15). Among clusters confirmed with fewer members, 7 are based on the detection of central galaxies (see comments in Table \ref{galaxy_clusters}). For these, cluster redshift estimates are considered to be quantitatively reliable, while velocity dispersions should be considered as qualitative estimates only. Out of the final three confirmed clusters, two (in the field 231500.8$-$001831.2) were considered blended at the same redshift  as it is not possible to separate them in photometric redshift and distance to the QSO LOS (see below). The remainder has a significant number of member galaxies but the center lies out of the field.

Our spectroscopic survey proved more effective in detecting galaxies (and hence, galaxy overdensities) at $z \lesssim$ 0.7. Out of 21 low-$z$ galaxy cluster/group candidates ($z_{clus}^{phot} \lesssim$ 0.7) we recovered 18, whilst from a total of 10 high-$z$ galaxy cluster/group candidates, we were able to confirm only 2, thanks to the redshifts retrieved from NED. 

The redshifts of the confirmed clusters can be compared against those estimated photometrically, as shown in Figure \ref{zcomp}. Also shown are the RCS1 galaxy cluster redshifts calculated in \citet{bg2004}, \citet{b2006}, \citet{gy2007} and \citet{gy2002}. We exclude all cluster redshifts found in the field 231500.81$-$001831.2 because we were unable to confirm the different cluster candidates individually. The resulting redshift differences $\delta z = | z_{clus}^{phot} - z_{clus}^{spec} |$ range between 0.004 $\lesssim \delta z \leq$ 0.079 for our confirmed clusters with N $\geq$ 15 members, and $\delta z \lesssim$ 0.070 for those with N $<$ 15; both following the one-to-one relation (dotted line) and within typical $\delta z$ values obtained in studies at similar redshift ranges \citep{b2006, gy2002}. 

Taking into account past investigations shown in Figure \ref{zcomp}, and data we present here (including our confirmations), we find an average redshift offset of $\delta z /(1 + z_{clus}^{spec}) = $ 0.036 $\pm$ 0.032  \citep{wi2001, gy2007}. This estimate implies $\delta z \sim$ 0.06 at $z =$ 0.5--0.7, reinforcing the empirical difficulty in confirming the overlapping clusters in the LOS 231500.81$-$001831.2 and 231958.70$-$002449.3 fields. 

We also estimate cluster masses $M_{200}$ and radii $r_{200}$ by using the virial theorem \citep{cy1997}. The mean virial radius $r_{200}$ of our clusters confirmed with N $\geq$ 15 members is $\overline{r_{200}} =$ 1.57 $\pm$ 0.58 $h_{71}^{-1}$Mpc and the mass enclosed within it is $\overline{M_{200}} =  6.92  \pm  4.15 \times 10 ^{14}$\hmsol. These values should be considered qualitatively. Thus, our sample of confirmed clusters appears to have mean masses consistent with $\overline{B_{gc}}$ typical of clusters of low-intermediate mass \citep{ye2003}. 

Comments on the individual cluster detections can be found on the Appendix.

\begin{figure*}[ht!]
\begin{center}
\includegraphics[scale=0.55]{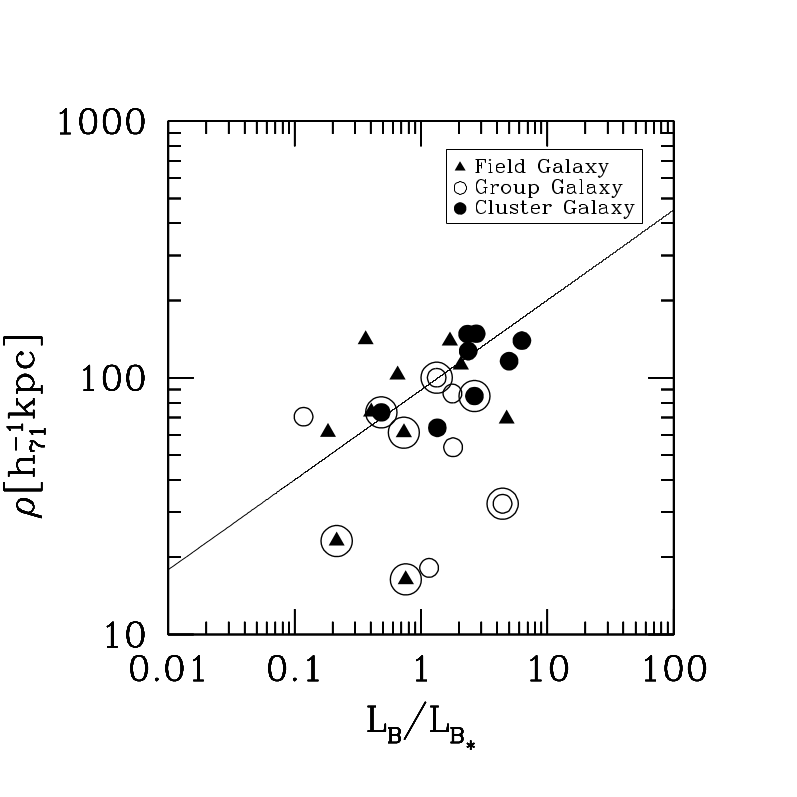}
\caption{Observed impact parameter versus $B$-band luminosity 
for our absorbing galaxies
(enclosed in a larger open circle) and interlopers (not enclosed 
in a larger open circle), associated to the 
field (filled triangles), galaxy groups (open circles) and galaxy 
clusters (closed circles). The line plotted is the Holmberg relation 
determined by \citet{ct2008} 
with $R^{\ast} \sim$ 89 $h_{71}^{-1}$kpc and 
$\beta =$ 0.35.}
\label{impact_parameter}
\end{center} 
\end{figure*}

\begin{figure*}[ht!]
\begin{center}
\includegraphics[scale=0.44]{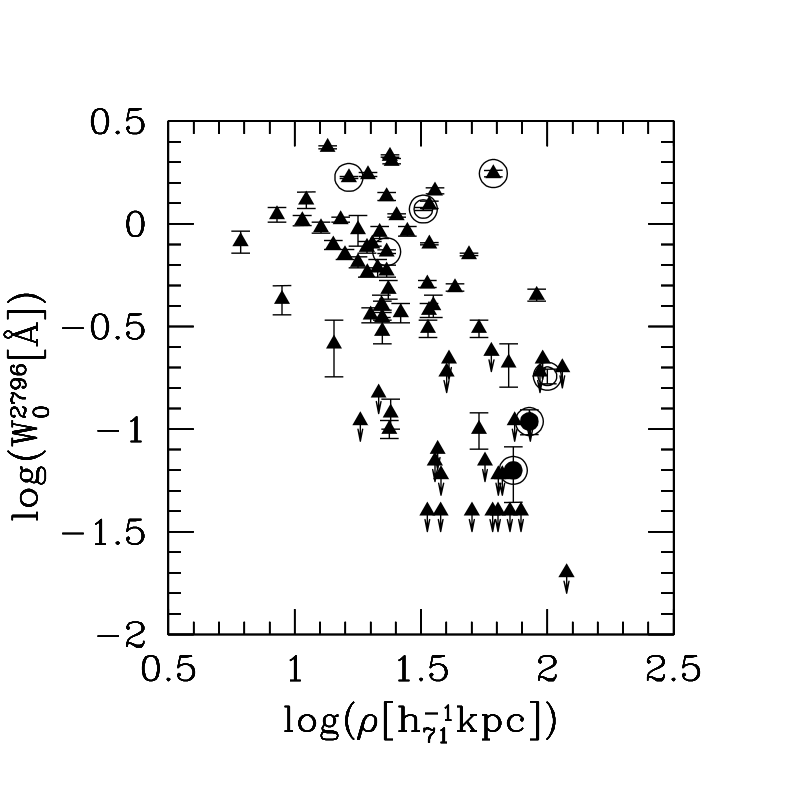}
\includegraphics[scale=0.44]{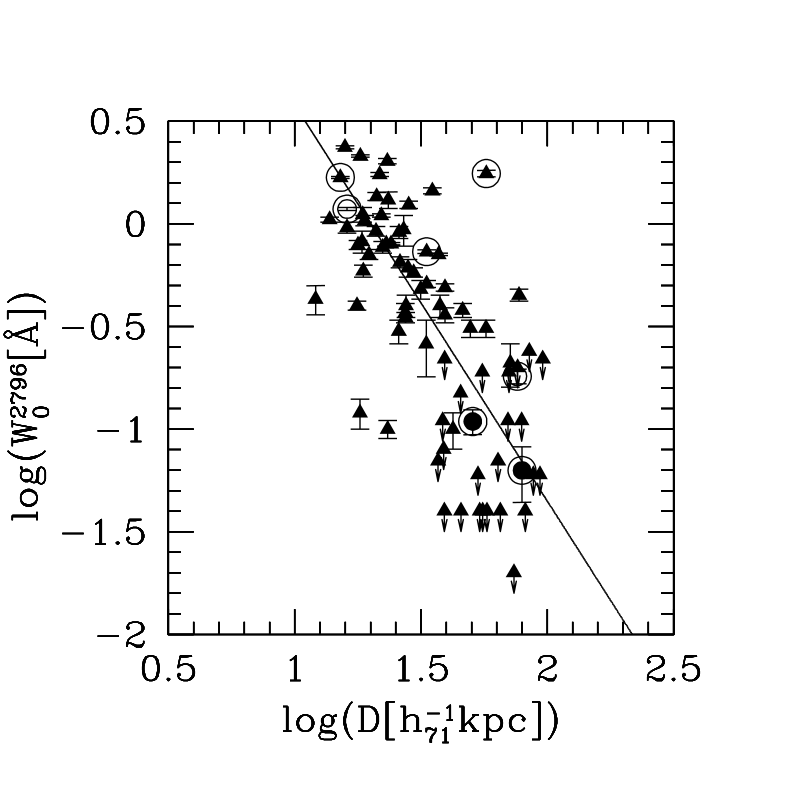}\\
\caption{{\it Left}: Absorption rest-frame equivalent width $W_{0}^{2796}$ [\AA] versus 
observed galaxy impact parameter to the LOS 
of our absorbing galaxies
(enclosed in a larger open circle) associated to the 
field (filled triangles), galaxy groups (open circles) and galaxy 
clusters (closed circles). Also the absorbing field-galaxies
published in the work of \citet{ch2010} 
are shown as a reference (filled triangles). {\it Right}: 
Absorption rest-frame equivalent width $W_{0}^{2796}$ [\AA] 
versus the impact parameter weighted by the $B$-band 
luminosity of the galaxies D $\equiv \rho \times (L_{B}/L_{B^{\ast}})^{-0.35}$ 
[$h_{71}^{-1}$kpc]. The plot shows our data (enclosed by a larger open circle) 
together with the absorbing field-galaxies of \citet{ch2010}. Again the 
symbols are the same as shown in the 
left panel. We include the anticorrelation 
found by \citet{ch2010} 
for absorbing field-galaxies (solid line).}
\label{equivalent_width}
\end{center} 
\end{figure*}

\subsection{Spectroscopic Hits}

\medskip

Combining the results from $\S$3.1 and $\S$3.2, we are now able to correlate \mgii\ absorbers to cluster/group environments, i.e., confirm photometric hits.

According to our definition in $\S$1, a spectroscopic hit is detected whenever either a \mgii\ absorption or an absorbing galaxy is found to reside in a spectroscopically confirmed galaxy cluster/group environment. More specifically, when $z_{abs}$ (or $z_{gal}$, if possible) falls within $\pm$ 1000 km/s of $z_{clus}^{spec}$, the confirmed cluster/group redshift.

Despite in some cases either being unable to confirm RCS1 cluster/group candidates or not expecting to find any, we did detect an agglomeration of few galaxies (N $<$ 10) with similar redshifts ($| \delta v | \lesssim$ 1000 km/s) separated at projected distances $\lesssim$ 0.5 $h_{71}^{-1}$Mpc from each other. We consider these kind of systems as groups of galaxies; absorbers inhabiting at such redshifts and closeness in space are identified as group absorbers: spectroscopic hits associated to groups instead of clusters of galaxies. 

We emphasize this distinction between group and cluster absorbers since galaxies in clusters may be affected by a history of extreme events, while the haloes of galaxies in our newly defined groups of galaxies may experience less aggressive (ongoing) interactions. Thus, it is important to distinguish them from those galaxies that, according to our follow-up, appear completely isolated.

All spectroscopic hits detected in our sample are listed in Table \ref{spectroscopic_hits}. The table shows the cluster/group redshift $z_{clus}^{spec}$, rest-frame velocity dispersion $\sigma_{v}$ [km/s], rest-frame velocity difference $\delta v_{gal}$ [km/s] (between the redshift of the \mgii\ absorbing galaxy $z_{gal}$ and the absorption redshift $z_{abs}$), and the rest-frame velocity difference $\delta v_{clus}$ [km/s] (between the cluster redshift $z_{clus}^{spec}$ and the absorber at $z_{abs}$ or absorbing galaxy at $z_{gal}$, if found). 

From a total of 24 absorption systems, only 14 are considered as photo-hits. From these 14 \mgii-cluster pair candidates, 8 are confirmed as spectroscopic hits associated to galaxy clusters, and 2 to galaxy groups. This makes a total of 10 confirmations out of 14, corresponding to a $\sim$ 71\% success rate in spectroscopically confirming photometric hits as \mgii\ absorbers in clusters/groups of galaxies. Additionally, 2 new spectroscopic hits were not classified as photometric hits as they were not matched to RCS1 clusters near $z_{abs}$. Based on our spectroscopic survey of galaxies, we hypothesize they are likely related to galaxy groups. Of the 12 absorption systems associated to galaxy clusters/groups, 10 are found at $z < 0.7$.

From the four photo-hits that could not be confirmed, we find two of those are at the redshift boundaries of the survey, one at $z=0.2507$ and the other one at $z=0.95$ (where the RCS1 cluster catalog is incomplete). The other two remaining could not be assigned clusters. If we considered these last two absorbers as contaminants in our \mgii -cluster pair study, then we have a contamination of 29\% (or 43\% if the group absorbers are not considered as spectroscopic hits). Similarly, if we exclude the photo-hits in redshift boundaries were RCS1 is highly incomplete, the contamination is only 17\% (33\% if the group absorbers are not considered hits).  



The median redshift of \mgii\ absorbers associated to galaxy cluster environments was $z_{med} =$ 0.60, and for those in galaxy group environments, $z_{med} =$ 0.62. The mean velocity difference of the whole sample of spectroscopic hits $\delta v_{clus}$ is $-$22.46 km/s and spans a range between $+$468.16 and $-$656.25 km/s, consistent with typical cluster velocity dispersions.

The 10 absorbers in Table \ref{spectroscopic_hits} that are not considered as photometric hits exhibit a wide redshift distribution, with four at $z > 1$, where the RCS1 cluster catalog is highly incomplete, and assignment therefore unreliable. 

\bigskip

\subsection{Characterization of Mg\,II Absorbers Associated with Cluster Galaxies }

Several studies have investigated the so-called "standard model of absorbers" established by the work of \citet{st1995}. According to this model, the gaseous extent of a galaxy can be described by a Holmberg luminosity scaling relation of the form $R(L) = R^{\ast}(L/L^{\ast})^{\beta}$. In the sample of \citet{st1995}, all \mgii\ absorbing galaxies fell below the relation, while non-absorbing galaxies fell above the relation. This result led \citet{st1995} to conclude that all normal galaxies with luminosities $L >$ 0.05$L_{K}^{\ast}$ should have \mgii\ gaseous haloes. However, other studies \citep{csk2004} have found otherwise, due to the presence of interlopers and misidentified absorbing galaxies within the original sample used by \citet{st1995}. 

Figure \ref{impact_parameter} shows the observed impact parameter versus $B$-band luminosity for our absorbing galaxies (enclosed in a large open circle) and interlopers, associated to the field (filled triangles), galaxy groups (open circles) and galaxy clusters (filled circles). $B$-band magnitudes were obtained by converting SDSS photometry with the transformation given in \citet{wb1991}. We show only 7 absorbing galaxies because we do not have photometry for the 3 absorbing galaxies found in the field centered on the LOS to HE2149$-$2745A. 

We adopted the Holmberg relation determined by \citet{ct2008} with $R^{\ast} \sim$ 89 $h_{71}^{-1}$kpc and $\beta =$ 0.35 for absorbing field-galaxies. Figure \ref{impact_parameter} shows that bright galaxies generally tend to have larger \mgii\ halo sizes than fainter ones. Our data (regardless of environment) do not follow the standard model since there are seven interlopers lying below the line indicating that the covering factor of \mgii\ haloes is less than unity, which can also be deduced from the large number of interlopers seen in our sample (see also Figure \ref{absolute_magnitudes}). However, in general, most absorbing galaxies confirmed in this work lie below the Holmberg relation derived in \citet{ct2008}, with interlopers falling above the line. Most importantly, all our absorbing galaxies confirmed in the field lie below the relation and most field interlopers lie above it. That our two absorbing cluster-galaxies lie above the relation may indicate the environment modifies the relation between galaxy luminosity and \mgii\ halo sizes.

Given the aforementioned contamination of \citet{st1995} data and the lack of additional information regarding their assignment to groups or clusters, it would be interesting to test the standard model with a larger sample of confirmed \mgii\ absorbers and interlopers inhabiting cluster/group/field environments.

Figure \ref{equivalent_width} (left panel) shows the rest-frame equivalent width of the \mgii\ absorptions against the impact parameter to the LOS of our absorbing galaxies (enclosed in a large open circle) confirmed in the field, groups of galaxies and clusters of galaxies (the symbols are the same as shown in Figure \ref{impact_parameter}). We also plot the sample of absorbing field-galaxies published in the work of \citet{ch2010} as a reference. From this figure we can see that regardless of their  environment, our sample of absorbing galaxies appears to follow the anti-correlation tendency found for absorbing galaxies in the field \citep[e.g.,][]{ch2010}. 

\citet{ch2010} found that the large scatter in this anti-correlation could be diminished when parametrizing the galaxy impact parameter $\rho$ and the $B$-band luminosity of absorbing field-galaxies in a scaling relation (D $= \rho \times (L_{B}/L_{B^{\ast}})^{-0.35}$ $h_{71}^{-1}$kpc). The right panel of Figure \ref{equivalent_width} shows the absorption rest-frame equivalent width $W_{0}^{2796}$ [\AA] versus the luminosity-weighted impact parameter D [$h_{71}^{-1}$kpc]. This plot shows our data (enclosed by a large open circle) compared to that of \citet{ch2010}. The symbols are the same as shown in the left panel. Of our three field points (filled triangles), one falls at $>$ 3$\sigma$ from the anti-correlation found by \citet{ch2010} (solid line). This outlier, from the 231509.34$+$001026.2 field, is the second strongest absorber in our sample($W_{0}^{2796} =$ 1.758 \AA), situated 61.19 $h_{71}^{-1}$kpc from the LOS, and appears to be the closest galaxy to the quasar sight-line up to the RCS1 limiting magnitude.

Our small sample of absorbing galaxies does not allow us to determine if a tight anti-correlation is valid for absorbing galaxies in clusters or groups of galaxies. When neglecting environment, they appear to follow the anti-correlation found for absorbing galaxies in the field (solid line). However a maximum likelihood method applied to a larger catalogue of absorbers would be useful to establish a specific scaling relation between $\rho$ [$h_{71}^{-1}$kpc] and galaxy luminosity for absorbing cluster/group-galaxies; the scaling relation found for absorbing field-galaxies (D $= \rho \times (L_{B}/L_{B^{\ast}})^{-0.35}$ [$h_{71}^{-1}$kpc]) does not necessarily hold for galaxies in clusters/groups. Applying this scaling relation could therefore be misleading when trying to define an analogous anti-correlation for absorbing cluster/group-galaxies.

Until a large sample of absorbing cluster/group-galaxies is available, the linear relation between both variables remains valid for isolated \mgii\ absorbing galaxies, yet remains unknown for those in clusters/groups.

\bigskip

\section{Discussion}

In our search for \mgii\ absorbers inhabiting dense environments, we detected 8 in clusters of galaxies and 2 in groups of galaxies, at a mean redshift $\overline{z} =$ 0.55. This corresponds to a success rate of 71\% in confirming spectroscopically photometric hits associated to clusters/groups of galaxies. Two additional \mgii\ absorbers have been associated to new groups identified in this study. 

We also confirmed \mgii\ absorbing galaxies from which 4 appear to belong to clusters of galaxies, 3 to our newly-defined groups of galaxies, and 3 to the field (i.e. isolated \mgii\ absorbing galaxies unconnected to groups of galaxies nearby).

Whilst the definition of absorbing galaxies we adopt is usually used in surveys searching for absorbing galaxies \citep[e.g.,][]{sd1994}, some caveats must be stated about the reliability of our considerations. 

The absorbing galaxies we report here may not be the only ones at $z_{abs}$ and $\rho <$ 150 $h_{71}^{-1}$kpc from the quasar sight-lines, and as such some of them may be misidentifications. In such cases, we would be dealing with at least a pair or a group of galaxies near the LOS in the foreground of the respective quasar. This would add more candidates to our sample of \mgii -group pairs.

Instead of absorbers residing in galaxies, some may have been expelled from the galactic halo, thus exhibiting relatively large impact parameters ($\rho >$ 100 $h_{71}^{-1}$kpc). In these cases, we would be tracing warm-cold gas reservoirs in the intergalactic medium or intracluster medium should the absorber reside in a cluster/group of galaxies. Despite this, we are not particularly interested in studying specifically where the absorption takes place within a galaxy. We merely state the assignment of  the absorber to a specific overdensity in order to build a sample of confirmed cluster/group absorbers.

The results of our survey also have some redshift dependence that is strong for the confirmation of galaxy hosts, but mild (if any) for the confirmation of absorber-cluster pairs. From the original 24 absorbers in the 9 LOS presented here, one third are at $z > 0.7$, while for the confirmed hosts only 1 of 10 is at $z > 0.7$. On the other hand, 2 out of 12 confirmed cluster-absorber pairs are at $z > 0.7$, while we had 3 out of 14 candidates in the same redshift regime. This difference is not surprising since it is easier to confirm a cluster of galaxies using spectra from their bright members than to identify one fainter galaxy responsible for the \mgii\ absorption.    

\bigskip

\subsection{Are The Absorbing Galaxies Typical Cluster/Group/Field Galaxies?}

When comparing the restframe $(B - R)$ color of our identified absorbing galaxies to those given in  \citet{fs1995}, out of the 4 we confirmed we found one with a color consistent with an S$_{bc}$ galaxy ($z_{abs} =$ 0.6013, 
$W_{0}^{2796} =$ 0.109 \AA).  Another has a color consistent with an irregular/star-forming galaxy, and is also associated to a weak absorption system ($z_{gal} =$ 0.5058, $W_{0}^{2796} =$ 0.063 \AA). Since the field where the latter is located shows a clear overdensity of galaxies at its redshift, it could be an infalling galaxy yet to lose a large amount of gas. Finally, for the remaining two confirmed absorbing galaxies (near the LOS to HE2149$-$2745A) we do not have photometry, yet their spectra characteristics (see our Figure \ref{spectra} and see Figure 12 in \citealt{ec2007}) are consistent with relatively early type galaxies, with both hosting weak absorbers ($W_{0}^{2796} =$ 0.015 and 0.175 \AA). 

Thus in general all of these galaxies host weak absorbers, they appear as cluster galaxies undergoing some level of ongoing star formation at $z \sim$ 0.60 (as shown by the presence of [O\,II]\,3\,727 \AA\ emission feature in their spectra); and lie at projected distances lower than 250 $h_{71}^{-1}$kpc from the cluster centre coordinates, with the exception of the absorbing galaxy associated to the cluster RCS231755$-$0011.3 where the projected distance is $\sim$ 1.6 $h_{71}^{-1}$Mpc.

On the other hand, two of our absorbing group-members appear disk-like and irregular, respectively producing absorptions of $W_{0}^{2796} =$ 1.181 \AA\ and 0.181 \AA\ (see Table \ref{absorbing_galaxies}); unlike absorbing galaxies in clusters they do not appear deficient in gas. Unfortunately we lack $B$-band photometry to secure magnitudes for the other absorbing galaxy found in a group of galaxies (at $z_{gal} =$ 0.4092), though this target similarly shows properties consistent with those of a star forming galaxy (see Figure \ref{spectra}). 

Moreover, our small sample of isolated absorbing galaxies are bright and have a color consistent with disk galaxies. They all produce strong absorptions between $W_{0}^{2796} =$ 0.732 \AA\ and 1.758 \AA, and appear as typical \mgii\ absorbing galaxies found in the field, as supported by previous works \citep{zm2007, sd1994, la1996, csk2005, kc2005, kc2007, ch2010, kmc2010}.

The fact that absorbing group-galaxies and field-galaxies are indistinguishable from those found in RCS1 clusters may reflect the  inclusion of small groups in our field sample. This potential cross-contamination must be investigated once further analysis is performed on those systems.

For now we can conclude, independent of the environment, absorbing galaxies reside in, that our catalogue heterogeneously samples the population of absorbing galaxies and is consistent with past findings. However, in considering environmental effects we find a dichotomy between our absorbing cluster-galaxies and those confirmed in the field. While the former show generally weak  ($W_{0}^{2796}$ $<$ 0.3 \AA) absorptions and consistently with early type galaxies, the latter produce strong absorptions and are consistent with bright star forming galaxies typically encountered in the field.

Although relations between \mgii\ absorption strength and galactic halo masses \citep{bm2006, gc2009} or galaxy type \citep{zm2007} have already been studied, these works do not probe rest-frame equivalent widths as weak as those found in Paper I and used in this work. 

\bigskip

\subsection{Implications For Papers I and II}

\medskip

The sample originally presented in Paper I and studied here includes a total of 11 photometric hits, so defined at $z <$ 0.9 with absorption systems having rest-frame equivalent widths $W_{0}^{2796} >$ 0.05 \AA.  Table \ref{spectroscopic_hits} shows we recovered 8 of these as spectroscopic hits, suggesting a 73 \% efficiency in the program. A closer look at these results finds that of 3 strong photometric hits ($W_{0}^{2796} >$ 1.0 \AA), 2 are recovered as spectroscopic hits, i.e. 2/3 of the strong \mgii\ system cluster member candidates genuinely reside in clusters. 

In Paper I we defined the weak systems as those having $W_{0}^{2796} <$ 1.0 \AA, and these show a $dN/dz$ in clusters consistent with what is expected for the field. Now, in order to deal with the small number statistics we classify weak systems as those having $W_{0}^{2796} <$ 0.3 \AA, and strong systems for $W_{0}^{2796} >$ 0.3 \AA. We see that of the 6 photometric hits for strong systems, we recover only 3 as spectroscopic hits, whereas for the weak systems we recover 5 out of 5. 

Although the overall numbers are consistent with expectations from Paper  I, their proposed interpretation is not supported by these new data. Here, we find that all weak absorbers are associated to clusters while this is true only for a fraction of the strong systems. The interpretation of the statistics in Paper I is that galaxies falling into the potential wells of clusters/groups loose their external shells where most of the weak systems are produced, and only the strong systems survive the interaction with the cluster. Here we see that all the weak systems are associated with clusters and only some of the strong systems. Despite dealing with small sample sizes, the results are at odds.  

Regarding the analysis of halo sizes performed in Paper II, our sample of photometric hits confirmed in clusters of galaxies (for which field and group absorbers should be considered contaminants) shows that, in general, Paper II  accounts for field contamination (their Table 1) in a manner consistent with our findings here (28\%), particularly when considering cluster-centric impact parameters between [0.5,1.0] $h_{71}^{-1}$Mpc. In this regime their field absorber fraction agrees with our findings when we also include photometric hits that could not be confirmed in the field or were unclassified. Therefore, the results of Paper II on baryonic halo sizes of gas in RCS1 clusters remain valid.

All these numbers can be increased by including additional systems identified within this study. For instance, a new analysis of the LOS to the quasar HE2149$-$2745A indicates we should have included the absorption system at $z_{abs} = 0.4464$ as a photometric hit, since M06 and W06 detected a red-sequence at a photometric redshift of $z^{phot}_{clus} = 0.40$, later confirmed spectroscopically as $z_{clus}^{spec} =$ 0.4465. By including this system, the statistics remain similar.

\bigskip

\section{Summary of Results}

\medskip

We have performed Gemini multi-object spectroscopy of fields around 9 distant background quasars, in whose spectra \mgii\ absorptions had previously been detected with rest-frame equivalent widths between 0.015 $\leq W_{0}^{2796} \leq$ 2.028 \AA. Our project sought to build a sample of \mgii\ absorbers in dense environments such as clusters or groups of galaxies. To achieve this, we use our spectroscopic survey to search for absorbing galaxies near the LOS and confirm RCS1 cluster/group candidates within the 9 fields ($d <$ 2 $h_{71}^{-1}$Mpc from the LOS) in order to analyze the environment where the \mgii\ absorbers and their galactic counterparts are located. In this study we include the analysis of spectra for 846 galaxies, 43 obtained from the SDSS database, 420 from NED and 383 from our own GMOS observations. The highest priority for our spectroscopic observations was to detect the absorbing galaxies and then to detect the cluster galaxies.  A total of 39 slits were devoted to identify the absorbing galaxies.
 
The main results of this work are summarized below:
 
Regarding our primary objective of identifying the absorbing galaxy:

\begin{itemize}
\item The identification of 10 out of 24 absorbing galaxies with redshifts between $0.2509 \leq z_{gal} \leq 1.0955$, up to an impact parameter of  $142 h_{71}^{-1}$ kpc and a maximum velocity difference of  $280$ km/s.

\item Our small sample of isolated absorbing galaxies appear as bright disk galaxies, exhibiting strong absorptions, consistent with typical \mgii\ absorbing galaxies in the field, as stated by previous works.

\item Regardless of the environment, we have found a heterogeneous sample of absorbing galaxies, in agreement with previous searches published in the literature.  

\end{itemize}

Regarding our objective of confirming the galaxy clusters at a maximum impact parameter of 2 Mpc from the QSO LOS:

\begin{itemize}
\item From a sample of 31 cluster/group candidates, we confirmed 20 spanning a redshift range 
0.2659 $\leq z_{spec} \leq$ 1.0152 with a mean mass consistent with $\overline{B_{gc}}$ typical of low-intermediate mass clusters. Non-confirmations of galaxy clusters/groups occurred mainly because we prioritized the detection of absorbing galaxies over the confirmation of galaxy cluster candidates.
\end{itemize}

Regarding the confirmation of our photometric hits and therefore confirming that the absorber galaxy belongs to the galaxy cluster:  
 
\begin{itemize}

\item We successfully determined the environment within which 12 of the absorbers reside: 8 inhabit clusters of galaxies and 4 appear in galaxy groups. Absorbers in this bound systems appear to lie between $+$468.16 and $-$656.25 km/s away from their cluster/group spectroscopic redshifts. This high success rate lends support to the results of paper I.

\item We applied our results to determine the degree of contamination due to field absorbers within the sample of \mgii\ absorbers matched to galaxy cluster/group environments. After spectroscopically confirming the systems, we found a contamination fraction of $\sim$ 17--29\% due to field absorbers, increasing to 33--46\% if we also classify group absorbers as contaminants. 

\item We did not confirm any cluster of galaxies in fields where no candidate at the absorber redshift was expected to be present.

\item We found two group absorbers where no cluster of galaxies had been previously detected within the RCS1 survey. Hence, including these absorbers adds two more non-isolated absorbers to our sample. 

\end{itemize}

To conclude, we prove the sample of photometric hits from Paper I is robust. Even though field absorbers may contaminate the sample (mostly due to misidentification of low redshift clusters, and instances where two absorptions fall within the photometric redshift uncertainty of  the same cluster), they have a minimal impact on the sample of photometric hits in Paper I . However for future work they should be taken into account.

\bigskip

\section{Acknowledgments}

We would like to thank Franz Bauer, C\'edric Ledoux and David Murphy for important suggestions included in the manuscript. SL, LFB, PL and NP were partly supported by the Chilean Centro de Astrof\'isica FONDAP No. 15010003. HA was supported by Proyecto CONICYT-ALMA 31070001 and Beca Mag\'ister CONICYT; SL was supported by FONDECYT grant No. 1100214; LFB was supported by FONDECYT grants No. 1085286 and 1120676; NDP was also supported by a FONDECYT grant No. 1110328; and IL was supported by MECESUP. The RCS project is supported by grants to H. Y. from the National Science and Engineering Research Council of Canada and the Canada Research Chair Program. This research has made use of the NASA/IPAC Extragalactic Database (NED), which is operated by the Jet Propulsion Laboratory, California Institute of Technology, under contract with the National Aeronautics and Space Administration. Funding for the SDSS and SDSS-II has been provided by the Alfred P. Sloan Foundation, the Participating Institutions, the National Science Foundation, the US Department of Energy, the National Aeronautics and Space Administration, the Japanese Monbukagakusho, the Max-Planck Society and the Higher Education Funding Council for England. The SDSS web site is http://www.sdss.org/.

\appendix

\bigskip

\section{Special Remarks}

\medskip

\subsection{On Individual Clusters}

\medskip

{\it LOS 231500.81$-$001831.2}:  As seen in Table \ref{galaxy_clusters}, the field around the LOS 231500.81$-$001831.2 presents 6 RCS1 cluster/group candidates that are spectroscopically confirmed by two cluster redshifts, one at $z_{clus}^{spec} =$ 0.5040 and the other at $z_{clus}^{spec} =$ 0.5870. All these RCS1 cluster/group candidates present low richness values ($B_{gc}$ $<$ 600) and low detection significance ($<$ 3.3$\sigma_{RCS}$), excepting RCS231506$-$0018.1 and RCS231515$-$0015.8) indicating that these candidates are much less likely to be confirmed with limited follow-up spectroscopy \textit{per se}. Also, since this survey is constrained by the number of slits we can observe, and our first priority was to find the absorbing galaxies, then it was not possible to put slits near all the cluster centers (see Figure \ref{camp231500}). Moreover, all candidates appear superimposed in redshift (as seen in Figure \ref{his231500}), color and angular space (see Table \ref{tab231500} and Figure \ref{camp231500}), which did not allow us to distinguish them individually. We observe a concentration of galaxies between redshifts $z \sim$ 0.502 and 0.507, centered on a notorious peak at $z \sim$ 0.504 (see Figure \ref{his231500}). These galaxies have ($R_{c} - z'$) colors between 0.6 and 1.0, and occupy similar locations in the color magnitude diagrams built from considering objects at a projected distance $d_{clus} <$ 0.5 $h_{71}^{-1}$Mpc from each cluster center. Therefore, it is not easy to discriminate between the different candidates. This, however, it is not a major issue since we are mostly interested in determining the velocity difference between the cluster redshift estimate and the absorption redshift, in order to establish whether the absorber resides in a cluster environment or not, and here the presence of a galaxy cluster in terms of redshift overdensity, color and angular coverage is clear.

\medskip

{\it LOS 022441.09$+$001547.9}: Among the 3 low-$z$ cluster/group candidates that were not confirmed, one is at $z_{clus}^{phot} =$ 0.173 near the LOS 022441.09$+$001547.9. This galaxy cluster candidate may be a false positive detection as suggested by its low richness value $B_{gc}$ $=$ 167 (3.25$\sigma_{RCS}$), and the fact that no redshift overdensity is observed in Figure \ref{his022441}, and a color magnitude relation is barely detected (and heavily polluted by galaxies at $z \sim$ 0.3--0.4) considering all extended sources within $d_{clus} <$ 500 $h_{71}^{-1}$kpc from the cluster center (near object No. 14, see Figure \ref{camp022441}). A possible explanation for such misidentification relies on the fact that RCS1 red-sequence finding algorithm works as long as $R_{c}$ (6\,500 \AA) and $z'$ (9\,100 \AA) bands sample the 4\,000 \AA\ break, being more effective at isolating the red-sequence at $z \gtrsim$ 0.4 \citep{gy2000, gy2005}, where the 4\,000 \AA\ break is entering the $R_{c}$ band ($\sim$ 5\,600 \AA). 

\medskip

{\it LOS to 022441.09$+$001547.9 and LOS to 231958.70$-$002449.3)}: About the other two unconfirmed low-$z$ cluster/group candidates ($z_{clus}^{phot}$ $=$ 0.511, LOS to 022441.09$+$001547.9; $z_{clus}^{phot}$ $=$ 0.651, LOS to 231958.70$-$002449.3) though their richness ($B_{gc}$ $\sim$ 400) and detection significance values ($>$ 3.6$\sigma_{RCS}$) support their presence, the number of redshifts available for those fields did not allow a proper confirmation of the systems. This is mainly because we prioritized to observe the objects near the LOS, instead of observing the objects near the cluster center positions.

\medskip

{\it LOS to 022839.32$+$004623.0}: Among the 10 cluster/group candidates at $z_{clus}^{phot} \gtrsim$ 0.7, only two were confirmed (the two high-$z$ cluster/group candidates near the LOS to 022839.32$+$004623.0, see Table \ref{galaxy_clusters}) and  this was only possible due to the redshifts taken from NED database.

\medskip

{\it LOS 022441.09$+$001547.9}: The problem encountered with the low-$z$ cluster candidate in the field around 022441.09$+$001547.9, is also found when looking at the redshift histogram of the field 022300.41$+$005250.0. In this field, we can see a redshift overdensity at $z \sim$ 0.19 (see Figure \ref{his022300}), that it is considered as a galaxy cluster at a photometric redshift $z_{clus}^{phot} =$ 0.1973 by \citet{gs2002}. The RCS1 did not detect the presence of a significant red-sequence at such redshift, probably because of its finding algorithm. However it does not matter for the purpose of this work, since the photometric hits we expect to confirm are at $z >$ 0.2. 

\medskip

{\it RCS022829$+$0045.8}: \citet{g2005} found a group of galaxies at a spectroscopic redshift $z_{clus}^{spec} =$ 0.7700 that we detected at $z_{clus}^{spec} =$ 0.7702 as the cluster redshift of RCS022829$+$0045.8. Though center positions differ ($\sim$ 1.5 $h_{71}^{-1}$Mpc), the population of galaxies at such redshift is seen throughout the entire GMOS pre-image field-of-view. 

\bigskip

\subsection{On Confirmed Photometric Hits}

\medskip

{\it  LOS 231500.81$-$001831.2}: There are two photometric hits confirmed with the same cluster environment in the field around the LOS 231500.81$-$001831.2 (see Tables \ref{galaxy_clusters} and \ref{spectroscopic_hits}). As mentioned in $\S$3.2, not being able to confirm individually superimposed galaxy cluster/group candidates in a single field is not a relevant problem in this case, because we are interested in highlighting that the environment where the absorption process took place was not that of an isolated galaxy, but a galaxy subjected to extreme events happening in clusters or groups of galaxies.

It should be noticed that even if the RCS1 detected different color magnitude relations, the overlap in color and angular space of several galaxies could be due to the presence of merging events or a filamentary structure just along the LOS where warm gas produced the absorption. If this were the case, the \mgii\ absorptions would be related to a dense environment too.

\medskip

{\it LOS 231958.70$-$002449.3}: One of the photometric hits detected in a group of galaxies is observed in the field centered on the LOS 231958.70$-$002449.3. There are 5 cluster/group candidates that could not be distinguished individually based on redshift information only. This field is crowded with high-$z$ galaxies ($z \sim$ 0.7--0.8) quite red in color (($R_{c} - z'$) $>$ 1.0) and sufficiently faint ($z' >$ 20.5), so that these galaxies may have been overlooked in the target selection algorithm or were observed but the S/N were not enough to retrieve a redshift from their spectra. Since the RCS1 cluster finding method has proven to work quite well at $z \sim$ 0.7--0.8  \citep{gy2007}, then the pair of galaxies at $z_{clus}^{spec} =$ 0.8486 (No. 10 and 11, see Table \ref{tab231958} and Figure \ref{camp231958}) could be tracing the external regions of the cluster candidate RCS231944$-$0026.8 at $z_{clus}^{phot} =$ 0.8440 ($d =$ 1.9--2 $h_{71}^{-1}$Mpc). This galaxy cluster candidate happens to be near the \mgii\ absorption redshift at $z_{abs} =$ 0.8463. Therefore, this system is now classified as a spectroscopic hit most probably associated with a group of galaxies. The unidentified \mgii\ absorbing galaxy could be located at the outskirts of this cluster/group of galaxies.

\medskip

{\it LOS 022441.09$+$001547.9}: The other spectroscopic hit in a group of galaxies was detected in the field 022441.09$+$001547.9 at $z_{clus}^{spec} =$ 0.3813 (see Table \ref{spectroscopic_hits}). Here the RCS1 cluster/group candidate near the absorption redshift $z_{abs} =$ 0.3791 was the cluster RCS022443$+$0017.6. This one was finally confirmed based on color, redshift and angular space information at $z_{clus}^{spec} =$ 0.3518 with 17 members (see the redshift histogram in Figure \ref{his022441} and Table \ref{galaxy_clusters}). However, the absorbing galaxy appears to have a companion (see Figure \ref{camp022441}), both in the background $>$ 5\,000 km/s from the confirmed cluster.

\bigskip

\subsection{On Unconfirmed Photometric Hits}

\medskip

Out of the 4 unconfirmed photometric hits, one is at a high redshift ($z_{abs} >$ 0.9) and we were unable to classify it, mostly because our spectroscopic survey turned out to be much more effective in detecting galaxies at $z \lesssim$ 0.7, and not to such high redshifts.

Among the other 3 photometric hits reported in Paper I, one remains unconfirmed and two are no longer valid according to our spectroscopic survey of the fields.

\medskip

{\it LOS 022553.59$+$005130.9}: This field presents two \mgii\ absorption systems ($z_{abs} =$ 0.6821 and 0.7500) that were associated to the same cluster/group candidate RCS022558$+$0051.8 at $z_{clus}^{phot} =$ 0.701 $\pm$ 0.1 (see Table \ref{galaxy_clusters}). This degeneracy is now broken because the galaxy cluster candidate was confirmed at $z_{clus}^{spec} =$ 0.7489, becoming a spectroscopic hit with the absorption at $z_{abs} =$ 0.7500. Since, no other RCS1 galaxy cluster candidate is expected to be found near $z_{abs} =$ 0.6821, and no overdensity of galaxies is seen in redshift nor angular space at that absorption redshift, we conclude its designation as a photometric hit does not longer apply. However, the existence of a group of galaxies cannot be ruled out considering the spectroscopic depth of this survey. Thus, this photometric hit remains unconfirmed.

\medskip

{\it LOS to 022441.09$+$001547.9}: The photometric hit of the cluster RCS022436$+$0014.2 at $z_{clus}^{phot} =$ 0.173 with the absorption system at $z_{abs} =$ 0.2507 in the LOS to 022441.09$+$001547.9, is now considered as a (confirmed) field absorber since this cluster/group candidate is most likely a false detection. Only 3 galaxies were detected within the cluster candidate photometric redshift uncertainty ($z_{min} =$ 0.2240 and $z_{max} =$ 0.2730, see Paper I) at $\overline{z} \sim$ 0.2377, more than 3\,000 km/s from the absorber redshift. 

\medskip

{\it LOS to 231509.34$+$001026.2}: The last unconfirmed photometric hit is now classified as a (confirmed) field absorber in the LOS to 231509.34$+$001026.2. Here we confirmed the (only) RCS1 cluster/group candidate near the LOS at $z_{clus}^{spec} =$ 0.4282, $>$ 3\,000 km/s away from the absorbing galaxy (at $z_{gal} =$ 0.4465). No overdensity of galaxies was found at the absorber redshift. 

\bigskip

\subsection{On Absorption Systems Not Classified As Photometric Hits}

\medskip

From a total of 9 absorption systems that are not classified as photometric hits in Paper I, we found 2 related to groups of galaxies and none associated to clusters of galaxies, which is consistent with their classification as ``not photometric hits''.

One \mgii\ absorber found at a redshift consistent with that of a group of galaxies is in the LOS HE2149$-$2745A at $z_{abs} =$ 0.4090. Although no notorious redshift peak is found at this redshift (see Figure \ref{hishe2149}), the fact that it has a companion may be inducing some modification in its halo size. 

The other spectroscopic hit related to a group of galaxies is in the field around the LOS 022441.09$+$001547.9 at $z_{clus}^{spec} =$ 0.6127. Even though this group of galaxies is undetected by the RCS1 color-based cluster finding algorithm, it does not mean that such group does not exist, recalling that the common definition of a group of galaxies primarily involves a high number density contrast with respect to the field; while the usual definition of a cluster admits the presence of a color magnitude relation for the brightest cluster members due to their coeval evolution since early times.

Among the other 7 absorption systems, we did not find clusters or groups of galaxies at the absorption redshifts. Four of these absorption systems are at high redshift $z_{abs} >$ 0.9 and as mentioned before, galaxies at these redshifts were difficult to observe with our spectroscopic survey. The other 3 systems are at redshifts $z_{abs} <$ 0.6, where our survey would have been able to detect galaxy overdensities. In fact in some cases we detected groups of galaxies, but all of them quite far from the absorber redshifts. 

Therefore it is quite likely that these 7 systems are actually field absorbers consistent with their classification given in Paper I. In other words, this small sample is not contaminated by cluster absorbers. Nevertheless, there are 2 spectroscopic hits in groups that require further spectroscopy.

\newpage
\begin{deluxetable}{lccccccc}
\tablewidth{0pc}
\tablecaption{Fields studied in this work.}
\tabletypesize{\tiny}
\tablehead{\multicolumn{8}{c}{\textit{The Sample}} \\
\hline
\\[0.05pt]
\multicolumn{1}{c}{LOS} & \multicolumn{1}{c}{$z_{em}$\tablenotemark{a}} & 
\multicolumn{1}{c}{N$_{\rm clus}$\tablenotemark{b}} & \multicolumn{1}{c}{$z_{abs}$\tablenotemark{c}} & 
\multicolumn{1}{c}{$W_{0}^{2796}$[\AA]\tablenotemark{d}} & \multicolumn{1}{c}{$\sigma_{W_{0}^{2796}}$[\AA]} & 
\multicolumn{1}{c}{Photo-hit}\tablenotemark{e} &\multicolumn{1}{c}{Photo-hit I}\tablenotemark{f}}
\startdata
\\[0.1pt]
022300.41$+$005250.0 & 1.248 & 2 & 0.9500 & 0.043 & 0.010 & yes & no\\ 
022441.09$+$001547.9 & 1.201 & 5 & 0.2507 & 0.732 & 0.037 & yes & yes\\
 &  &  & 0.3791 & 1.181 & 0.043 & yes & yes\\
 &  &  & 0.6152 & 0.181 & 0.016 & no & no\\
 &  &  & 0.9402 & 0.080 & 0.020 & no & no\\
 &  &  & 1.0560 & 0.881 & 0.036 & no & no\\ 
022553.59$+$005130.9 & 1.815 & 5 & 0.6821 & 0.333 & 0.019 & yes & yes\\
 &  &  & 0.7500 & 0.159 & 0.015 & yes & yes\\
 &  &  & 1.0951 & 1.685 & 0.065 & no & no\\
 &  &  & 1.2258 & 0.177 & 0.032 & no & no\\
022839.32$+$004623.0 & 1.288 & 5 & 0.6548 & 0.597 & 0.016 & yes & yes\\ 
HE2149$-$2745A & 2.030 & 1 & 0.4090 & 0.228 & 0.008 & no & no \\
 &  &  & 0.4464 & 0.016 & 0.005 & yes$^{*}$ & no\\
 &  &  & 0.5144 & 0.028 & 0.003 & no & no\\
 &  &  & 0.6012 & 0.175 & 0.006 & yes & yes\\ 
 &  &  & 0.6032 & 0.015 & 0.004 & yes & no\\
 &  &  & 1.0189 & 0.219 & 0.013 & no & no\\
231500.81$-$001831.2 & 1.324 & 6 & 0.5043 & 0.148 & 0.009 & yes & yes\\
 &  &  & 0.5072 & 0.063 & 0.009 & yes & yes\\
231509.34$+$001026.2 & 0.848 & 1 & 0.4473 & 1.758 & 0.009 & yes & yes\\
231759.63$-$000733.2 & 1.148 & 1 & 0.6013 & 0.109 & 0.016 & yes & yes\\
231958.70$-$002449.3 & 1.891 & 5 & 0.4071 & 0.151 & 0.017 & no & no\\
 &  &  & 0.4158 & 0.192 & 0.021 & no & no\\ 
 &  &  & 0.8463 & 2.028 & 0.024 & yes & yes\\
\enddata
\tablenotetext{a}{Redshift of the QSO.}
\tablenotetext{b}{Number of RCS1 cluster/group candidates lying at $d <$ 2 $h_{71}^{-1}$Mpc from the LOS.}
\tablenotetext{c}{Absorption redshift with $\delta z_{abs} \sim$ 10$^{-4}$.}
\tablenotetext{d}{Rest-frame equivalent width determined in Paper I.}
\tablenotetext{e}{Absorption system that is considered as a photometric hit or not according to Tables 2 and 3 from Paper I, except for $^{*}$ which corresponds to a new photo-hit. These are used in the analysis of the present work.}
\tablenotetext{f}{Absorption system that is considered as a photometric hit used in the analysis of Paper I, that is those having $z < 0.9$ and $W_0^{2796} > 0.05$ \AA}
\label{absorption_systems}
\end{deluxetable}

\newpage
\begin{deluxetable}{lccc}
\tablewidth{0pc}
\tablecaption{Spectroscopic Targets observed with GMOS.}
\tabletypesize{\tiny}
\tablehead{\multicolumn{4}{c}{\textit{The Sample}} \\
\hline
\\[0.05pt]
\multicolumn{1}{c}{LOS} & \multicolumn{1}{c}{$N_{\rm tot}$\tablenotemark{a}} & 
\multicolumn{1}{c}{$N_{\rho\,<\,\rm 150\,\,\,h_{71}^{-1}kpc}$\tablenotemark{b}} & 
\multicolumn{1}{c}{$R_{\rm faint}$\tablenotemark{c}} }
\startdata
\\[0.1pt]
022300.41$+$005250.0 & 43 & 3 & 24.75 \\ 
022441.09$+$001547.9 & 51 & 8 & 23.64 \\
022553.59$+$005130.9 & 48 & 2 & 22.30 \\
022839.32$+$004623.0 & 50 & 3 & 23.00 \\ 
HE2149$-$2745A & 49 & 4 & \nodata \\ 
231500.81$-$001831.2 & 55 & 7 & 22.67 \\
231509.34$+$001026.2 & 49 & 6 & 22.91 \\
231759.63$-$000733.2 & 49 & 2 & 21.96 \\
231958.70$-$002449.3 & 46 & 4 & 25.73 \\ 
\enddata
\tablenotetext{a}{Number of GMOS targets per field. A total of 440 slits (439 objects) were observed with GMOS.}
\tablenotetext{b}{Number of GMOS targets at an impact parameter $\rho <$ 150 $h_{71}^{-1}$kpc from the LOS.}
\tablenotetext{c}{Magnitude of the faintest target at an impact parameter $\rho <$ 150 $h_{71}^{-1}$kpc from the LOS.}
\label{spectroscopy_targets}
\end{deluxetable}

\newpage
\begin{deluxetable}{p{1.0in}cccccccp{0.005in}p{1.5in}c}
\tablecaption{Confirmation of Absorbing Galaxies.}
\tabletypesize{\tiny}
\tablehead{\multicolumn{11}{c}{\textit{The Sample}} \\
\hline
\multicolumn{1}{c}{LOS} & 
\multicolumn{1}{c}{$z_{abs}$} & 
\multicolumn{1}{c}{$W_{0}^{2796}$ [\AA]} & 
\multicolumn{1}{c}{$\sigma_{W_{0}^{2796}}$ [\AA]} & 
\multicolumn{1}{c}{$z_{gal}$\tablenotemark{a}} & 
\multicolumn{1}{c}{$\sigma_{z_{gal}}$} &
\multicolumn{1}{c}{$\delta {v} $ [km s$^{-1}$]} &
\multicolumn{1}{c}{$\rho$ [$h_{71}^{-1}$kpc]\tablenotemark{b}} & 
\multicolumn{1}{c}{Photo-hit} & 
\multicolumn{1}{c}{Spectral Lines} & 
\multicolumn{1}{c}{$M_{R_{c}}$} }
\startdata
022300.41$+$005250.0 & 0.9500 & 0.043 & 0.010 &\nodata &\nodata &\nodata &\nodata & yes & \nodata & \nodata \\
022441.09$+$001547.9 & 0.2507 & 0.732 & 0.037 & 0.2509 & 0.0001 & $-$48   & 23.13 & yes & [O\,II]\,3727,Ca\,II\,K,Ca\,II\,H,H$_{\gamma}$,H$_{\beta}$, & $-$18.78 \\
  & & & & & & & & & [O\,III]\,4959,[O\,III]\,5007,[N\,II]\,6548--6583,H$_{\alpha}$,[S\,II]\,6716--6730 & \\
  & 0.3791 & 1.181 & 0.043 & 0.3793 & 0.0001 & $-$44 & 32.31 & yes & [O\,II]\,3727,Ca\,II\,K,Ca\,II\,H,H$_{\delta}$,H$_{\beta}$ & $-$22.07 \\
  & 0.6152 & 0.181 & 0.016 & 0.6152 & 0.0008 & 0 & 100.17 & no & [O\,II]\,3727,Ca\,II\,K,Ca\,II\,H & $-$20.82 \\ 
  & 0.9402 & 0.080 & 0.020 & \nodata & \nodata & \nodata & \nodata & no & \nodata & \nodata \\
  & 1.0560 & 0.881 & 0.036 & \nodata & \nodata & \nodata & \nodata & no & \nodata & \nodata \\
022553.59$+$005130.9 & 0.6821 & 0.333 & 0.019 & \nodata & \nodata & \nodata & \nodata & yes & \nodata & \nodata \\
  & 0.7500 & 0.159 & 0.015 & \nodata & \nodata & \nodata & \nodata & yes & \nodata & \nodata \\
  & 1.0951 & 1.685 & 0.065 & 1.0955 & 0.0008 & $-$57 & 16.38 & no & [O\,II]\,3727 & $-$20.21 \\ 
  & 1.2258 & 0.177 & 0.032 & \nodata & \nodata & \nodata & \nodata & no & \nodata & \nodata \\ 
022839.32$+$004623.0 & 0.6548 & 0.597 & 0.016 & \nodata & \nodata & \nodata & \nodata & yes & \nodata & \nodata \\ 
HE2149$-$2745A & 0.4090 & 0.228 & 0.008 & 0.4092 & 0.0001 & $-$43 & 142.71 & no & [O\,II]\,3727,Ca\,II\,K,Ca\,II\,H,G\,band & \nodata \\
  & 0.4464 & 0.016 & 0.005 & \nodata & \nodata & \nodata & \nodata & yes & \nodata & \nodata \\ 
  & 0.5144 & 0.028 & 0.003 & \nodata & \nodata & \nodata & \nodata & no & \nodata & \nodata \\ 
  & 0.6012 & 0.175 & 0.006 & 0.6005 & 0.0002 & 131 & 65.78 & yes & [O\,II]\,3727,Ca\,II\,K,Ca\,II\,H,G\,band & \nodata \\
  & 0.6032 & 0.015 & 0.004 & 0.6030 & 0.0010 & 37 & 5.72 & yes & Ca\,II\,K,Ca\,II\,H,G\,band & \nodata \\
  & 1.0189 & 0.219 & 0.013 & \nodata & \nodata & \nodata & \nodata & no & \nodata & \nodata \\
231500.81$-$001831.2 & 0.5043 & 0.148 & 0.009 & \nodata &\nodata & \nodata & \nodata & yes & \nodata & \nodata \\
  & 0.5072 & 0.063 & 0.009 & 0.5058 & 0.0008 & 279 & 73.36 & yes & [O\,II]\,3727,Ca\,II\,K,Ca\,II\,H & $-$19.72 \\
231509.34$+$001026.2 & 0.4473 & 1.758 & 0.009 & 0.4465 & 0.0008 & 166 & 61.19 & yes & [O\,II]\,3727,H$_{\delta}$,H$_{\gamma}$,H$_{\beta}$,[O\,III]\,4959, & $-$20.17 \\ 
  & & & & & & & & & [O\,III]\,5007 & \\
231759.63$-$000733.2 & 0.6013 & 0.109 & 0.016 & 0.6009 & 0.0002 & 75 & 84.92 & yes & [O\,II]\,3727,Ca\,II\,K,Ca\,II\,H,G\,band & $-$21.49 \\ 
231958.70$-$002449.3 & 0.4071 & 0.151 & 0.017 & \nodata & \nodata & \nodata & \nodata & no & \nodata & \nodata \\ 
  & 0.4158 & 0.192 & 0.021 & \nodata & \nodata & \nodata & \nodata & no & \nodata & \nodata \\
  & 0.8463 & 2.028 & 0.024 & \nodata & \nodata & \nodata & \nodata & yes & \nodata & \nodata \\
\enddata
\tablecomments{The LOS to the QSO 022553.59$+$005130.9 also presents a high redshift 
absorption at $z_{abs} =$ 1.2258 as shown in Table \ref{absorption_systems}. However, at such high redshift no absorbing galaxy could 
have been detected due to problems encountered with our observations. Therefore in the following 
it is removed from our sample (see $\S$3.1). 
}
\tablenotetext{a}{Redshift of the absorbing galaxy.}
\tablenotetext{b}{Galaxy impact parameter to the LOS.}
\label{absorbing_galaxies}
\end{deluxetable}

\newpage
\begin{deluxetable}{llccccp{3.2in}}
\tablewidth{0pc}
\tablecaption{Confirmation of RCS1 Cluster/Group Candidates.}
\tabletypesize{\tiny}
\tablehead{\multicolumn{7}{c}{\textit{The Sample}} \\
\hline
\\[0.05pt]
\multicolumn{1}{c}{Field} & \multicolumn{1}{c}{Cluster/Group} & 
\multicolumn{1}{c}{$z_{clus}^{phot}$} & \multicolumn{1}{c}{$z_{clus}^{spec}$} & 
\multicolumn{1}{c}{$\sigma_{v}$ [km/s]} & \multicolumn{1}{c}{N\tablenotemark{a}} & 
\multicolumn{1}{c}{Comments}}
\startdata
\\[0.05pt]\\ 
022300.41$+$005250.0 & RCS022302$+$0052.9 & 0.509 & 0.4386 & 300 $\pm$ 90 & 8 & Central galaxies. There is a group at $z \sim$ 0.4740 but do not follow a red-sequence.\\ 
 & RCS022253$+$0055.1 & 0.939 & \nodata & \nodata & \nodata & \nodata \\
022441.09$+$001547.9 & RCS022436$+$0014.2 & 0.173 & \nodata & \nodata & \nodata & Possible false positive detection.\\
 & RCS022443$+$0017.6 & 0.431 & 0.3518 & 646 $\pm$ 93 & 17 & Central galaxies. \\
 & RCS022431$+$0018.0 & 0.480 & 0.4338 & 234 $\pm$ 73 & 8 & Central galaxies. Confirmation possible because of SDSS data.\\
 & RCS022454$+$0013.3 & 0.511 & \nodata & \nodata & \nodata & \nodata \\
 & RCS022449$+$0016.2 & 0.818 & \nodata & \nodata & \nodata & There is one early type galaxy at $\sim z_{clus}^{phot}$ at $d_{clus} \sim$ 1 $h_{71}^{-1}$Mpc from the cluster center.\\
022553.59$+$005130.9 & RCS022602$+$0055.5 & 0.352 & 0.3974 & 606 $\pm$ 40 & 31 & Center outside field-of-view. Galaxies are at $d_{clus} >$ 1 $h_{71}^{-1}$Mpc from the cluster center.\\
 & RCS022553$+$0052.5 & 0.423 & 0.4172 & 965 $\pm$ 38 & 25 & Central galaxies. Immersed in the background of galaxies at $z \sim$ 0.3956.\\
 & RCS022558$+$0051.8 & 0.701 & 0.7489 & 833 $\pm$ 48 & 5 & Central galaxies. Galaxies follow a red-sequence.\\
 & RCS022546$+$0050.0 & 0.873 & \nodata & \nodata & \nodata & \nodata \\
 & RCS022556$+$0052.7 & 0.928 & \nodata & \nodata & \nodata & \nodata \\
022839.32$+$004623.0 & RCS022841$+$0044.9 & 0.271 & 0.2659 & 944 $\pm$ 80 & 6 & Central galaxies. Confirmation possible because of SDSS data.\\
 & RCS022844$+$0047.7 & 0.516 & 0.4934 & 314 $\pm$ 49 & 5 & Central galaxies. \\
 & RCS022832$+$0046.5 & 0.629 & 0.6559 & 319 $\pm$ 67 & 19 & We found one galaxy at the cluster center.\\
 & RCS022829$+$0045.8 & 0.774 & 0.7702 & 560 $\pm$ 86 & 23 & Center outside field-of-view. These are at 1.5--2 $h_{71}^{-1}$Mpc from the cluster center.\\
 & RCS022828$+$0044.9 & 1.032 & 1.0152 & 742 $\pm$ 57 & 22 & Confirmation possible because of NED data. \\
HE2149$-$2745A & Group/Cluster & 0.603 & 0.6030 & 287 $\pm$ 69 & 9 & Central galaxies. In agreement with M06. According to W06 these galaxies follow a red-sequence.\\
231500.81$-$001831.2 & RCS231515$-$0015.8 & 0.496 & 0.5040 & 214 $\pm$ 58 & 16 & Center outside field-of-view.\\
 & RCS231512$-$0020.1 & 0.517 & 0.5040 & 214 $\pm$ 58 & 16 & Center outside field-of-view.\\
 & RCS231459$-$0018.9 & 0.522 & 0.5040 & 214 $\pm$ 58 & 16 & This is a preliminary result.\\
 & RCS231501$-$0013.6 & 0.557 & 0.5870 & 699 $\pm$ 42 & 14 & The groups at $z \sim$ 0.5040 and 0.5925 also could confirm this cluster.\\ 
 & RCS231506$-$0018.1 & 0.560 & 0.5040 & 214 $\pm$ 58 & 16 & We may have detected the central galaxy at $z =$ 0.5046.\\
 & RCS231515$-$0015.6 & 0.566 & 0.5870 & 699 $\pm$ 42 & 14 & The groups at $z \sim$ 0.5040 and 0.5925 also could confirm this cluster.\\
231509.34$+$001028.2 & RCS231510$+$0012.1 & 0.420 & 0.4282 & 735 $\pm$ 46 & 10 & Central Galaxies. There is a group of galaxies with $z =$ 0.3726 and 0.2 mag bluer in color.\\
231759.63$-$000733.2 & RCS231755$-$0011.3 & 0.573 & 0.5974 & 631 $\pm$ 25 & 12 & Center outside field-of-view. There is a group of galaxies at $z \sim$ 0.4730 but do not follow a red-sequence.\\
231958.70$-$002449.3 & RCS231947$-$0028.3 & 0.651 & \nodata & \nodata & \nodata & There is a group of galaxies at $z \sim$ 0.5614 near the cluster center.\\
 & RCS231958$-$0025.1 & 0.789 & \nodata & \nodata & \nodata & There are two galaxies at $z \sim$ 0.7196. We may have observed central galaxies. \\
 & RCS231958$-$0023.2 & 0.796 & \nodata & \nodata & \nodata & The groups at $z \sim$ 0.8092 or 0.7194 may also confirm it.\\
 & RCS231944$-$0027.0 & 0.805 & \nodata & \nodata & \nodata & There are two galaxies at $z \sim$ 0.8090. \\
 & RCS231944$-$0026.8 & 0.844 & \nodata & \nodata & \nodata & Center outside field-of-view. There are two galaxies at $z \sim$ 0.8486 and $>$ 1.5 $h_{71}^{-1}$Mpc from the LOS.\\
\enddata
\tablenotetext{a}{Number of galaxy cluster members.}
\tablecomments{The field centered on HE2149$-$2745A could not be studied photometrically.
However, according to M06 and W06 there appear to be photometrically and spectroscopically 
two clusters of galaxies at $z =$ 0.6030 and $z =$ 0.4465.} 
\label{galaxy_clusters}
\end{deluxetable}

\newpage
\begin{deluxetable}{cccccccccccccc}
\tablewidth{0.0pc}
\tablecaption{Detection of Spectroscopic Hits.}
\tabletypesize{\tiny}
\tablehead{\colhead{Field} & \colhead{$z_{abs}$} & 
\colhead{$W_{0}^{2796}$ [\AA]} & \colhead{$\sigma_{W_{0}^{2796}}$ [\AA]} & 
\colhead{$z_{gal}$} & \colhead{$\sigma_{z_{gal}}$} & \colhead{$\delta v_{gal}$ [km s$^{-1}$]} &
\colhead{$\rho$ [$h_{71}^{-1}$kpc]} & \colhead{Photo-hit} & \colhead{Photo-hit I} & \colhead{$z_{clus}^{spec}$} & 
\colhead{$\sigma_{v}$ [km/s]} & \colhead{N} & \colhead{$\delta v_{clus}$ [km/s]}}
\startdata
022300.41$+$005250.0 & 0.9500 & 0.043 & 0.010 & \nodata & \nodata & \nodata & \nodata & yes & no & \nodata & \nodata & \nodata & \nodata \\ 
022441.09$+$001547.9 & 0.2507\tablenotemark{a} & 0.732 & 0.037 & 0.2509 & 0.0001 & $-$48 & 23.13 & yes & yes & \nodata & \nodata & \nodata & \nodata \\
		   & 0.3791 & 1.181 & 0.043 & 0.3793 & 0.0001 & $-$44 & 32.31  & yes & yes & 0.3791$^{\ast}$ & \nodata\tablenotemark{c} & 2 & $-$43.50 \\
 		   & 0.6152 & 0.181 & 0.016 & 0.6152 & 0.0008 & 0 & 100.17 & no & no & 0.6127$^{\ast}$ & \nodata\tablenotemark{c} & 3 & $-$473.36 \\
		   & 0.9402 & 0.080 & 0.020 & \nodata & \nodata & \nodata & \nodata & no & no & \nodata & \nodata & \nodata & \nodata \\
		   & 1.0560 & 0.881 & 0.036 & \nodata & \nodata & \nodata & \nodata & no & no & \nodata & \nodata & \nodata & \nodata \\
022553.59$+$005130.9 & 0.6821 & 0.333 & 0.019 & \nodata & \nodata & \nodata & \nodata & yes & yes & \nodata & \nodata & \nodata & \nodata \\
 		   & 0.7500 & 0.159 & 0.015 & \nodata & \nodata & \nodata & \nodata & yes & yes & 0.7489 & 833 $\pm$ 48 & 5 & $-$188.63 \\
		   & 1.0951 & 1.685 & 0.065 & 1.0955 & 0.0008 & $-$57 & 16.38 & no & no & \nodata & \nodata & \nodata & \nodata \\
		   & 1.2258 & 0.177& 0.032 & \nodata & \nodata & \nodata & \nodata & no & no & \nodata & \nodata & \nodata & \nodata \\
022839.32$+$004623.0 & 0.6548 & 0.597 & 0.016 & \nodata & \nodata & \nodata & \nodata & yes & yes & 0.6559 & 319 $\pm$ 67 & 19 & $+$199.35 \\ 
HE2149$-$2745A	   & 0.4090 & 0.228 & 0.008 & 0.4092 & 0.0001 & $-$43 & 142.71 & no & no & 0.4095$^{\ast}$ & \nodata\tablenotemark{c} & 2 & $+$59.60 \\
		   & 0.4464 & 0.016 & 0.005 & \nodata & \nodata & \nodata & \nodata & yes\tablenotemark{b} & no & 0.4482 & 292 $\pm$ 32 & 5 & $+$373.11 \\
		   & 0.5144 & 0.028 & 0.003 & \nodata & \nodata & \nodata & \nodata & no & no & \nodata & \nodata & \nodata & \nodata \\
	           & 0.6012 & 0.175 & 0.006 & 0.6005 & 0.0002 & 131 & 65.78 & yes & yes & 0.6030 & 287 $\pm$ 69 & 9 & $+$468.24 \\ 
		   & 0.6032 & 0.015 & 0.004 & 0.6030 & 0.0010 & 37 & 5.72 & yes & no & 0.6030 & 287 $\pm$ 69 & 9 & $+$37.43 \\
		   & 1.0189 & 0.219 & 0.013 & \nodata & \nodata & \nodata & \nodata & no & no & \nodata & \nodata & \nodata & \nodata \\
231500.81$-$001831.2 & 0.5043 & 0.148 & 0.009 & \nodata & \nodata & \nodata & \nodata & yes & yes & 0.5040 & 214 $\pm$ 58 & 16 & $-$59.83 \\
                     & 0.5072 & 0.063 & 0.009 & 0.5058 & 0.0008 & 279 & 73.36 & yes & yes & 0.5040 & 214 $\pm$ 58 & 16 & $-$358.83 \\
231509.34$+$001026.2 & 0.4473 & 1.758 & 0.009 & 0.4465 & 0.0008 & 166 & 61.19 & yes & yes & \nodata & \nodata & \nodata & \nodata \\
231759.63$-$000733.2 & 0.6013 & 0.109 & 0.016 & 0.6009 & 0.0002 & 75 & 84.92 & yes & yes & 0.5974 & 631 $\pm$ 25 & 12 & $-$656.60 \\
231958.70$-$002449.3 & 0.4071 & 0.151 & 0.017 & \nodata & \nodata & \nodata & \nodata & no & no & \nodata & \nodata & \nodata & \nodata \\
		   & 0.4158 & 0.192 & 0.021 & \nodata & \nodata & \nodata & \nodata & no & no & \nodata & \nodata & \nodata & \nodata \\ 
                   & 0.8463 & 2.028 & 0.024 & \nodata & \nodata & \nodata & \nodata & yes & yes & 0.8486$^{\ast}$ & \nodata\tablenotemark{c} & 2 & $+$373.49 \\
\enddata
\tablecomments{All spectroscopic hits most probably associated 
with a galaxy group are denoted by $\ast$ symbols.}
\tablenotetext{a}{As mentioned in $\S$3.2, the galaxy cluster 
at $z_{clus}^{phot} =$ 0.173 in the field 022441.09$+$001547.9 (Table \ref{galaxy_clusters}) seems to be a misidentification 
and as such its association to the absorption at $z_{abs} =$ 0.2507 as 
a photometric hit is no longer valid. Only 3 galaxies were detected 
within the cluster candidate photometric 
redshift uncertainty ($z_{min} =$ 0.2240 and $z_{max} =$ 0.2730, see Paper I) at 
$\overline{z} \sim$ 0.2377 at a rest-frame velocity $>$ 3\,000 km/s from the absorber redshift.} 
\tablenotetext{b}{A red-sequence at $z_{clus}^{phot} =$ 0.40 was detected by W06 and later 
spectroscopically confirmed in M06 at $z_{clus}^{spec} =$ 0.4465. Here we consider it as a  spectroscopic hit associated with a cluster of 
galaxies, even though we did not detect a considerable overdensity expected at that 
redshift.}
\tablenotetext{c}{No velocity dispersion estimate is given in this case as there are only 
2--3 group members.}
\label{spectroscopic_hits}
\end{deluxetable}

\newpage
\begin{deluxetable}{ccccccccp{3.2in}}
\tablewidth{0pc}
\tablecaption{Spectroscopic Targets of Field Centered on 022300.41$+$005250.0.}
\tabletypesize{\tiny}
\tablehead{ \colhead{No.} & \colhead{RA(J2000)} & \colhead{DEC(J2000)} & 
\colhead{$z_{gal}$} & \colhead{$\sigma_{z_{gal}}$} & \colhead{Flag\tablenotemark{a}} & 
\colhead{$z'$} & \colhead{$R_{c}$ - $z'$} & \colhead{Comments}}
\startdata
1 & 02 22 56.77 & $+$00 54 57.56 & 0.43834 & 0.00007 & 1 & 20.26 & 0.81 & CaII H,CaII K \\
2 & 02 22 51.75 & $+$00 55 10.67 & 0.19624 & 0.00057 & 3 & 18.72 & 0.63 & MgI 5176,Na D 5892 \\
3 & 02 22 57.55 & $+$00 54 19.48 & 0.47211 & 0.00010 & 1 & 19.00 & 0.55 & [OII]3727,CaII H,CaII K,H$_{\beta}$ \\
4 & 02 23 03.31 & $+$00 54 01.04 & 0.41877 & 0.00012 & 2 & 21.72 & 0.50 & [OII]3727,H$_{\beta}$,[OIII]5007 \\
5 & 02 22 52.92 & $+$00 53 53.23 & 0.29379 & 0.00005 & 1 & 21.28 & 0.24 & H$_{\gamma}$,H$_{\beta}$,[OIII]4959,[OIII]5007 \\
6 & 02 22 59.48 & $+$00 53 40.52 & \nodata & \nodata & 0 & 20.72 & 1.84 & \nodata \\
7 & 02 22 59.34 & $+$00 53 34.01 & \nodata & \nodata & 0 & 21.51 & 1.54 & \nodata \\
8 & 02 23 02.07 & $+$00 53 15.97 & 0.43740 & 0.00007 & 1 & 20.18 & 0.83 & CaII H,CaII K \\
9 & 02 23 01.56 & $+$00 52 57.65 & \nodata & \nodata & 0 & 23.29 & 1.20 & \nodata \\
10 & 02 23 00.02 & $+$00 53 03.73 & 0.83062 & 0.00080 & 3 & 22.17 & 1.29 & [OII]3727 \\
11 & 02 23 00.18 & $+$00 52 44.47 & \nodata & \nodata & 0 & 22.76 & 0.49 & \nodata \\
12 & 02 22 58.99 & $+$00 52 30.83 & \nodata & \nodata & 0 & 20.16 & 1.46 & \nodata \\
13 & 02 23 09.71 & $+$00 51 55.22 & 0.70911 & 0.00012 & 1 & 21.21 & 1.00 & [OII]3727,CaII H,CaII K \\
14 & 02 22 59.61 & $+$00 52 05.38 & 0.24733 & 0.00005 & 1 & 21.63 & 0.15 & H$_{\beta}$,[OIII]4959,[OIII]5007,H$_{\alpha}$ \\
15 & 02 22 57.32 & $+$00 52 14.20 & 0.48234 & 0.00010 & 1 & 20.41 & 0.64 & [OII]3727,CaII H,H$_{\beta}$,[OIII]5007 \\
16 & 02 23 09.31 & $+$00 50 56.22 & 0.30679 & 0.00010 & 1 & 20.19 & 0.32 & CaII H,H$_{\beta}$,[OIII]4959,[OIII]5007 \\
17 & 02 23 08.63 & $+$00 50 31.06 & 0.82949 & 0.00085 & 2 & 20.80 & 1.28 & [OII]3727,CaII H \\
18 & 02 23 06.60 & $+$00 51 06.01 & 0.31065 & 0.00010 & 1 & 20.87 & 0.37 & [OII]3727,[SII]4068,[OIII]4959,[OIII]5007 \\
19 & 02 23 06.42 & $+$00 51 30.28 & 0.19584 & 0.00012 & 2 & 21.02 & 0.42 & H$_{\beta}$,[OIII]4959,[OIII]5007 \\
20 & 02 22 50.39 & $+$00 50 39.98 & 0.19500 & 0.00007 & 1 & 20.81 & 0.31 & H$_{\beta}$,[OIII]5007 \\
21 & 02 22 52.79 & $+$00 51 45.54 & 0.61172 & 0.00021 & 3 & 19.00 & 0.46 & H$_{\beta}$,[OIII]4959 \\
22 & 02 22 50.60 & $+$00 51 14.26 & 0.08511 & 0.00004 & 1 & 19.14 & 0.24 & HeI5875,[NII]6548,H$_{\alpha}$,[NII]6583,[SII]6716,[SII]6730 \\
23 & 02 22 56.34 & $+$00 50 15.50 & 0.19658 & 0.00006 & 1 & 20.88 & 0.29 & H$_{\beta}$,[OIII]5007,H$_{\alpha}$ \\
24 & 02 23 09.66 & $+$00 55 06.56 & 0.42789 & 0.00006 & 1 & 19.52 & 0.69 & CaII H,CaII K,G band \\
25 & 02 23 04.30 & $+$00 54 28.26 & 0.42024 & 0.00012 & 1 & 20.46 & 0.41 & [OII]3727,[OIII]4959,[OIII]5007 \\
26 & 02 23 03.39 & $+$00 54 49.18 & 0.47211 & 0.00009 & 1 & 19.79 & 0.79 & [OII]3727,CaII H,CaII K,H$_{\delta}$,G band \\
27 & 02 22 57.08 & $+$00 54 38.34 & 0.44022 & 0.00009 & 1 & 18.92 & 0.64 & [OII]3727,CaII H,CaII K,H$_{\beta}$,[OIII]5007 \\
28 & 02 22 50.15 & $+$00 54 19.26 & 0.19301 & 0.00042 & 2 & 18.37 & 0.59 & MgI 5176,Na D 5892 \\
29 & 02 22 54.17 & $+$00 55 21.11 & 0.38917 & 0.00012 & 1 & 20.20 & 0.64 & CaII H,CaII K,G band \\
30 & 02 23 06.04 & $+$00 53 52.08 & 0.12877 & 0.00003 & 1 & 17.93 & 0.58 & [NII]6548,[NII]6583,H$_{\alpha}$,[SII]6716 \\
31 & 02 23 09.74 & $+$00 53 17.81 & 0.12794 & 0.00004 & 1 & 18.35 & 0.58 & [OII]3727,H$_{\beta}$,[OIII]5007,[NII]6548,H$_{\alpha}$,[NII]6583,[SII]6716 \\
32 & 02 23 04.57 & $+$00 53 34.40 & 0.42039 & 0.00014 & 2 & 19.12 & 0.60 & [OII]3727,CaII K \\
33 & 02 23 04.60 & $+$00 52 01.31 & 0.47146 & 0.00006 & 1 & 20.32 & 0.87 & CaII H,CaII K,G band \\
34 & 02 23 02.57 & $+$00 52 32.66 & 0.43393 & 0.00012 & 1 & 19.65 & 0.81 & CaII H,CaII K,G band \\
35 & 02 23 02.00 & $+$00 53 03.30 & 0.43615 & 0.00012 & 1 & 19.97 & 0.81 & CaII H,CaII K,G band \\
36 & 02 23 01.78 & $+$00 52 44.69 & 0.43908 & 0.00012 & 1 & 19.16 & 0.87 & CaII H,CaII K,G band \\
37 & 02 22 52.30 & $+$00 52 19.52 & 0.12706 & 0.00057 & 2 & 16.44 & 0.59 & MgI 5176,Na D 5892 \\
38 & 02 23 00.43 & $+$00 51 36.79 & 0.43924 & 0.00012 & 1 & 19.83 & 0.85 & CaII H,CaII K,G band \\
39 & 02 22 53.21 & $+$00 51 20.84 & 0.19601 & 0.00049 & 3 & 18.33 & 0.61 & MgI 5176,Na D 5892 \\
40 & 02 22 59.56 & $+$00 51 03.46 & 0.43909 & 0.00005 & 1 & 20.19 & 0.61 & [OII]3727,CaII H,CaII K,H$_{\beta}$ \\
41 & 02 22 53.76 & $+$00 50 54.85 & 0.61774 & 0.00040 & 2 & 20.11 & 0.80 & [OII]3727,CaII H,CaII K \\
42 & 02 23 09.89 & $+$00 50 39.52 & 0.19435 & 0.00007 & 1 & 18.60 & 0.66 & CaII H,CaII K \\
43 & 02 23 08.45 & $+$00 50 22.31 & 0.47197 & 0.00007 & 2 & 20.47 & 0.82 & [OII]3727,CaII K \\
\enddata
\tablenotetext{a}{Redshift reliability classifier.}
\tablecomments{Whenever information could not be obtained for a specific target, a 
symbol '...' is used. Stars found in our data have zero redshift.}
\label{tab022300}
\end{deluxetable}

\newpage
\begin{figure}
\begin{center}
\includegraphics[height=14cm,width=15cm]{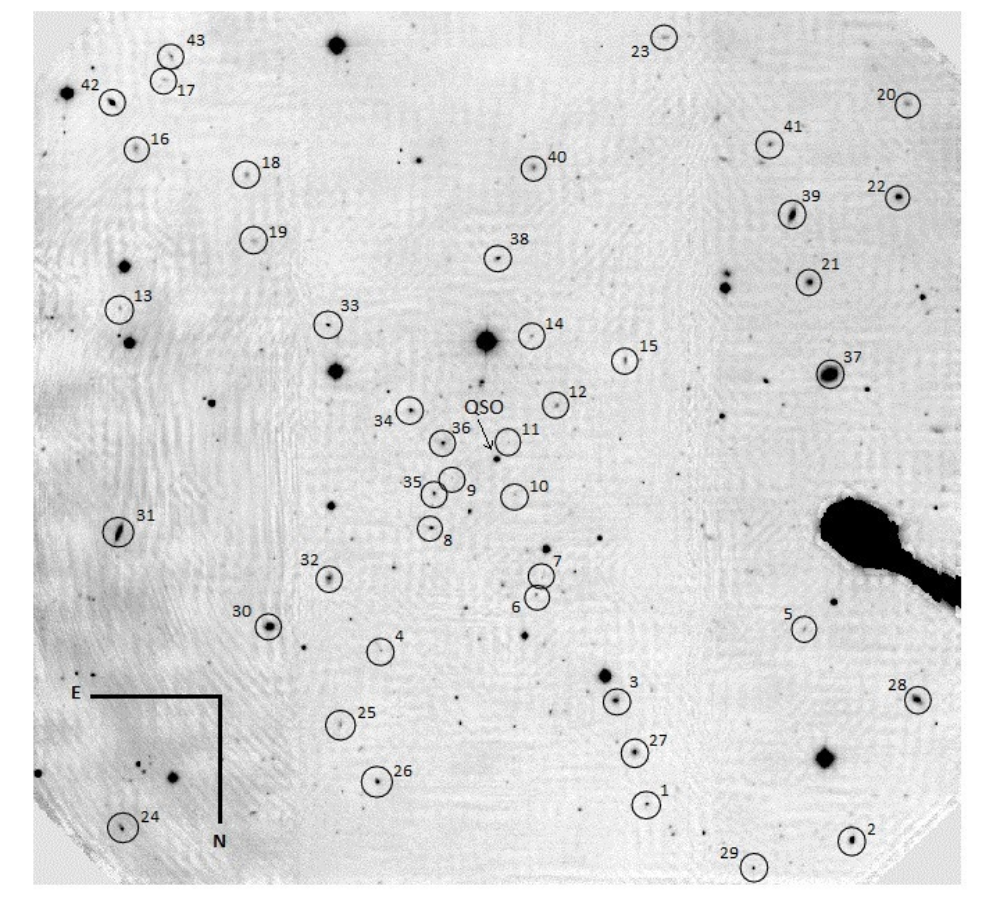}
\end{center}
\begin{flushright}
\includegraphics[height=5cm,width=7cm]{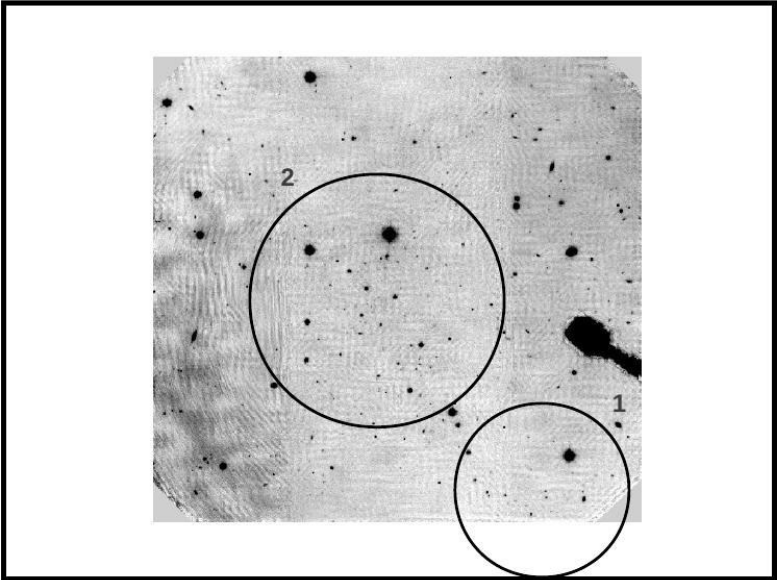} 
\end{flushright}
\begin{center}
\caption{{\it Top}: 5.5$\arcmin\times$5.5$\arcmin$ image of the field centered on the 
SDSS quasar 022300.41$+$005250.0. Galaxies are labeled according to the 
identification number given in Table \ref{tab022300}.  {\it Bottom}: A zoom-out of the image shown at the 
top. Center coordinates of each RCS1 cluster/group candidate are shown in circles. Each 
cluster is labeled according to their identification numbers 
given in the redshift histogram of Figure \ref{his022300}.}
\label{camp022300}
\end{center}
\end{figure}
\clearpage

\newpage
\begin{figure}
\begin{center}
\includegraphics[height=16cm,width=16cm]{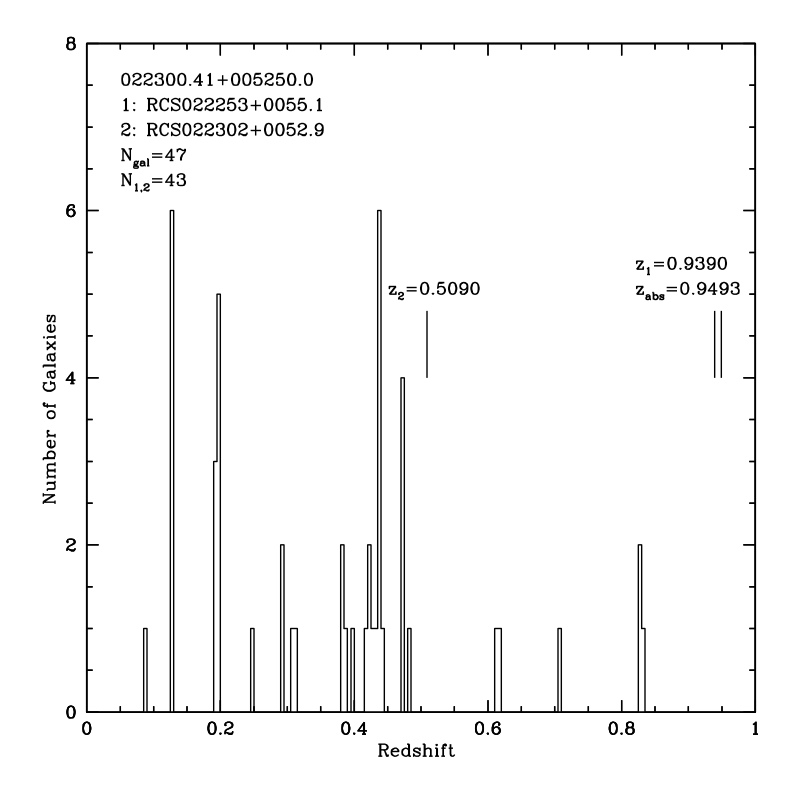}
\caption{Redshift histogram of the field centered on the SDSS quasar 022300.41$+$005250.0. The 
bin size is of 0.005 in redshift space, which translates in $\Delta v \sim$ 1000 km/s at 
$z =$ 0.5. The total number of redshifts available for this field is given by N$_{gal}$, 
from which N$_{1,2}$ is the number of redshifts classified with reliability flag 1 or 2.}
\label{his022300}
\end{center}
\end{figure}
\clearpage

\newpage
\begin{deluxetable}{ccccccccp{3.2in}}
\tablewidth{0pc}
\tablecaption{Spectroscopic Targets of Field Centered on 022441.09$+$001547.9.}
\tabletypesize{\tiny}
\tablehead{ \colhead{No.} & \colhead{RA(J2000)} & \colhead{DEC(J2000)} & 
\colhead{$z_{gal}$} & \colhead{$\sigma_{z_{gal}}$} & \colhead{Flag\tablenotemark{a}} & 
\colhead{$z'$} & \colhead{$R_{c}$ - $z'$} & \colhead{Comments}}
\startdata
1 & 02 24 31.66 & $+$00 15 45.32 & 0.35485 & 0.00005 & 2 & 18.63 & 0.70 & CaII H,CaII K,G band,H$_{\gamma}$ \\
2 & 02 24 31.02 & $+$00 15 40.36 & 0.35510 & 0.00022 & 3 & 20.78 & 0.70 & CaII H,CaII K,G band \\
3 & 02 24 34.25 & $+$00 15 11.77 & 0.64546 & 0.00010 & 1 & 20.72 & 0.66 & [OII]3727,CaII H,CaII K,H$_{\delta}$ \\
4 & 02 24 35.04 & $+$00 14 37.57 & 0.38374 & 0.00010 & 1 & 20.94 & 0.54 & H$_{\beta}$,[OIII]5007,H$_{\alpha}$,[NII]6583 \\
5 & 02 24 32.44 & $+$00 13 32.30 & 0.63028 & 0.00033 & 3 & 23.40 & 0.25 & [OII]3727,[OIII]5007 \\
6 & 02 24 33.33 & $+$00 13 55.92 & 0.35481 & 0.00012 & 1 & 21.67 & 0.51 & H$_{\beta}$,[OIII]4959,[OIII]5007 \\
7 & 02 24 39.05 & $+$00 17 33.97 & 0.56784 & 0.00021 & 3 & 22.39 & 0.63 & [OII]3727,CaII H \\
8 & 02 24 39.78 & $+$00 16 19.31 & 0.81839 & 0.00016 & 2 & 20.35 & 1.66 & CaII H,CaII K \\
9 & 02 24 41.41 & $+$00 16 07.39 & 0.93172 & 0.00080 & 3 & 23.41 & 0.17 & [OII]3727 \\
10 & 02 24 37.07 & $+$00 15 46.76 & 0.35191 & 0.00022 & 3 & 21.72 & 0.47 & [OII]3727,CaII H,CaII K,H$_{\beta}$,[OIII]5007 \\
11 & 02 24 40.83 & $+$00 15 44.68 & 0.25088 & 0.00011 & 1 & 21.23 & 0.50 & [OII]3727,H$_{\gamma}$,H$_{\beta}$,[OIII]4959,[OIII]5007,H$_{\alpha}$,[NII]6583 \\
12 & 02 24 36.50 & $+$00 15 28.44 & 0.35217 & 0.00009 & 1 & 20.35 & 0.65 & [OII]3727,CaII H,CaII K,H$_{\beta}$,[OIII]5007 \\
13 & 02 24 42.07 & $+$00 15 13.46 & 0.68534 & 0.00005 & 2 & 22.06 & 0.67 & CaII H,CaII K,[OII]3727,H$_{\delta}$ \\
14 & 02 24 37.82 & $+$00 14 23.68 & 0.43117 & 0.00002 & 1 & 21.35 & 0.42 & [OII]3727,CaII H,CaII K,H$_{\beta}$,[OIII]5007 \\
15 & 02 24 38.27 & $+$00 14 23.75 & 0.77796 & 0.00080 & 2 & 21.18 & 0.45 & [OII]3727 \\
16 & 02 24 40.27 & $+$00 13 55.70 & \nodata & \nodata & 0 & 20.47 & 0.51 & \nodata \\
17 & 02 24 43.87 & $+$00 17 27.38 & \nodata & \nodata & 0 & 21.72 & 0.49 & \nodata \\
18 & 02 24 43.13 & $+$00 15 10.01 & 0.98009 & 0.00080 & 2 & 22.12 & 0.90 & [OII]3727 \\
19 & 02 24 46.16 & $+$00 18 03.06 & 0.10770 & 0.00012 & 1 & 21.31 & 0.13 & H$_{\beta}$,[OIII]5007,H$_{\alpha}$ \\
20 & 02 24 45.53 & $+$00 14 38.94 & 0.36228 & 0.00003 & 1 & 20.99 & 0.31 & H$_{\gamma}$,H$_{\beta}$,[OIII]4959,[OIII]5007 \\
21 & 02 24 47.76 & $+$00 17 25.66 & 1.22330 & 0.00018 & 1 & 23.67 & 0.16 & FeII2374$-$2382,MnII2576,FeII2586,MnII2594,FeII2600 \\
22 & 02 24 46.76 & $+$00 17 09.28 & 0.34656 & 0.00013 & 1 & 20.83 & 0.62 & [OII]3727,CaII H,CaII K,H$_{\beta}$,[OIII]5007 \\
23 & 02 24 48.50 & $+$00 16 20.78 & 0.37887 & 0.00006 & 1 & 21.28 & 0.31 & [OII]3727,CaII H,CaII K,H$_{\beta}$,[OIII]5007,H$_{\gamma}$ \\
24 & 02 24 50.62 & $+$00 16 02.06 & 0.61225 & 0.00080 & 2 & 20.82 & 0.80 & [OII]3727 \\
25 & 02 24 49.69 & $+$00 15 32.40 & 0.30797 & 0.00012 & 1 & 20.41 & 0.37 & [OII]3727,CaII K,H$_{\beta}$,[OIII]5007 \\
26 & 02 24 51.31 & $+$00 14 19.57 & 0.38888 & 0.00006 & 1 & 21.97 & 0.49 & H$_{\beta}$,[OIII]5007,H$_{\alpha}$ \\
27 & 02 24 31.89 & $+$00 17 49.09 & 0.43430 & 0.00021 & 2 & 19.19 & 0.83 & CaII H,CaII K \\
28 & 02 24 31.19 & $+$00 14 49.38 & 0.39753 & 0.00007 & 1 & 20.76 & 0.56 & [OII]3727,[OIII]5007 \\
29 & 02 24 35.35 & $+$00 14 37.57 & 0.20110 & 0.00042 & 3 & 20.94 & 0.54 & [OIII]4959,[OIII]5007 \\
30 & 02 24 34.54 & $+$00 14 36.78 & 0.34920 & 0.00020 & 3 & 20.46 & 0.75 & CaII H,CaII K \\
31 & 02 24 33.68 & $+$00 13 43.86 & 0.74531 & 0.00060 & 3 & 20.72 & 0.80 & [OII]3727 \\
32 & 02 24 32.86 & $+$00 13 38.35 & 0.75270 & 0.00014 & 3 & 20.35 & 1.51 & [OII]3727,CaII H \\
33 & 02 24 36.21 & $+$00 13 56.06 & 0.77845 & 0.00080 & 3 & 20.66 & 0.95 & [OII]3727 \\
34 & 02 24 38.89 & $+$00 16 46.52 & 0.35460 & 0.00035 & 3 & 19.77 & 0.69 & CaII H,CaII K,G band \\
35 & 02 24 41.29 & $+$00 15 43.09 & 0.37929 & 0.00012 & 1 & 19.00 & 0.51 & [OII]3727,H$_{\beta}$,H$_{\delta}$ \\
36 & 02 24 41.85 & $+$00 15 38.81 & 0.61516 & 0.00080 & 3 & 21.42 & 0.52 & [OII]3727 \\
37 & 02 24 42.38 & $+$00 15 18.94 & 0.68548 & 0.00080 & 3 & 21.86 & 0.80 & [OII]3727 \\
38 & 02 24 40.62 & $+$00 15 12.02 & 0.55876 & 0.00010 & 1 & 20.46 & 0.59 & [OII]3727,CaII H,CaII K \\
39 & 02 24 40.01 & $+$00 15 11.20 & 0.35013 & 0.00015 & 1 & 19.84 & 0.78 & CaII H,CaII K,G band,H$_{\gamma}$ \\
40 & 02 24 37.90 & $+$00 14 49.70 & 0.35554 & 0.00012 & 3 & 20.19 & 0.53 & H$_{\beta}$,[OIII]4959,[OIII]5007 \\
41 & 02 24 45.98 & $+$00 17 54.02 & 0.35270 & 0.00023 & 2 & 19.49 & 0.73 & CaII H,CaII K,G band \\
42 & 02 24 44.13 & $+$00 17 45.60 & 0.35144 & 0.00004 & 1 & 16.68 & 0.75 & CaII H,CaII K,H$_{\delta}$,G band,H$_{\beta}$,MgI5175 \\
43 & 02 24 46.39 & $+$00 17 35.77 & 0.68621 & 0.00023 & 2 & 19.54 & 1.30 & CaII H,CaII K \\
44 & 02 24 49.54 & $+$00 17 23.53 & 0.34470 & 0.00042 & 3 & 19.61 & 0.69 & CaII K,G band \\
45 & 02 24 45.15 & $+$00 13 52.86 & \nodata & \nodata & 0 & 20.26 & 0.77 & \nodata \\
46 & 02 24 48.87 & $+$00 16 13.44 & 0.61090 & 0.00076 & 2 & 20.51 & 1.25 & CaII H,CaII K,G band \\
47 & 02 24 48.36 & $+$00 16 11.89 & 0.61190 & 0.00080 & 3 & 20.20 & 1.35 & CaII K \\
48 & 02 24 47.70 & $+$00 13 47.78 & 0.43289 & 0.00013 & 1 & 18.75 & 0.62 & CaII H,CaII K,G band,H$_{\delta}$,H$_{\beta}$ \\
49 & 02 24 51.37 & $+$00 17 20.54 & 0.77240 & 0.00064 & 3 & 23.75 & 0.92 & [OII]3727,H$_{\gamma}$ \\
50 & 02 24 50.76 & $+$00 17 17.38 & 0.43410 & 0.00030 & 3 & 20.35 & 0.56 & [OII]3727,H$_{\beta}$ \\
51 & 02 24 50.11 & $+$00 15 31.00 & 0.43202 & 0.00006 & 1 & 21.07 & 0.38 & [OII]3727,H$_{\beta}$,[OIII]5007 \\
\enddata
\tablenotetext{a}{Redshift reliability classifier.}
\tablecomments{Whenever information could not be obtained for a specific target, a 
symbol '...' is used. Stars found in our data have zero redshift.}
\label{tab022441}
\end{deluxetable}

\newpage
\begin{figure}
\begin{center}
\includegraphics[height=14cm,width=15cm]{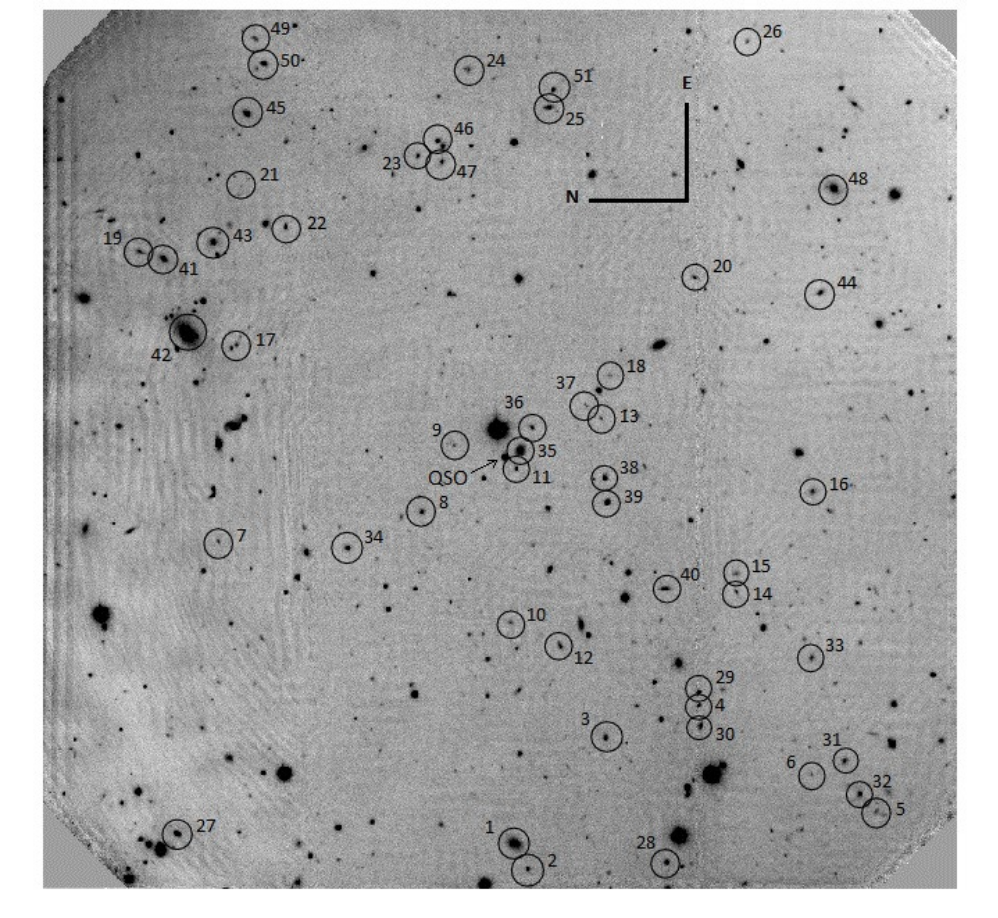}
\end{center}
\begin{flushright}
\includegraphics[height=5cm,width=7cm]{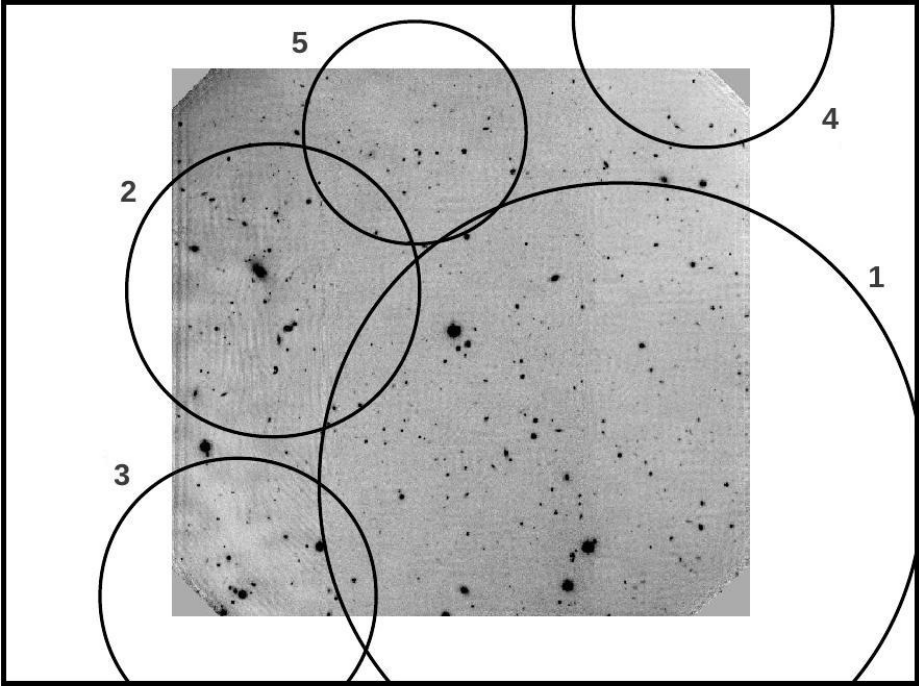} 
\end{flushright}
\begin{center}
\caption{{\it Top}: 5.5$\arcmin\times$5.5$\arcmin$ image of the field centered on the 
SDSS quasar 022441.09$+$001547.9. Galaxies are labeled according to the 
identification number given in Table \ref{tab022441}.  {\it Bottom}: A zoom-out of the image shown at the 
top. Center coordinates of each RCS1 cluster/group candidate are shown in circles. Each 
cluster is labeled according to their identification numbers 
given in the redshift histogram of Figure \ref{his022441}.}
\label{camp022441}
\end{center}
\end{figure}
\clearpage

\newpage
\begin{figure}
\begin{center}
\includegraphics[height=16cm,width=16cm]{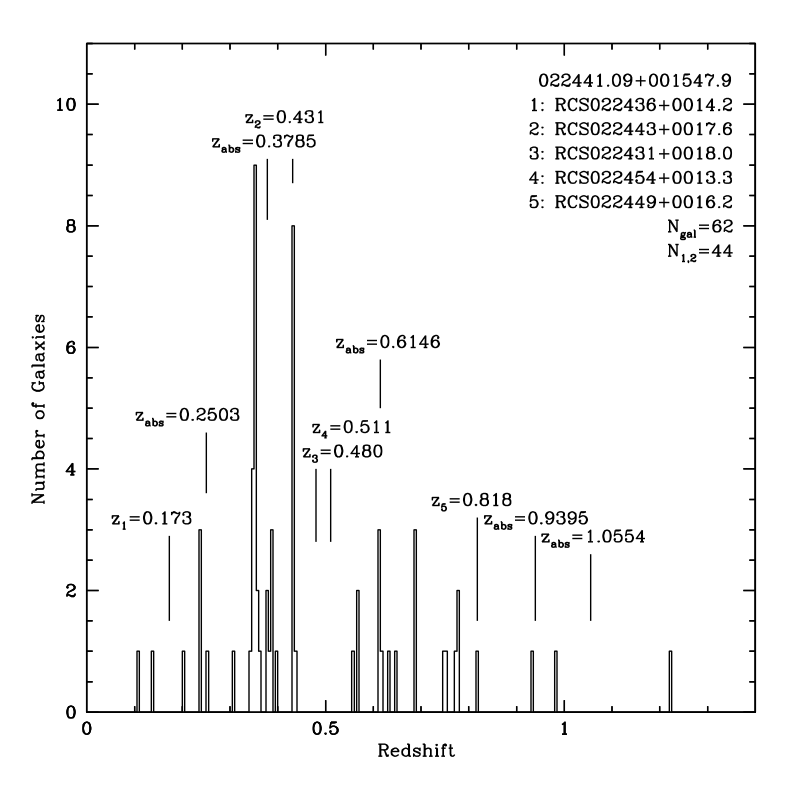}
\caption{Redshift histogram of the field centered on the SDSS quasar 022441.09$+$001547.9. The 
bin size is of 0.005 in redshift space, which translates in $\Delta v \sim$ 1000 km/s at 
$z =$ 0.5. The total number of redshifts available for this field is given by N$_{gal}$, 
from which N$_{1,2}$ is the number of redshifts classified with reliability flag 1 or 2.}
\label{his022441}
\end{center}
\end{figure}
\clearpage

\newpage
\begin{deluxetable}{ccccccccp{3.2in}}
\tablewidth{0pc}
\tablecaption{Spectroscopic Targets of Field Centered on 022553.59$+$005130.9.}
\tabletypesize{\tiny}
\tablehead{ \colhead{No.} & \colhead{RA(J2000)} & \colhead{DEC(J2000)} & 
\colhead{$z_{gal}$} & \colhead{$\sigma_{z_{gal}}$} & \colhead{Flag\tablenotemark{a}} & 
\colhead{$z'$} & \colhead{$R_{c}$ - $z'$} & \colhead{Comments}}
\startdata
1 & 02 25 50.89 & $+$00 49 08.83 & 0.40633 & 0.00011 & 1 & 20.36 & 0.53 & [OII]3727,CaII K,H$_{\beta}$,[OIII]5007 \\
2 & 02 25 54.06 & $+$00 49 01.42 & 0.39400 & 0.00008 & 1 & 21.80 & 0.26 & [OII]3727,H$_{\beta}$,[OIII]4959,[OIII]5007 \\
3 & 02 25 57.11 & $+$00 49 30.68 & 0.27039 & 0.00007 & 1 & 21.13 & 0.27 & [OII]3727,H$_{\beta}$,[OIII]4959,[OIII]5007 \\
4 & 02 25 57.42 & $+$00 49 59.84 & 0.51153 & 0.00007 & 1 & 21.96 & 1.00 & [OII]3727,H$_{\gamma}$,H$_{\beta}$,[OIII]5007 \\
5 & 02 25 44.97 & $+$00 49 54.80 & 0.09798 & 0.00004 & 1 & 19.52 & 0.21 & [OII]3727,H$_{\beta}$,[OIII]4959,[OIII]5007,H$_{\alpha}$,[NII]6583,[SII]6716,[SII]6730 \\
6 & 02 25 46.36 & $+$00 49 32.16 & 0.43080 & 0.00045 & 3 & 21.40 & 0.75 & CaII H,CaII K \\
7 & 02 25 46.48 & $+$00 50 14.71 & 0.19232 & 0.00011 & 1 & 21.05 & 0.13 & [OII]3727,H$_{\beta}$,[OIII]4959,[OIII]5007 \\
8 & 02 25 47.47 & $+$00 50 31.96 & 0.58654 & 0.00018 & 1 & 20.31 & 1.15 & [OII]3727,CaII H,CaII K \\
9 & 02 25 50.06 & $+$00 49 39.65 & 0.09824 & 0.00022 & 1 & 22.39 & 0.33 & H$_{\beta}$,[OIII]4959,[OIII]5007 \\
10 & 02 25 52.69 & $+$00 50 25.22 & 0.74548 & 0.00080 & 2 & 21.51 & 0.80 & [OII]3727 \\
11 & 02 25 54.12 & $+$00 51 28.22 & 0.57103 & 0.00035 & 1 & 21.01 & 1.01 & CaII H,CaII K \\
12 & 02 25 55.72 & $+$00 51 27.58 & 0.68241 & 0.00012 & 1 & 20.68 & 0.63 & [OII]3727,CaII H,CaII K \\
13 & 02 25 59.99 & $+$00 52 09.52 & 0.39532 & 0.00003 & 1 & 20.27 & 0.37 & H$_{\gamma}$,H$_{\beta}$,[OIII]4959,[OIII]5007 \\
14 & 02 25 43.39 & $+$00 51 03.13 & 0.40418 & 0.00008 & 1 & 19.58 & 0.55 & [OII]3727,CaII H,CaII K,H$_{\delta}$ \\
15 & 02 25 49.06 & $+$00 52 23.74 & 0.19824 & 0.00006 & 1 & 20.48 & 0.24 & [OII]3727,H$_{\beta}$,[OIII]4959,[OIII]5007 \\
16 & 02 25 53.77 & $+$00 52 21.36 & 0.39449 & 0.00005 & 1 & 20.09 & 0.71 & [OII]3727,H$_{\beta}$,[OIII]4959,[OIII]5007 \\
17 & 02 25 53.60 & $+$00 52 30.86 & 0.74775 & 0.00080 & 3 & 21.79 & 0.91 & [OII]3727 \\
18 & 02 25 54.17 & $+$00 52 41.99 & 0.53291 & 0.00006 & 1 & 19.72 & 0.79 & [OII]3727,CaII H,CaII K,H$_{\gamma}$ \\
19 & 02 25 55.54 & $+$00 52 19.49 & 0.23591 & 0.00006 & 1 & 20.93 & 0.32 & H$_{\gamma}$,H$_{\beta}$,[OIII]4959,[OIII]5007,H$_{\alpha}$,[NII]6583 \\
20 & 02 25 57.39 & $+$00 52 06.49 & 0.39743 & 0.00013 & 1 & 20.29 & 0.59 & CaII H,CaII K,H$_{\delta}$,G band \\
21 & 02 25 43.42 & $+$00 52 34.86 & 0.40693 & 0.00012 & 1 & 21.02 & 0.59 & [OII]3727,H$_{\beta}$,[OIII]5007 \\
22 & 02 25 44.82 & $+$00 52 20.71 & 0.42169 & 0.00008 & 1 & 21.91 & -0.21 & [OII]3727,H$_{\gamma}$,H$_{\beta}$,[OIII]4959,[OIII]5007 \\
23 & 02 25 44.49 & $+$00 53 24.58 & 0.13893 & 0.00004 & 1 & 21.86 & 0.59 & H$_{\beta}$,[OIII]4959,[OIII]5007,H$_{\alpha}$ \\
24 & 02 25 47.02 & $+$00 53 03.62 & 0.40838 & 0.00034 & 1 & 20.62 & 0.68 & [OII]3727,H$_{\delta}$,H$_{\gamma}$,H$_{\beta}$,[OIII]5007 \\
25 & 02 25 48.33 & $+$00 53 25.15 & 0.12931 & 0.00005 & 1 & 21.27 & 0.53 & H$_{\gamma}$,H$_{\beta}$,[OIII]4959,[OIII]5007,H$_{\alpha}$ \\
26 & 02 25 50.75 & $+$00 53 03.98 & 0.73801 & 0.00023 & 1 & 20.94 & 1.45 & [OII]3727,CaII H,CaII K \\
27 & 02 25 53.60 & $+$00 54 03.31 & 0.66596 & 0.00004 & 1 & 24.94 & -0.30 & [OII]3727,H$_{\gamma}$,H$_{\beta}$,[OIII]4959 \\
28 & 02 25 49.29 & $+$00 48 29.16 & 0.29930 & 0.00085 & 2 & 19.98 & 0.82 & CaII K,G band \\
29 & 02 25 49.18 & $+$00 48 56.09 & 0.75676 & 0.00025 & 3 & 21.16 & 1.12 & [OII]3727,CaII K \\
30 & 02 25 47.58 & $+$00 49 07.97 & 0.56685 & 0.00010 & 1 & 20.22 & 0.34 & [OII]3727,CaII H,CaII K \\
31 & 02 25 48.28 & $+$00 49 39.54 & 0.31383 & 0.00012 & 1 & 19.83 & 0.64 & CaII H,CaII K,G band \\
32 & 02 25 48.09 & $+$00 49 48.61 & 0.40926 & 0.00019 & 1 & 20.61 & 0.73 & [OII]3727,CaII H,CaII K,H$_{\delta}$,H$_{\beta}$ \\
33 & 02 25 48.37 & $+$00 50 03.44 & 0.41511 & 0.00009 & 1 & 20.02 & 0.67 & [OII]3727,CaII H,CaII K,H$_{\delta}$,H$_{\beta}$ \\
34 & 02 25 48.18 & $+$00 50 40.42 & 0.41906 & 0.00022 & 1 & 19.62 & 0.74 & [OII]3727,CaII H,CaII K \\
35 & 02 25 50.78 & $+$00 50 36.74 & 0.16005 & 0.00013 & 1 & 16.95 & 0.54 & [OII]3727,CaII H,CaII K,H$_{\delta}$,H$_{\beta}$,[OIII]5007 \\
36 & 02 25 53.67 & $+$00 51 28.91 & 1.09550 & 0.00080 & 1 & 23.72 & 0.83 & [OII]3727 \\
37 & 02 25 59.22 & $+$00 51 50.72 & 0.56748 & 0.00012 & 1 & 19.86 & 0.69 & [OII]3727,CaII H,CaII K \\
38 & 02 25 51.90 & $+$00 51 36.72 & 0.39817 & 0.00009 & 1 & 18.81 & 0.64 & [OII]3727,CaII H,CaII K,H$_{\delta}$,H$_{\beta}$ \\
39 & 02 25 50.75 & $+$00 52 26.76 & 0.42031 & 0.00017 & 1 & 19.84 & 0.85 & [OII]3727,CaII H,CaII K,G band \\
40 & 02 25 53.79 & $+$00 51 57.49 & 0.56961 & 0.00014 & 1 & 20.67 & 1.02 & [OII]3727,CaII H,CaII K \\
41 & 02 25 52.54 & $+$00 52 59.84 & 0.46596 & 0.00005 & 2 & 21.20 & 0.69 & [OII]3727,CaII K \\
42 & 02 25 53.04 & $+$00 52 51.92 & 0.75386 & 0.00029 & 2 & 20.84 & 1.33 & [OII]3727,CaII H,CaII K \\
43 & 02 25 54.32 & $+$00 52 16.68 & 0.39518 & 0.00013 & 1 & 18.50 & 0.54 & [OII]3727,CaII H,CaII K,H$_{\gamma}$,H$_{\beta}$,[OIII]4959,[OIII]5007 \\
44 & 02 25 54.08 & $+$00 52 32.38 & 0.39510 & 0.00010 & 1 & 19.81 & 0.76 & CaII H,CaII K \\
45 & 02 25 54.54 & $+$00 53 38.54 & 0.39550 & 0.00047 & 2 & 20.75 & 0.50 & [OII]3727,CaII H,CaII K \\
46 & 02 25 56.98 & $+$00 53 36.60 & 0.41939 & 0.00002 & 1 & 20.19 & 0.48 & H$_{\beta}$,[OIII]5007 \\
47 & 02 25 47.15 & $+$00 53 30.19 & 0.40621 & 0.00004 & 1 & 19.29 & 0.68 & [OII]3727,CaII H,H$_{\beta}$,[OIII]4959,[OIII]5007 \\
48 & 02 25 47.49 & $+$00 53 17.84 & 0.56936 & 0.00046 & 2 & 20.17 & 1.20 & CaII H,CaII K,G band \\
\enddata
\tablenotetext{a}{Redshift reliability classifier.}
\tablecomments{Whenever information could not be obtained for a specific target, a 
symbol '...' is used. Stars found in our data have zero redshift.}
\label{tab022553}
\end{deluxetable}

\clearpage
\begin{figure}
\begin{center}
\includegraphics[height=14cm,width=15cm]{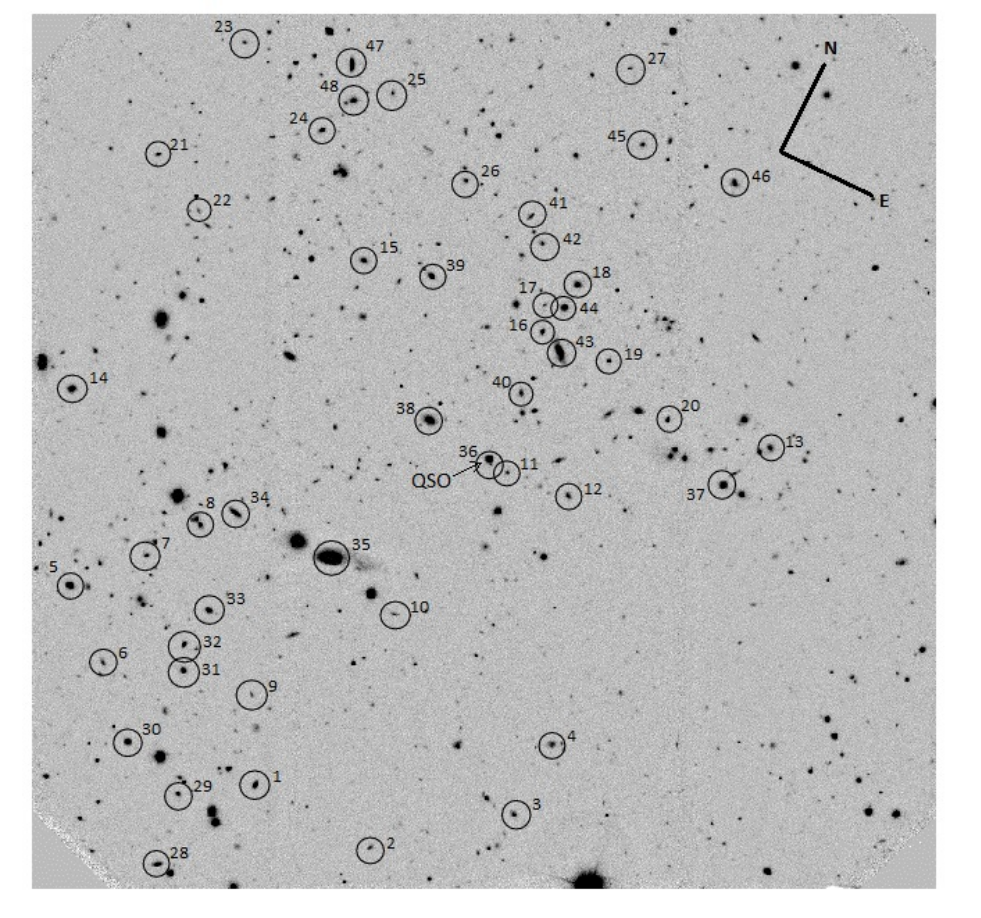}
\end{center}
\begin{flushright}
\includegraphics[height=5cm,width=7cm]{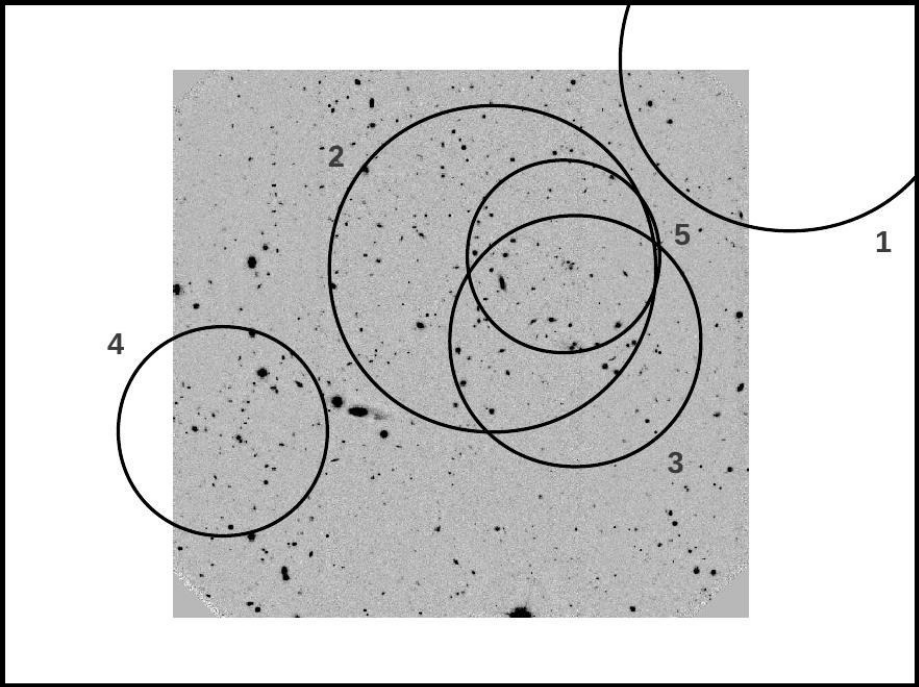} 
\end{flushright}
\begin{center}
\caption{{\it Top}: 5.5$\arcmin\times$5.5$\arcmin$ image of the field centered on the 
SDSS quasar 022553.59$+$005130.9. Galaxies are labeled according to the 
identification number given in Table \ref{tab022553}.  {\it Bottom}: A zoom-out of the image shown at the 
top. Center coordinates of each RCS1 cluster/group candidate are shown in circles. Each 
cluster is labeled according to their identification numbers 
given in the redshift histogram of Figure \ref{his022553}.}
\label{camp022553}
\end{center}
\end{figure}
\clearpage

\newpage
\begin{figure}
\begin{center}
\includegraphics[height=16cm,width=16cm]{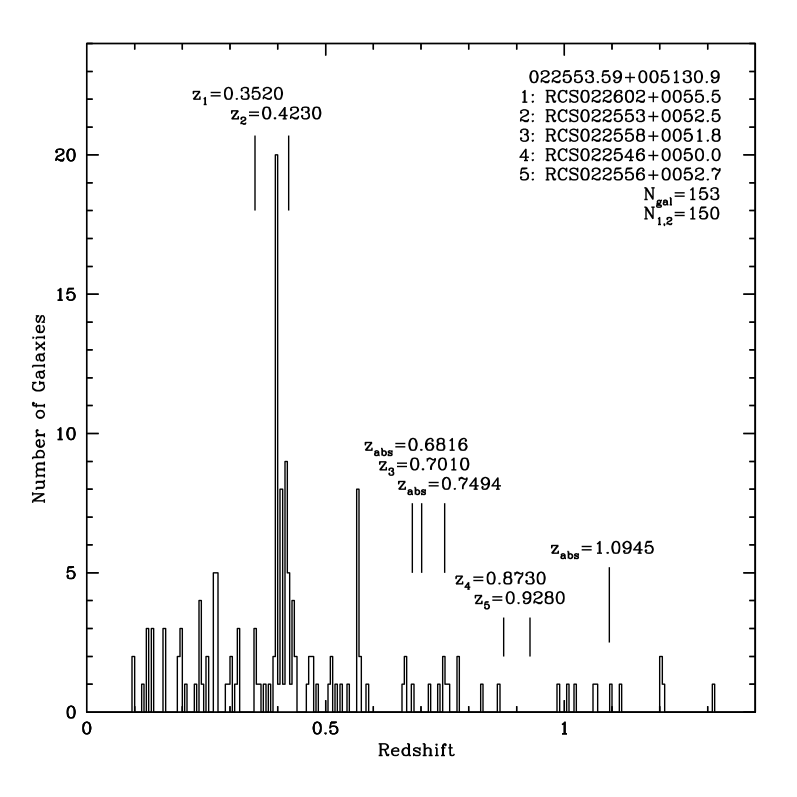}
\caption{Redshift histogram of the field centered on the SDSS quasar 022553.59$+$005130.9. The 
bin size is of 0.005 in redshift space, which translates in $\Delta v \sim$ 1000 km/s at 
$z =$ 0.5. The total number of redshifts available for this field is given by N$_{gal}$, 
from which N$_{1,2}$ is the number of redshifts classified with reliability flag 1 or 2.}
\label{his022553}
\end{center}
\end{figure}
\clearpage

\newpage
\begin{deluxetable}{ccccccccp{3.2in}}
\tablewidth{0pc}
\tablecaption{Spectroscopic Targets of Field Centered on 022839.32$+$004623.0.}
\tabletypesize{\tiny}
\tablehead{ \colhead{No.} & \colhead{RA(J2000)} & \colhead{DEC(J2000)} & 
\colhead{$z_{gal}$} & \colhead{$\sigma_{z_{gal}}$} & \colhead{Flag\tablenotemark{a}} & 
\colhead{$z'$} & \colhead{$R_{c}$ - $z'$} & \colhead{Comments}}
\startdata
1 & 02 28 34.14 & $+$00 43 00.59 & 0.68320 & 0.00020 & 2 & 21.41 & 0.49 & [OII]3727,CaII H,CaII K \\
2 & 02 28 41.27 & $+$00 43 16.68 & 0.41586 & 0.00004 & 1 & 22.45 & 0.02 & [OII]3727,H$_{\gamma}$,H$_{\beta}$,[OIII]4959,[OIII]5007 \\
3 & 02 28 36.23 & $+$00 43 28.96 & 0.65169 & 0.00002 & 3 & 20.97 & 0.92 & [OII]3727,CaII H \\
4 & 02 28 36.72 & $+$00 43 21.00 & 0.46362 & 0.00010 & 1 & 20.47 & 0.96 & CaII H,CaII K \\
5 & 02 28 33.25 & $+$00 43 41.48 & \nodata & \nodata & 0 & 21.54 & 1.49 & \nodata \\
6 & 02 28 34.94 & $+$00 43 35.80 & 0.56726 & 0.00015 & 1 & 21.14 & 0.50 & [OII]3727,H$_{\gamma}$ \\
7 & 02 28 33.28 & $+$00 44 12.91 & 0.66299 & 0.00025 & 3 & 21.60 & 1.33 & CaII H,CaII K \\
8 & 02 28 33.49 & $+$00 44 22.42 & 0.80617 & 0.00080 & 2 & 22.19 & 0.53 & [OII]3727 \\
9 & 02 28 37.79 & $+$00 44 46.28 & 0.56164 & 0.00005 & 3 & 21.88 & 0.43 & [OII]3727,[OIII]5007 \\
10 & 02 28 42.37 & $+$00 44 38.69 & 0.26592 & 0.00003 & 1 & 18.86 & 0.71 & CaII H,CaII K,G band \\
11 & 02 28 31.84 & $+$00 44 36.49 & 0.32159 & 0.00008 & 1 & 19.09 & 0.68 & [OII]3727,CaII H,CaII K,G band,H$_{\beta}$,[OIII]4959,[OIII]5007 \\
12 & 02 28 36.29 & $+$00 45 31.03 & 0.13341 & 0.00005 & 1 & 21.76 & 0.27 & H$_{\beta}$,[OIII]4959,[OIII]5007,H$_{\alpha}$ \\
13 & 02 28 36.98 & $+$00 45 23.04 & 0.65660 & 0.00080 & 3 & 21.10 & 1.12 & Ca3933 \\
14 & 02 28 36.90 & $+$00 45 43.34 & 0.67060 & 0.00080 & 3 & 22.52 & 0.20 & [OII]3727 \\
15 & 02 28 37.60 & $+$00 45 12.49 & \nodata & \nodata & 0 & 21.40 & 0.85 & \nodata \\ 
16 & 02 28 37.08 & $+$00 46 02.39 & 0.65371 & 0.00017 & 1 & 20.88 & 0.77 & [OII]3727,CaII H,CaII K \\
17 & 02 28 39.79 & $+$00 46 32.41 & 0.36409 & 0.00006 & 1 & 23.80 & -1.07 & [OII]3727,[OIII]4959,[OIII]5007 \\
18 & 02 28 47.11 & $+$00 47 10.18 & 0.77262 & 0.00045 & 3 & 20.63 & 1.28 & CaII H,CaII K \\
19 & 02 28 45.00 & $+$00 46 42.20 & 0.76862 & 0.00080 & 3 & 22.78 & 0.36 & [OII]3727 \\
20 & 02 28 34.72 & $+$00 46 43.28 & 0.25626 & 0.00006 & 1 & 19.95 & 0.42 & [OII]3727,CaII K,H$_{\beta}$,[OIII]4959,[OIII]5007 \\
21 & 02 28 44.25 & $+$00 47 34.58 & 0.46480 & 0.00030 & 3 & 20.37 & 0.87 & CaII H,CaII K \\
22 & 02 28 45.04 & $+$00 47 10.10 & 0.77258 & 0.00080 & 3 & 22.00 & 0.68 & [OII]3727 \\
23 & 02 28 29.02 & $+$00 46 52.28 & \nodata & \nodata & 0 & 20.81 & 1.79 & \nodata \\
24 & 02 28 35.75 & $+$00 47 38.94 & 0.65156 & 0.00027 & 3 & 22.22 & 0.24 & [OII]3727,CaII H,CaII K \\
25 & 02 28 43.61 & $+$00 47 59.21 & 0.30259 & 0.00003 & 1 & 20.40 & 0.42 & H$_{\beta}$,[OIII]4959,[OIII]5007 \\
26 & 02 28 36.26 & $+$00 48 08.75 & \nodata & \nodata & 0 & 21.68 & 0.74 & \nodata \\
27 & 02 28 46.67 & $+$00 48 33.19 & 0.91790 & 0.00080 & 3 & 21.31 & 0.95 & [OII]3727 \\
28 & 02 28 39.72 & $+$00 43 32.74 & 0.26685 & 0.00008 & 1 & 20.40 & 0.44 & [OII]3727,CaII H,CaII K,H$_{\beta}$ \\
29 & 02 28 44.84 & $+$00 44 03.26 & 0.30669 & 0.00007 & 1 & 18.71 & 0.66 & CaII H,CaII K,G band,H$_{\gamma}$ \\
30 & 02 28 49.61 & $+$00 44 00.96 & 0.21074 & 0.00015 & 1 & 19.32 & 0.39 & H$_{\beta}$,[OIII]4959,[OIII]5007 \\
31 & 02 28 48.07 & $+$00 43 42.17 & 0.12718 & 0.00005 & 1 & 20.78 & 0.22 & H$_{\beta}$,[OIII]4959,[OIII]5007,H$_{\alpha}$,[NII]6583,[SII]6716,[SII]6730 \\
32 & 02 28 40.92 & $+$00 44 10.57 & 0.65726 & 0.00015 & 2 & 21.25 & 1.27 & CaII H,CaII K \\
33 & 02 28 49.42 & $+$00 44 33.22 & 0.49164 & 0.00031 & 3 & 21.13 & 0.29 & HeII3781,[OIII]4959,[OIII]5007 \\
34 & 02 28 37.46 & $+$00 44 30.26 & 0.25637 & 0.00006 & 1 & 19.95 & 0.32 & [OII]3727,H$_{\delta}$,H$_{\gamma}$,H$_{\beta}$,[OIII]4959,[OIII]5007 \\
35 & 02 28 40.95 & $+$00 44 55.14 & 0.77796 & 0.00080 & 2 & 24.59 & -0.49 & [OII]3727 \\
36 & 02 28 43.33 & $+$00 44 42.11 & 0.26799 & 0.00015 & 3 & 19.03 & 0.92 & CaII H,CaII K \\
37 & 02 28 34.75 & $+$00 44 44.77 & 0.65924 & 0.00025 & 2 & 19.78 & 1.23 & [OII]3727,CaII H \\
38 & 02 28 38.33 & $+$00 45 17.06 & 0.25622 & 0.00006 & 1 & 24.22 & 0.55 & [OII]3727,H$_{\gamma}$,H$_{\beta}$,[OIII]4959,[OIII]5007 \\
39 & 02 28 30.42 & $+$00 45 23.80 & 0.62621 & 0.00080 & 2 & 21.79 & 0.10 & [OII]3727 \\
40 & 02 28 32.65 & $+$00 45 15.59 & 0.18719 & 0.00006 & 1 & 17.80 & 0.52 & [OII]3727,CaII H,CaII K,H$_{\beta}$,[OIII]5007,[NII]6548,H$_{\alpha}$,[NII]6583 \\
41 & 02 28 30.57 & $+$00 46 43.18 & 0.65553 & 0.00055 & 1 & 20.26 & 1.02 & [OII]3727,CaII H,CaII K \\
42 & 02 28 31.72 & $+$00 47 28.10 & 0.65552 & 0.00080 & 2 & 21.25 & 0.80 & [OII]3727 \\
43 & 02 28 33.58 & $+$00 47 26.56 & 0.16356 & 0.00007 & 3 & 19.77 & 0.50 & [OIII]5007,H$_{\alpha}$ \\
44 & 02 28 35.27 & $+$00 46 47.14 & 0.16276 & 0.00006 & 1 & 19.49 & 0.42 & [OII]3727,H$_{\gamma}$,H$_{\beta}$,[OIII]4959,[OIII]5007 \\
45 & 02 28 36.41 & $+$00 46 35.18 & 0.30340 & 0.00011 & 1 & 19.67 & 0.43 & [OII]3727,CaII H,CaII K,H$_{\beta}$,[OIII]5007 \\
46 & 02 28 39.08 & $+$00 46 27.91 & 0.15198 & 0.00008 & 1 & 18.14 & 0.62 & H$_{\beta}$,[NII]6548,H$_{\alpha}$,[NII]6583 \\
47 & 02 28 39.72 & $+$00 46 17.94 & \nodata & \nodata & 0 & 19.62 & 0.43 & \nodata \\
48 & 02 28 43.85 & $+$00 47 45.67 & 0.46858 & 0.00019 & 1 & 19.19 & 0.90 & [OII]3727,CaII H,CaII K,G band \\
49 & 02 28 46.64 & $+$00 48 07.16 & \nodata & \nodata & 0 & 21.16 & 1.38 & \nodata \\
50 & 02 28 39.84 & $+$00 48 09.32 & 0.25590 & 0.00028 & 3 & 19.79 & 0.38 & CaII K,G band \\
\enddata
\tablenotetext{a}{Redshift reliability classifier.}
\tablecomments{Whenever information could not be obtained for a specific target, a 
symbol '...' is used. Stars found in our data have zero redshift.}
\label{tab022839}
\end{deluxetable}

\newpage
\begin{figure}
\begin{center}
\includegraphics[height=14cm,width=15cm]{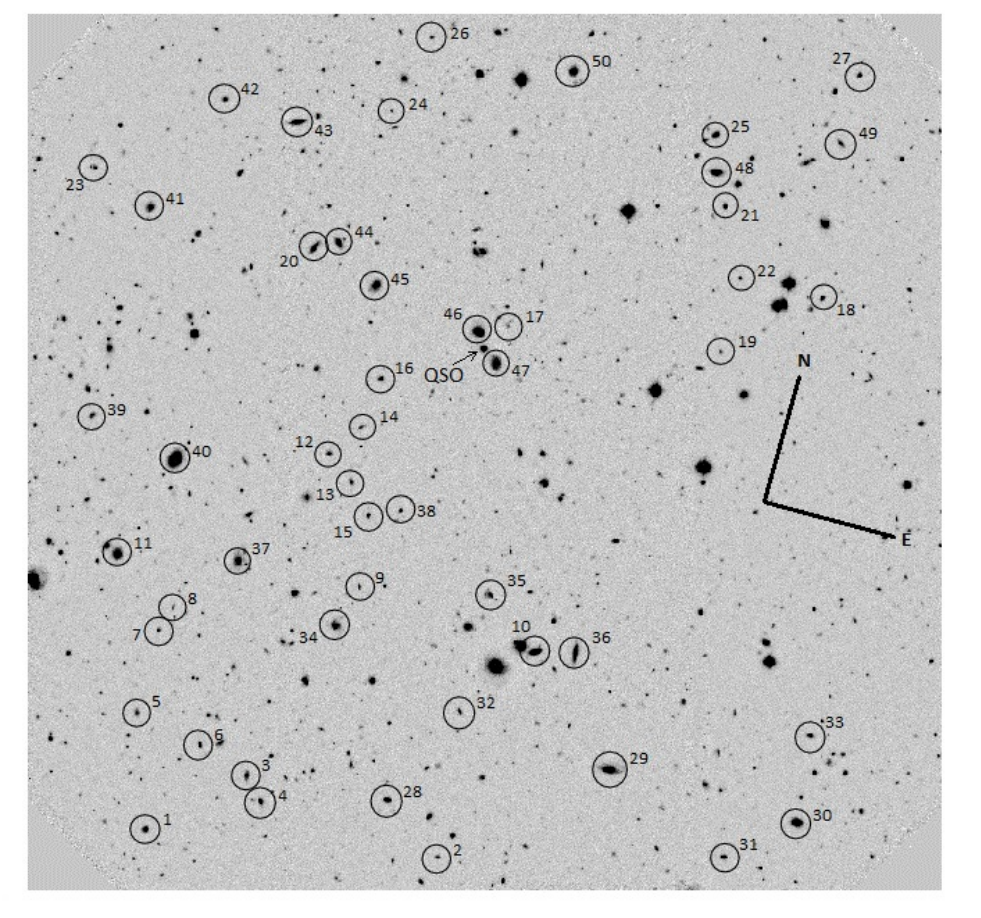}
\end{center}
\begin{flushright}
\includegraphics[height=5cm,width=7cm]{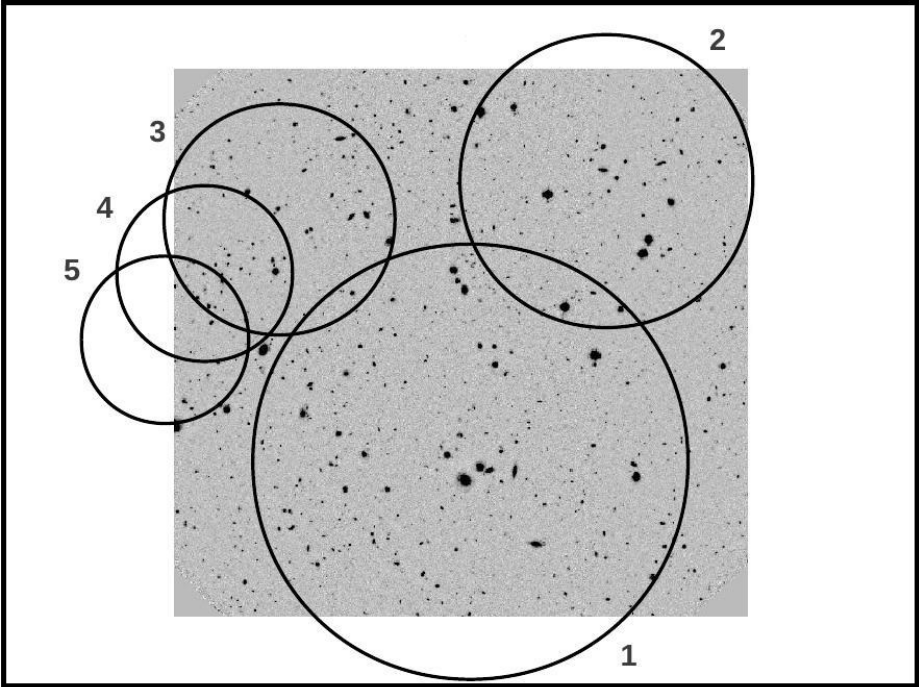} 
\end{flushright}
\begin{center}
\caption{{\it Top}: 5.5$\arcmin\times$5.5$\arcmin$ image of the field centered on the 
SDSS quasar 022839.32$+$004623.0. Galaxies are labeled according to the 
identification number given in Table \ref{tab022839}.  {\it Bottom}: A zoom-out of the image shown at the 
top. Center coordinates of each RCS1 cluster/group candidate are shown in circles. Each 
cluster is labeled according to their identification numbers 
given in the redshift histogram of Figure \ref{his022839}.}
\label{camp022839}
\end{center}
\end{figure}
\clearpage

\newpage
\begin{figure}
\begin{center}
\includegraphics[height=16cm,width=16cm]{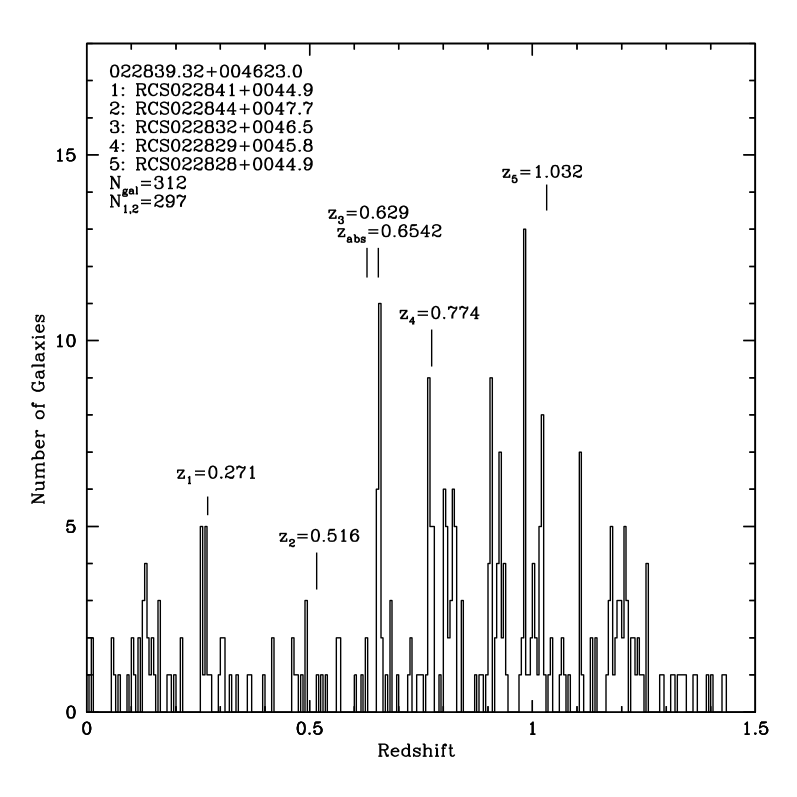}
\caption{Redshift histogram of the field centered on the SDSS quasar 022839.32$+$004623.0. The 
bin size is of 0.005 in redshift space, which translates in $\Delta v \sim$ 1000 km/s at 
$z =$ 0.5. The total number of redshifts available for this field is given by N$_{gal}$, 
from which N$_{1,2}$ is the number of redshifts classified with reliability flag 1 or 2.}
\label{his022839}
\end{center}
\end{figure}
\clearpage

\newpage
\begin{deluxetable}{ccccccccp{3.2in}}
\tablewidth{0pc}
\tablecaption{Spectroscopic Targets of Field Centered on HE2149$-$2745A.}
\tabletypesize{\tiny}
\tablehead{ \colhead{No.} & \colhead{RA(J2000)} & \colhead{DEC(J2000)} & 
\colhead{$z_{gal}$} & \colhead{$\sigma_{z_{gal}}$} & \colhead{Flag\tablenotemark{a}} & 
\colhead{$z'$} & \colhead{$R_{c}$ - $z'$} & \colhead{Comments}}
\startdata
1 & 21 52 19.30 & $-$27 30 19.29 & 2.59530 & 0.00474 & 3 & \nodata & \nodata & CIV 1550,CIII 1909 \\
2 & 21 52 18.44 & $-$27 30 09.25 & 0.00000 & 0.00000 & 0 & \nodata & \nodata & \nodata \\
3 & 21 52 15.07 & $-$27 31 23.88 & 0.73932 & 0.00012 & 2 & \nodata & \nodata & [OII]3727,CaII K,G band \\
4 & 21 52 15.74 & $-$27 31 19.20 & 0.73952 & 0.00040 & 2 & \nodata & \nodata & CaII H,CaII K,G band \\
5 & 21 52 12.88 & $-$27 33 03.71 & 0.00000 & 0.00000 & 0 & \nodata & \nodata & \nodata \\
6 & 21 52 14.57 & $-$27 30 02.92 & 0.00000 & 0.00000 & 0 & \nodata & \nodata & \nodata \\
7 & 21 52 10.51 & $-$27 33 47.38 & \nodata & \nodata & 0 & \nodata & \nodata & \nodata \\
8 & 21 52 09.02 & $-$27 32 40.88 & 0.49600 & 0.00006 & 2 & \nodata & \nodata & [OII]3727,CaII H,CaII K \\
9 & 21 52 08.91 & $-$27 32 20.11 & 0.00000 & 0.00000 & 0 & \nodata & \nodata & \nodata \\
10 & 21 52 10.34 & $-$27 32 02.94 & 0.81198 & 0.00006 & 1 & \nodata & \nodata & [OII]3727,CaII H,CaII K \\
11 & 21 52 04.90 & $-$27 32 19.93 & 0.59416 & 0.00021 & 1 & \nodata & \nodata & [OII]3727,H$_{\gamma}$ \\
12 & 21 52 08.23 & $-$27 31 58.62 & \nodata & \nodata & 0 & \nodata & \nodata & \nodata \\
13 & 21 52 07.05 & $-$27 31 53.58 & \nodata & \nodata & 0 & \nodata & \nodata & \nodata \\
14 & 21 52 06.33 & $-$27 31 25.36 & 0.60271 & 0.00006 & 2 & \nodata & \nodata & [OII]3727,CaII H,CaII K \\ 
15 & 21 52 07.24 & $-$27 31 08.01 & 0.80307 & 0.00021 & 2 & \nodata & \nodata & [OII]3727,[OIII]5007 \\
16 & 21 51 56.50 & $-$27 33 07.38 & 0.60998 & 0.00010 & 1 & \nodata & \nodata & [OII]3727,CaII H,CaII K,H$_{\gamma}$ \\
17 & 21 51 57.70 & $-$27 32 35.09 & 0.48246 & 0.00005 & 1 & \nodata & \nodata & [OII]3727,H$_{\gamma}$,H$_{\beta}$,[OIII]5007 \\
18 & 21 51 59.70 & $-$27 32 25.05 & 0.73870 & 0.00012 & 1 & \nodata & \nodata & [OII]3727,CaII H,CaII K \\
19 & 21 52 01.74 & $-$27 32 06.72 & 0.60433 & 0.00015 & 1 & \nodata & \nodata & [OII]3727,CaII K,H$_{\gamma}$,H$_{\beta}$ \\
20 & 21 52 05.17 & $-$27 30 59.90 & 0.46006 & 0.00010 & 1 & \nodata & \nodata & [OII]3727,H$_{\beta}$,[OIII]4959,[OIII]5007 \\
21 & 21 52 01.46 & $-$27 29 25.12 & 0.47426 & 0.00007 & 3 & \nodata & \nodata & H$_{\beta}$,[OIII]5007 \\
22 & 21 52 04.87 & $-$27 29 29.26 & 0.60338 & 0.00012 & 3 & \nodata & \nodata & H$_{\beta}$,[OIII]4959,[OIII]5007 \\
23 & 21 52 02.73 & $-$27 29 13.27 & \nodata & \nodata & 0 & \nodata & \nodata & \nodata \\
24 & 21 51 55.45 & $-$27 33 04.18 & \nodata & \nodata & 0 & \nodata & \nodata & \nodata \\
25 & 21 51 57.60 & $-$27 29 21.88 & 0.81260 & 0.00040 & 2 & \nodata & \nodata & [OII]3727,CaII H,CaII K \\
26 & 21 52 18.32 & $-$27 31 27.19 & 0.65148 & 0.00012 & 1 & \nodata & \nodata & CaII H,CaII K,[OII]3727 \\
27 & 21 52 19.74 & $-$27 30 32.79 & 0.80710 & 0.00078 & 3 & \nodata & \nodata & CaII H,CaII K \\
28 & 21 52 18.36 & $-$27 30 19.18 & 0.53370 & 0.00014 & 3 & \nodata & \nodata & [OII]3727,CaII K \\
29 & 21 52 14.48 & $-$27 33 12.13 & 0.27426 & 0.00009 & 1 & \nodata & \nodata & [OII]3727,CaII H,CaII K,H$_{\beta}$,[OIII]5007 \\
30 & 21 52 12.30 & $-$27 32 40.99 & \nodata & \nodata & 0 & \nodata & \nodata & \nodata \\
31 & 21 52 15.07 & $-$27 31 23.88 & 0.73889 & 0.00012 & 2 & \nodata & \nodata & [OII]3727,CaII K,G band \\
32 & 21 52 14.30 & $-$27 31 02.97 & 0.73953 & 0.00046 & 2 & \nodata & \nodata & [OII]3727,CaII H,CaII K \\
33 & 21 52 10.31 & $-$27 33 37.44 & 0.35786 & 0.00006 & 1 & \nodata & \nodata & [OII]3727,H$_{\beta}$,[OIII]5007 \\
34 & 21 52 07.91 & $-$27 33 42.95 & 0.27272 & 0.00010 & 1 & \nodata & \nodata & [OII]3727,CaII H,CaII K,H$_{\beta}$ \\
35 & 21 52 03.40 & $-$27 32 47.87 & 0.32849 & 0.00004 & 1 & \nodata & \nodata & [OII]3727,H$_{\gamma}$,H$_{\beta}$,[OIII]4959,[OIII]5007 \\
36 & 21 52 04.66 & $-$27 32 28.68 & 0.47360 & 0.00007 & 3 & \nodata & \nodata & CaII H,CaII K \\
37 & 21 52 10.22 & $-$27 32 10.14 & 0.19677 & 0.00006 & 1 & \nodata & \nodata & H$_{\beta}$,[OIII]4959,[OIII]5007 \\
38 & 21 52 08.18 & $-$27 31 53.62 & 0.60045 & 0.00023 & 1 & \nodata & \nodata & [OII]3727,CaII K,G band \\
39 & 21 52 07.25 & $-$27 31 35.79 & 0.27610 & 0.00035 & 3 & \nodata & \nodata & CaII H,CaII K \\
40 & 21 52 06.81 & $-$27 31 25.47 & 0.40924 & 0.00012 & 1 & \nodata & \nodata & [OII]3727,CaII H,CaII K \\
41 & 21 52 06.70 & $-$27 30 51.48 & 0.59591 & 0.00022 & 1 & \nodata & \nodata & [OII]3727,CaII H,CaII K,G band,H$_{\gamma}$ \\
42 & 21 52 05.17 & $-$27 29 36.45 & 0.29960 & 0.00007 & 2 & \nodata & \nodata & H$_{\beta}$,[OIII]5007 \\
43 & 21 51 57.49 & $-$27 33 30.46 & 0.40688 & 0.00009 & 1 & \nodata & \nodata & [OII]3727,CaII H,H$_{\gamma}$,H$_{\beta}$,[OIII]5007 \\
44 & 21 51 59.52 & $-$27 33 20.49 & 2.55150 & 0.00757 & 3 & \nodata & \nodata & CIV 1550,CIII 1909 \\
45 & 21 52 01.28 & $-$27 32 41.57 & 0.27412 & 0.00012 & 1 & \nodata & \nodata & CaII H,CaII K,G band \\
46 & 21 51 56.86 & $-$27 32 29.94 & 0.49543 & 0.00017 & 3 & \nodata & \nodata & [OII]3727,CaII H,CaII K \\ 
47 & 21 51 56.46 & $-$27 31 59.85 & 0.27373 & 0.00012 & 1 & \nodata & \nodata & [OII]3727,H$_{\beta}$,[OIII]5007 \\
48 & 21 51 55.71 & $-$27 31 54.91 & 0.38975 & 0.00010 & 1 & \nodata & \nodata & [OII]3727,H$_{\gamma}$,H$_{\beta}$,[OIII]5007 \\
49 & 21 52 01.98 & $-$27 29 16.12 & 0.54816 & 0.00005 & 1 & \nodata & \nodata & [OII]3727,H$_{\gamma}$,H$_{\beta}$,[OIII]4959 \\
\enddata
\tablenotetext{a}{Redshift reliability classifier.}
\tablecomments{Whenever information could not be obtained for a specific target, a 
symbol '...' is used. Stars found in our data have zero redshift.}
\label{tabhe2149}
\end{deluxetable}

\newpage
\begin{figure}
\begin{center}
\includegraphics[height=14cm,width=15cm]{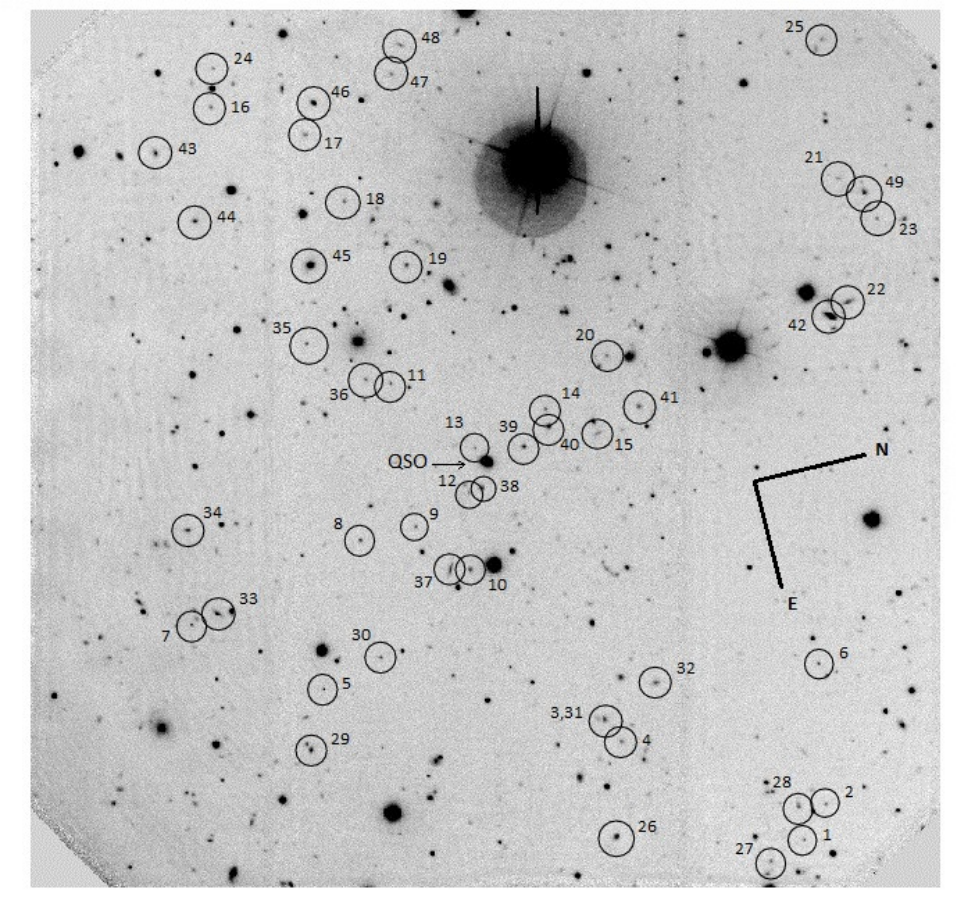}
\end{center}
\begin{flushright}
\includegraphics[height=5cm,width=7cm]{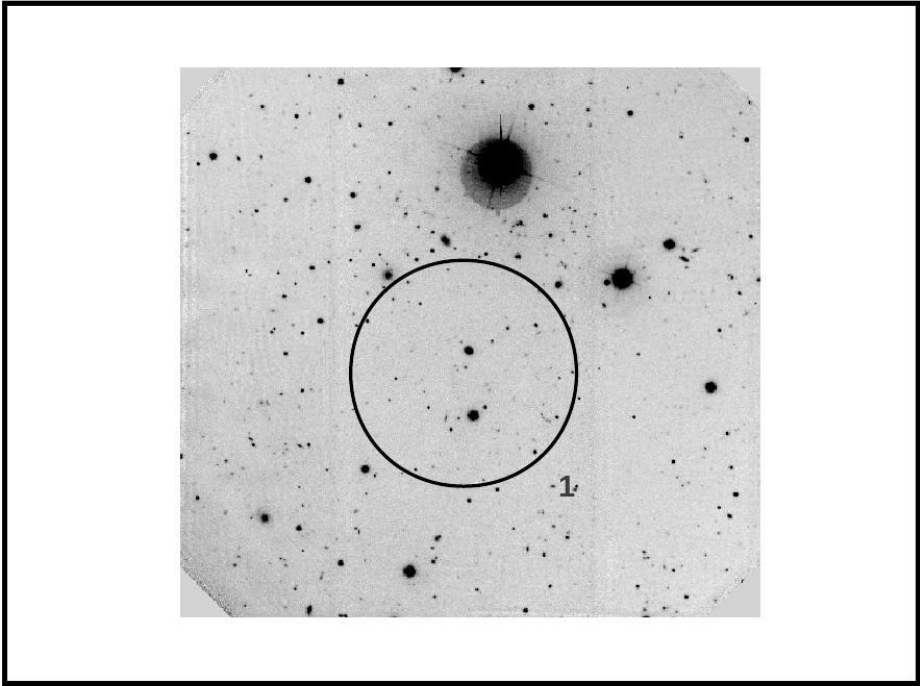} 
\end{flushright}
\begin{center}
\caption{{\it Top}: 5.5$\arcmin\times$5.5$\arcmin$ image of the field centered on the 
SDSS quasar HE2149$-$2745A. Galaxies are labeled according to the 
identification number given in Table \ref{tabhe2149}.  {\it Bottom}: A zoom-out of the image shown at the 
top. Center coordinates of each RCS1 cluster/group candidate are shown in circles. Each 
cluster is labeled according to their identification numbers 
given in the redshift histogram of Figure \ref{hishe2149}.}
\label{camphe2149}
\end{center}
\end{figure}
\clearpage

\newpage
\begin{figure}
\begin{center}
\includegraphics[height=16cm,width=16cm]{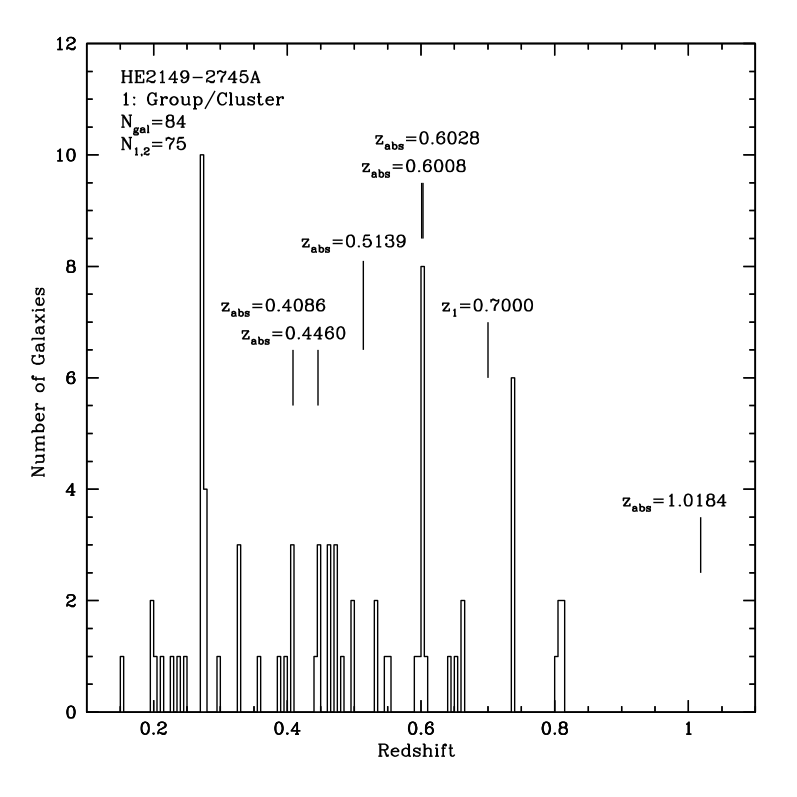}
\caption{Redshift histogram of the field centered on the SDSS quasar HE2149$-$2745A. The 
bin size is of 0.005 in redshift space, which translates in $\Delta v \sim$ 1000 km/s at 
$z =$ 0.5. The total number of redshifts available for this field is given by N$_{gal}$, 
from which N$_{1,2}$ is the number of redshifts classified with reliability flag 1 or 2.}
\label{hishe2149}
\end{center}
\end{figure}
\clearpage

\newpage
\begin{deluxetable}{ccccccccp{3.2in}}
\tablewidth{0pc}
\tablecaption{Spectroscopic Targets of Field Centered on 231500.81$-$001831.2.}
\tabletypesize{\tiny}
\tablehead{ \colhead{No.} & \colhead{RA(J2000)} & \colhead{DEC(J2000)} & 
\colhead{$z_{gal}$} & \colhead{$\sigma_{z_{gal}}$} & \colhead{Flag\tablenotemark{a}} & 
\colhead{$z'$} & \colhead{$R_{c}$ - $z'$} & \colhead{Comments}}
\startdata
1 & 23 15 05.95 & $-$00 16 06.92 & \nodata & \nodata & 0 & 20.94 & 1.48 & \nodata \\
2 & 23 15 00.86 & $-$00 15 58.32 & 0.22311 & 0.00006 & 1 & 21.98 & 0.11 & H$_{\beta}$,[OIII]4959,[OIII]5007 \\
3 & 23 15 01.58 & $-$00 16 46.02 & 0.99420 & 0.00080 & 3 & 22.16 & 0.83 & [OII]3727 \\
4 & 23 14 59.20 & $-$00 16 18.37 & \nodata & \nodata & 0 & 21.02 & 1.65 & \nodata \\
5 & 23 14 56.03 & $-$00 16 58.80 & 0.58952 & 0.00007 & 3 & 21.06 & 0.52 & [OII]3727,CaII H \\
6 & 23 14 55.68 & $-$00 16 54.52 & 0.74977 & 0.00080 & 2 & 22.33 & 0.81 &  [OII]3727 \\
7 & 23 14 53.88 & $-$00 17 24.40 & 0.74160 & 0.00007 & 3 & 20.96 & 0.67 & [OII]3727,CaII K \\
8 & 23 15 01.89 & $-$00 16 55.20 & 0.59264 & 0.00017 & 1 & 21.16 & 0.70 & [OII]3727,CaII H,CaII K \\
9 & 23 15 01.63 & $-$00 17 09.53 & 0.74944 & 0.00012 & 1 & 21.73 & 1.16 & [OII]3727,CaII H,CaII K \\
10 & 23 15 03.43 & $-$00 17 17.74 & 0.50683 & 0.00012 & 2 & 21.08 & 0.84 & CaII H,CaII K,G band \\
11 & 23 15 05.64 & $-$00 18 53.28 & 0.50470 & 0.00025 & 1 & 20.75 & 0.94 & [OII]3727,CaII H,CaII K,G band \\
12 & 23 15 04.06 & $-$00 18 56.59 & 0.31196 & 0.00005 & 1 & 21.10 & 0.22 & [OII]3727,H$_{\beta}$,[OIII]4959,[OIII]5007 \\
13 & 23 15 00.92 & $-$00 18 14.98 & 0.29255 & 0.00005 & 1 & 20.99 & 0.45 & [OII]3727,H$_{\beta}$,[OIII]4959,[OIII]5007 \\
14 & 23 15 00.96 & $-$00 18 24.52 & 0.41367 & 0.00003 & 1 & 22.47 & -0.03 & [OII]3727,[OIII]4959,[OIII]5007 \\
15 & 23 15 00.21 & $-$00 18 08.06 & 0.41360 & 0.00004 & 1 & 20.23 & 0.39 & [OII]3727,H$_{\delta}$,H$_{\gamma}$,H$_{\beta}$,[OIII]4959,[OIII]5007 \\
16 & 23 15 01.29 & $-$00 18 49.86 & 0.58493 & 0.00080 & 3 & 20.65 & 1.16 & [OII]3727 \\
17 & 23 15 00.62 & $-$00 18 43.42 & 0.50576 & 0.00080 & 1 & 22.07 & 0.42 & [OII]3727 \\
18 & 23 15 00.31 & $-$00 18 38.92 & 0.49888 & 0.00042 & 2 & 21.27 & 0.07 & [OII]3727,CaII H \\
19 & 23 15 00.90 & $-$00 19 06.82 & 0.57703 & 0.00007 & 3 & 22.02 & 0.55 & [OII]3727,CaII K \\
20 & 23 14 56.02 & $-$00 18 20.38 & \nodata & \nodata & 0 & 22.37 & 0.74 & \nodata \\
21 & 23 15 06.88 & $-$00 19 24.53 & 0.68810 & 0.00014 & 2 & 21.52 & 0.72 & [OII]3727,CaII K \\
22 & 23 15 06.73 & $-$00 19 33.85 & 0.68793 & 0.00028 & 2 & 20.05 & 1.33 & CaII H,CaII K \\
23 & 23 15 03.98 & $-$00 19 25.72 & 0.59259 & 0.00029 & 2 & 20.58 & 0.94 & [OII]3727,CaII H,CaII K \\
24 & 23 15 03.17 & $-$00 19 44.40 & 0.50138 & 0.00080 & 3 & 22.00 & 0.57 & [OII]3727 \\
25 & 23 15 01.42 & $-$00 19 33.10 & \nodata & \nodata & 0 & 20.17 & 1.63 & \nodata \\
26 & 23 15 03.15 & $-$00 20 12.73 & 0.22222 & 0.00012 & 1 & 20.13 & 0.70 & H$_{\beta}$,[OIII]4959,[OIII]5007 \\
27 & 23 15 02.82 & $-$00 20 22.13 & 0.69295 & 0.00029 & 1 & 21.72 & 0.60 & [OII]3727,CaII H,CaII K \\
28 & 23 15 05.81 & $-$00 20 17.59 & 0.59108 & 0.00021 & 2 & 20.13 & 0.84 & [OII]3727,CaII K \\ 
29 & 23 15 09.82 & $-$00 20 10.07 & 0.58384 & 0.00007 & 1 & 20.19 & 1.17 & CaII H,CaII K \\
30 & 23 14 56.27 & $-$00 16 22.91 & 0.37639 & 0.00012 & 1 & 21.34 & 0.46 & [OII]3727,H$_{\beta}$,[OIII]5007 \\
31 & 23 14 55.54 & $-$00 16 55.56 & \nodata & \nodata & 0 & 22.33 & 0.81 & \nodata \\
32 & 23 14 55.19 & $-$00 16 49.98 & 0.38087 & 0.00007 & 2 & 21.06 & 0.54 & [OII]3727,H$_{\beta}$ \\ 
33 & 23 14 53.03 & $-$00 16 55.34 & \nodata & \nodata & 0 & 20.29 & 1.70 & \nodata \\
34 & 23 15 03.22 & $-$00 17 04.52 & 0.91013 & 0.00080 & 1 & 21.74 & 0.66 & [OII]3727 \\
35 & 23 15 01.68 & $-$00 16 29.32 & 0.77120 & 0.00080 & 3 & 22.52 & 0.76 & [OII]3727 \\
36 & 23 15 01.89 & $-$00 16 55.20 & 0.59352 & 0.00021 & 1 & 21.16 & 0.70 & [OII]3727,CaII K \\
37 & 23 14 58.80 & $-$00 17 07.40 & 0.50386 & 0.00012 & 2 & 19.50 & 1.03 & [OII]3727,CaII K,G band \\
38 & 23 14 52.06 & $-$00 18 12.74 & 0.69403 & 0.00080 & 3 & 22.24 & 0.85 & [OII]3727 \\
39 & 23 14 50.80 & $-$00 18 40.97 & 0.50149 & 0.00014 & 1 & 19.43 & 0.96 & CaII H,CaII K \\
40 & 23 15 03.39 & $-$00 17 24.00 & 0.50529 & 0.00007 & 1 & 23.12 & 0.05 & CaII H,CaII K \\
41 & 23 15 02.99 & $-$00 17 37.07 & 0.50416 & 0.00007 & 2 & 21.15 & 0.47 & [OII]3727,CaII K \\
42 & 23 15 04.39 & $-$00 18 49.82 & \nodata & \nodata & 0 & 21.94 & 0.70 & \nodata \\
43 & 23 15 02.88 & $-$00 18 25.85 & 0.50343 & 0.00080 & 3 & 21.11 & 0.77 & [OII]3727 \\
44 & 23 15 02.31 & $-$00 18 41.51 & 0.50527 & 0.00007 & 1 & 20.21 & 0.87 & CaII K,G band \\
45 & 23 15 00.69 & $-$00 18 16.78 & \nodata & \nodata & 0 & 20.32 & 1.71 & \nodata \\
46 & 23 15 01.53 & $-$00 19 18.66 & 0.50250 & 0.00035 & 3 & 20.82 & 0.80 & CaII H,CaII K,H$_{\delta}$,G band \\
47 & 23 14 59.97 & $-$00 18 09.72 & 0.50278 & 0.00080 & 1 & 20.15 & 0.88 & [OII]3727 \\
48 & 23 14 59.14 & $-$00 19 19.49 & 0.57724 & 0.00080 & 3 & 20.76 & 0.46 & [OII]3727 \\
49 & 23 14 57.46 & $-$00 18 53.93 & 0.50294 & 0.00006 & 1 & 19.61 & 0.93 & CaII H,CaII K,G band \\
50 & 23 15 09.36 & $-$00 19 09.23 & \nodata & \nodata & 0 & 20.63 & 1.16 & \nodata \\
51 & 23 15 04.99 & $-$00 19 26.80 & 0.43890 & 0.00012 & 1 & 21.88 & 0.68 & [OII]3727,H$_{\beta}$,[OIII]5007 \\
52 & 23 15 02.17 & $-$00 19 33.46 & 0.59303 & 0.00007 & 1 & 19.87 & 1.19 & CaII H,CaII K \\
53 & 23 15 11.60 & $-$00 19 29.17 & 0.47277 & 0.00042 & 1 & 19.45 & 0.89 & CaII H,CaII K \\
54 & 23 15 02.67 & $-$00 20 15.86 & 0.57669 & 0.00021 & 2 & 20.71 & 0.86 & CaII H,CaII K \\
55 & 23 15 11.01 & $-$00 20 02.76 & 0.58508 & 0.00007 & 1 & 20.19 & 1.05 & CaII H,CaII K \\
\enddata
\tablenotetext{a}{Redshift reliability classifier.}
\tablecomments{Whenever information could not be obtained for a specific target, a 
symbol '...' is used. Stars found in our data have zero redshift.}
\label{tab231500}
\end{deluxetable}

\newpage 
\begin{figure}
\begin{center}
\includegraphics[height=14cm,width=15cm]{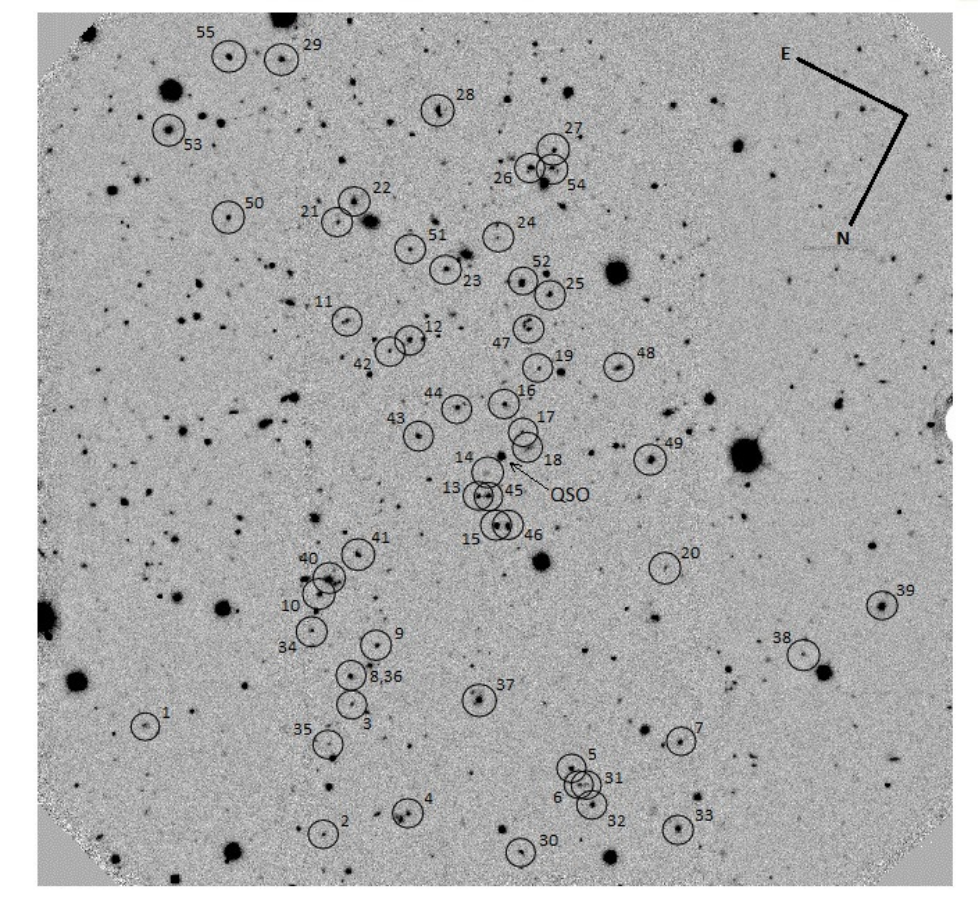}
\end{center}
\begin{flushright}
\includegraphics[height=5cm,width=7cm]{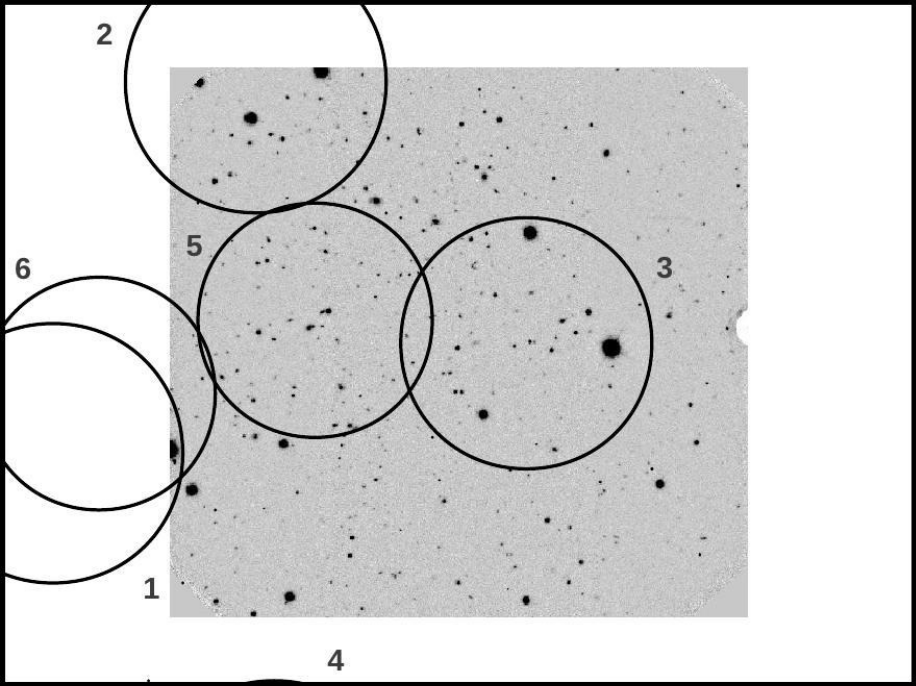} 
\end{flushright}
\begin{center}
\caption{{\it Top}: 5.5$\arcmin\times$5.5$\arcmin$ image of the field centered on the 
SDSS quasar 231500.81$-$001831.2. Galaxies are labeled according to the 
identification number given in Table \ref{tab231500}.  {\it Bottom}: A zoom-out of the image shown at the 
top. Center coordinates of each RCS1 cluster/group candidate are shown in circles. Each 
cluster is labeled according to their identification numbers 
given in the redshift histogram of Figure \ref{his231500}.}
\label{camp231500}
\end{center}
\end{figure}
\clearpage

\newpage
\begin{figure}
\begin{center}
\includegraphics[height=16cm,width=16cm]{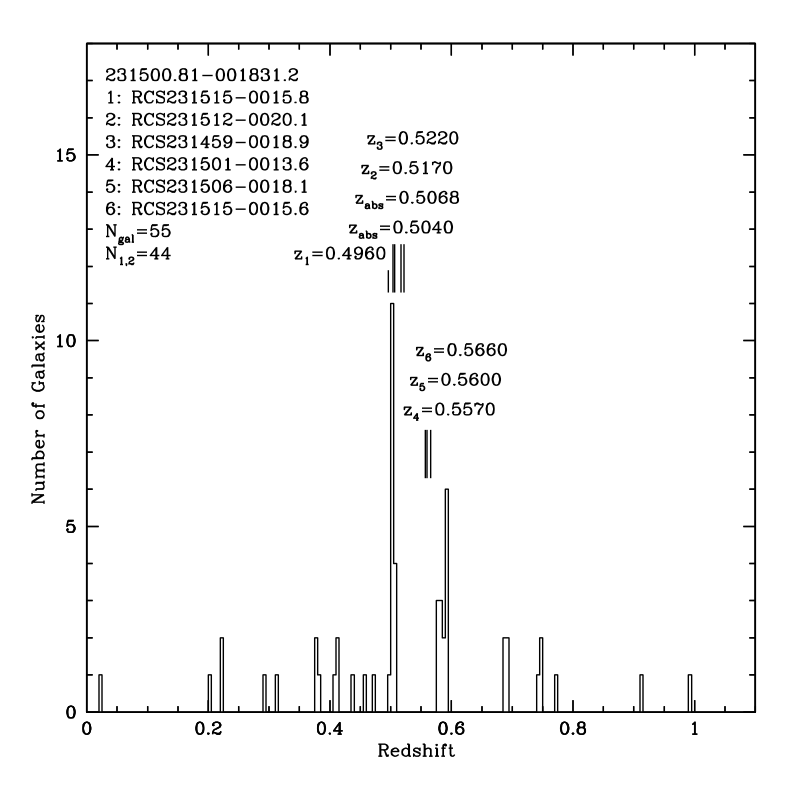}
\caption{Redshift histogram of the field centered on the SDSS quasar 
231500.81$-$001831.2. The bin size is of 0.005 in redshift space, which translates 
in $\Delta v \sim$ 1000 km/s at $z =$ 0.5. The total number of redshifts available 
for this field is given by N$_{gal}$, from which N$_{1,2}$ is the number of redshifts 
classified with reliability flag 1 or 2.}
\label{his231500}
\end{center}
\end{figure}
\clearpage

\newpage
\begin{deluxetable}{ccccccccp{3.2in}}
\tablewidth{0pc}
\tablecaption{Spectroscopic Targets of Field Centered on 231509.34$+$001026.2.}
\tabletypesize{\tiny}
\tablehead{ \colhead{No.} & \colhead{RA(J2000)} & \colhead{DEC(J2000)} & 
\colhead{$z_{gal}$} & \colhead{$\sigma_{z_{gal}}$} & \colhead{Flag\tablenotemark{a}} & 
\colhead{$z'$} & \colhead{$R_{c}$ - $z'$} & \colhead{Comments}}
\startdata
1 & 23 15 09.52 & $+$00 08 32.89 & \nodata & \nodata & 0 & 20.97 & 0.30 & \nodata \\
2 & 23 15 10.24 & $+$00 08 22.74 & 0.20110 & 0.00003 & 1 & 21.59 & 0.21 & H$_{\beta}$,[OIII]4959,[OIII]5007 \\
3 & 23 15 19.93 & $+$00 09 54.97 & 0.38483 & 0.00015 & 3 & 21.54 & 0.58 & H$_{\beta}$,[OIII]5007 \\
4 & 23 15 03.91 & $+$00 08 42.72 & 0.37344 & 0.00005 & 2 & 21.59 & 0.42 & [OII]3727,H$_{\beta}$ \\
5 & 23 15 07.31 & $+$00 09 40.68 & 0.47830 & 0.00001 & 1 & 19.89 & 0.69 & [OII]3727,CaII H,CaII K \\
6 & 23 15 08.39 & $+$00 10 06.96 & 0.49733 & 0.00015 & 2 & 22.23 & 0.41 & [OII]3727,[OIII]5007 \\
7 & 23 15 10.24 & $+$00 09 24.91 & 0.29651 & 0.00009 & 1 & 20.76 & 0.36 & [OII]3727,H$_{\beta}$,[OIII]4959,[OIII]5007 \\
8 & 23 15 13.93 & $+$00 10 25.36 & 0.37115 & 0.00005 & 1 & 22.17 & 0.08 & [OII]3727,H$_{\beta}$,[OIII]4959,[OIII]5007 \\
9 & 23 15 19.53 & $+$00 10 12.18 & \nodata & \nodata & 0 & 21.18 & 0.64 & \nodata \\
10 & 23 15 17.87 & $+$00 10 15.13 & \nodata & \nodata & 0 & 21.75 & 0.27 & \nodata \\
11 & 23 15 05.30 & $+$00 10 48.86 & 0.43027 & 0.00007 & 1 & 22.08 & -0.05 & [OII]3727,[OIII]5007,H$_{\beta}$,[OIII]4959 \\
12 & 23 15 07.22 & $+$00 10 54.62 & 0.41790 & 0.00005 & 2 & 20.65 & 1.39 & H$_{\beta}$,[OIII]4959 \\
13 & 23 15 07.88 & $+$00 10 53.36 & 0.37240 & 0.00023 & 2 & 20.79 & 0.63 & CaII H,CaII K,H$_{\gamma}$ \\
14 & 23 15 08.50 & $+$00 10 40.76 & 0.59149 & 0.00080 & 3 & 22.85 & -0.13 & [OII]3727 \\
15 & 23 15 08.47 & $+$00 10 47.06 & \nodata & \nodata & 0 & 21.71 & 1.05 & \nodata \\
16 & 23 15 13.16 & $+$00 10 55.34 & 0.79466 & 0.00080 & 3 & 21.92 & 0.33 & [OII]3727 \\
17 & 23 15 07.26 & $+$00 11 36.96 & 0.37010 & 0.00005 & 3 & 20.05 & 0.50 & H$_{\beta}$,[OIII]5007 \\
17 & 23 15 07.26 & $+$00 11 36.96 & 0.72158 & 0.00080 & 3 & 21.39 & 0.39 & [OII]3727 \\
18 & 23 15 08.77 & $+$00 11 39.70 & 0.75464 & 0.00080 & 3 & 21.64 & 1.44 & [OII]3727 \\
19 & 23 15 09.24 & $+$00 12 24.52 & \nodata & \nodata & 0 & 21.40 & 1.12 & \nodata \\
20 & 23 15 11.83 & $+$00 12 33.26 & 0.43011 & 0.00001 & 1 & 19.30 & 0.80 & CaII H,CaII K,G band \\
21 & 23 15 11.36 & $+$00 12 49.18 & 0.37113 & 0.00006 & 1 & 20.51 & 0.26 & H$_{\beta}$,[OIII]4959,[OIII]5007,H$_{\gamma}$ \\
22 & 23 15 06.56 & $+$00 12 41.69 & 0.40355 & 0.00012 & 2 & 22.60 & 0.26 & [OII]3727,OIII4363,H$_{\beta}$,[OIII]5007 \\
23 & 23 15 12.16 & $+$00 13 19.02 & \nodata & \nodata & 0 & 21.83 & 0.50 & \nodata \\
24 & 23 15 09.92 & $+$00 08 08.38 & 0.42953 & 0.00075 & 3 & 24.19 & 0.10 & H$_{\beta}$,[OIII]5007 \\
25 & 23 15 10.57 & $+$00 08 26.02 & 0.20009 & 0.00003 & 1 & 22.67 & 0.25 & [OII]3727,H$_{\beta}$,[OIII]4959,[OIII]5007 \\
26 & 23 15 04.04 & $+$00 08 17.81 & 0.41585 & 0.00012 & 2 & 19.87 & 0.78 & [OII]3727,CaII H,CaII K \\
27 & 23 15 09.61 & $+$00 08 49.52 & 0.37335 & 0.00011 & 1 & 18.81 & 0.87 & CaII H,CaII K,G band,MgI 5176 \\
28 & 23 15 10.01 & $+$00 09 20.81 & \nodata & \nodata & 0 & 21.86 & 0.34 & \nodata \\
29 & 23 15 10.04 & $+$00 09 57.64 & 0.36937 & 0.00011 & 1 & 21.39 & 0.48 & [OII]3727,CaII K,G band,H$_{\beta}$,[OIII]5007 \\
30 & 23 15 10.88 & $+$00 09 39.02 & 0.31363 & 0.00006 & 1 & 20.25 & 0.70 & H$_{\beta}$,[OIII]4959,[OIII]5007,[OII]3727,H$_{\gamma}$ \\
31 & 23 15 10.21 & $+$00 10 12.43 & 0.84680 & 0.00080 & 3 & 20.09 & 1.56 & [OII]3727 \\
32 & 23 15 09.95 & $+$00 10 32.27 & 0.44652 & 0.00008 & 1 & 21.69 & 0.01 & [OIII]5007,[OIII]4959,H$_{\beta}$,H$_{\gamma}$,H$_{\delta}$,[OII]3727 \\
33 & 23 15 10.78 & $+$00 10 30.00 & \nodata & \nodata & 0 & 21.23 & 0.84 & \nodata \\
34 & 23 15 12.18 & $+$00 10 05.38 & 0.49620 & 0.00012 & 1 & 22.38 & -0.22 & CaII H,CaII K,H$_{\delta}$,H$_{\gamma}$ \\
35 & 23 15 16.39 & $+$00 09 26.57 & 0.63448 & 0.00010 & 1 & 20.97 & 0.37 & [OII]3727,[OIII]5007,[OIII]4959 \\
36 & 23 15 04.75 & $+$00 10 14.16 & 0.47947 & 0.00008 & 1 & 19.69 & 0.64 & [OII]3727,CaII H,CaII K,G band \\
37 & 23 15 06.72 & $+$00 10 22.01 & 0.26136 & 0.00006 & 1 & 19.50 & 0.54 & H$_{\alpha}$,[OIII]5007,H$_{\beta}$,H$_{\gamma}$,[OII]3727 \\
38 & 23 15 07.05 & $+$00 10 49.58 & 0.37117 & 0.00012 & 2 & 21.13 & 0.60 & CaII H,CaII K,G band \\
39 & 23 15 05.09 & $+$00 10 54.05 & 0.47045 & 0.00017 & 1 & 20.54 & 0.53 & CaII H,CaII K,[OII]3727,H$_{\beta}$,[OIII]5007 \\
40 & 23 15 08.10 & $+$00 11 48.26 & 0.37283 & 0.00013 & 1 & 19.56 & 0.71 & CaII H,CaII K,G band \\
41 & 23 15 08.35 & $+$00 12 01.80 & 0.43074 & 0.00001 & 1 & 19.52 & 0.55 & CaII H,CaII K \\
42 & 23 15 09.54 & $+$00 11 42.65 & 0.37023 & 0.00006 & 1 & 21.50 & 0.60 & [OII]3727,H$_{\beta}$,FeII 5284 \\
43 & 23 15 09.91 & $+$00 11 54.06 & 0.37290 & 0.00050 & 3 & 20.85 & 0.60 & CaII H,CaII K \\
44 & 23 15 08.90 & $+$00 12 28.15 & 0.42986 & 0.00025 & 2 & 19.62 & 0.82 & CaII H,CaII K \\
45 & 23 15 13.50 & $+$00 12 54.32 & 0.79003 & 0.00089 & 2 & 19.72 & 0.63 & [OII]3727,CaII H,CaII K \\
46 & 23 15 06.37 & $+$00 12 38.45 & 0.54693 & 0.00031 & 2 & 22.01 & 0.33 & [OII]3727,[OIII]4959,H$_{\beta}$,[OIII]5007 \\
47 & 23 15 07.53 & $+$00 12 38.30 & 0.43340 & 0.00012 & 2 & 19.29 & 0.74 & CaII H,CaII K,G band \\ 
48 & 23 15 06.89 & $+$00 12 55.58 & \nodata & \nodata & 0 & 20.30 & 1.24 & \nodata \\
\enddata
\tablenotetext{a}{Redshift reliability classifier.}
\tablecomments{Whenever information could not be obtained for a specific target, a 
symbol '...' is used. Stars found in our data have zero redshift.}
\label{tab231509}
\end{deluxetable}

\newpage 
\begin{figure}
\begin{center}
\includegraphics[height=14cm,width=15cm]{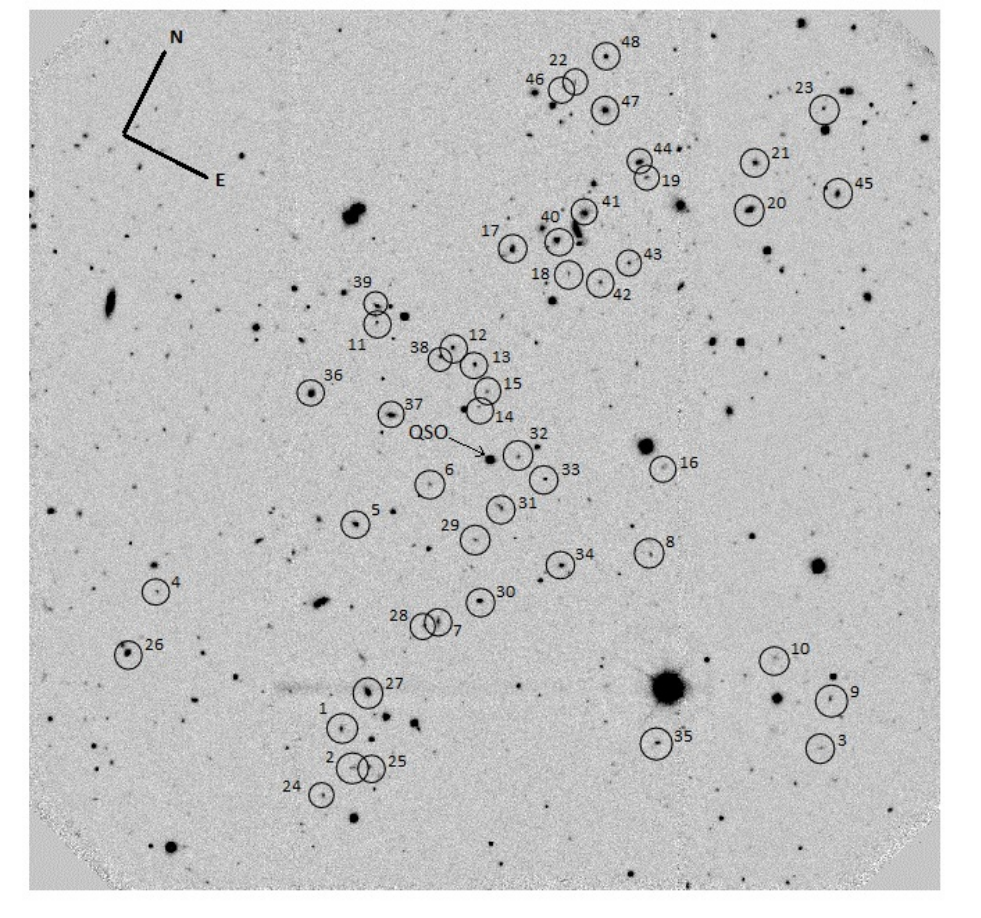}
\end{center}
\begin{flushright}
\includegraphics[height=5cm,width=7cm]{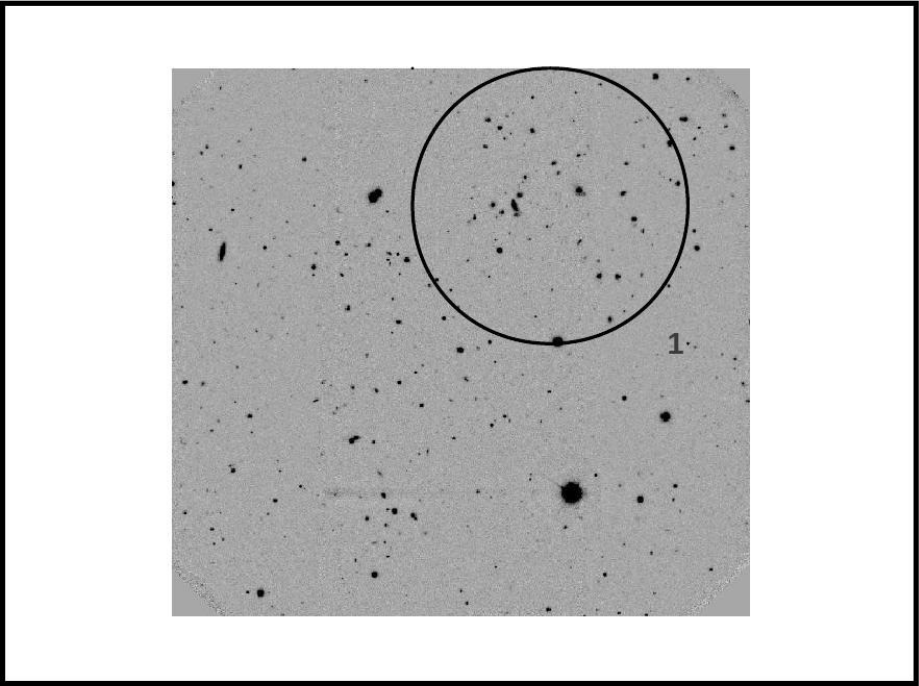} 
\end{flushright}
\begin{center}
\caption{{\it Top}: 5.5$\arcmin\times$5.5$\arcmin$ image of the field centered on the 
SDSS quasar 231509.34$+$001026.2. Galaxies are labeled according to the 
identification number given in Table \ref{tab231509}.  {\it Bottom}: A zoom-out of the image shown at the 
top. Center coordinates of each RCS1 cluster/group candidate are shown in circles. Each 
cluster is labeled according to their identification numbers 
given in the redshift histogram of Figure \ref{his231509}.}
\label{camp231509}
\end{center}
\end{figure}
\clearpage

\newpage
\begin{figure}
\begin{center}
\includegraphics[height=16cm,width=16cm]{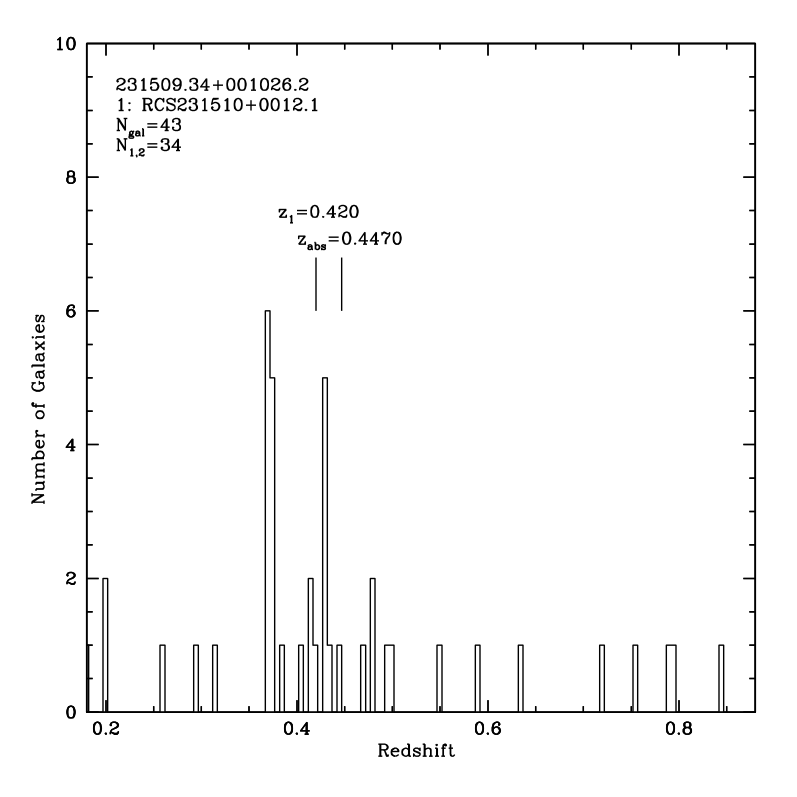}
\caption{Redshift histogram of the field centered on the SDSS quasar 
231509.34$+$001026.2. The bin size is of 0.005 in redshift space, which translates 
in $\Delta v \sim$ 1000 km/s at $z =$ 0.5. The total number of redshifts available 
for this field is given by N$_{gal}$, from which N$_{1,2}$ is the number of redshifts 
classified with reliability flag 1 or 2.}
\label{his231509}
\end{center}
\end{figure}
\clearpage

\newpage
\begin{deluxetable}{ccccccccp{3.2in}}
\tablewidth{0pc}
\tablecaption{Spectroscopic Targets of Field Centered on 231759.63$-$000733.2.}
\tabletypesize{\tiny}
\tablehead{ \colhead{No.} & \colhead{RA(J2000)} & \colhead{DEC(J2000)} & 
\colhead{$z_{gal}$} & \colhead{$\sigma_{z_{gal}}$} & \colhead{Flag\tablenotemark{a}} & 
\colhead{$z'$} & \colhead{$R_{c}$ - $z'$} & \colhead{Comments}}
\startdata
1 & 23 18 06.15 & $-$00 05 27.64 & 0.69533 & 0.00029 & 1 & 22.40 & 0.24 & [OII]3727,CaII H,CaII K \\
2 & 23 17 50.90 & $-$00 05 16.98 & 0.70286 & 0.00009 & 1 & 21.43 & 0.30 & [OII]3727,[NeIII]3868,HeI 3888,H$_{\delta}$,H$_{\gamma}$ \\
3 & 23 17 59.34 & $-$00 05 43.69 & 0.46756 & 0.00010 & 1 & 22.49 & 0.29 & [OII]3727,H$_{\beta}$,[OIII]4959,[OIII]5007 \\
4 & 23 17 57.61 & $-$00 06 04.93 & 0.46890 & 0.00006 & 2 & 21.52 & 0.95 & [OII]3727,CaII H,CaII K \\
5 & 23 17 57.33 & $-$00 05 54.60 & 0.31495 & 0.00005 & 1 & 21.82 & 0.35 & [OII]3727,H$_{\beta}$,[OIII]4959,[OIII]5007 \\
6 & 23 18 02.59 & $-$00 07 39.32 & 1.41520 & 0.00028 & 1 & 22.50 & 0.30 & FeII 2344,FeII 2374,FeII 2382,FeII 2586,FeII 2600,MnII 2594,MgII 2796,MgII 2803 \\
7 & 23 18 01.39 & $-$00 06 30.17 & 0.50117 & 0.00023 & 2 & 21.36 & 0.69 & [OII]3727,CaII H,CaII K \\
8 & 23 18 01.20 & $-$00 07 13.08 & 0.59330 & 0.00007 & 1 & 22.46 & 0.30 & [OII]3727,H$_{\gamma}$ \\
9 & 23 18 01.04 & $-$00 07 22.08 & \nodata & \nodata & 0 & 22.33 & 1.57 & \nodata \\
10 & 23 18 00.65 & $-$00 06 49.72 & \nodata & \nodata & 0 & 22.65 & 0.58 & \nodata \\
11 & 23 18 00.65 & $-$00 06 59.22 & 0.59770 & 0.00030 & 3 & 23.07 & 0.38 & [OII]3727,CaII H,CaII K,G band \\
12 & 23 18 00.17 & $-$00 06 42.16 & \nodata & \nodata & 0 & 22.23 & 1.98 & \nodata \\
13 & 23 17 56.81 & $-$00 07 28.60 & \nodata & \nodata & 0 & 22.42 & 0.94 & \nodata \\
14 & 23 18 02.34 & $-$00 08 05.60 & 0.47300 & 0.00012 & 1 & 21.57 & 0.35 & [OII]3727,CaII H,CaII K,H$_{\gamma}$,H$_{\beta}$,[OIII]5007 \\
15 & 23 18 02.26 & $-$00 07 52.10 & \nodata & \nodata & 0 & 21.72 & 2.00 & \nodata \\
16 & 23 17 56.41 & $-$00 08 45.13 & \nodata & \nodata & 0 & 21.99 & 1.64 & \nodata \\
17 & 23 17 52.75 & $-$00 08 58.88 & 0.38222 & 0.00006 & 1 & 22.36 & 0.25 & H$_{\beta}$,[OIII]4959,[OIII]5007 \\
18 & 23 17 52.97 & $-$00 08 30.59 & 0.56368 & 0.00035 & 2 & 20.97 & 0.39 & [OII]3727,CaII H,CaII K \\
19 & 23 17 53.25 & $-$00 08 24.04 & 0.59975 & 0.00012 & 1 & 19.72 & 1.20 & CaII H,CaII K,G band \\
20 & 23 17 52.53 & $-$00 09 10.48 & \nodata & \nodata & 0 & 21.89 & 1.03 & \nodata \\
21 & 23 18 04.23 & $-$00 09 27.40 & \nodata & \nodata & 0 & 21.96 & 3.25 & \nodata \\
22 & 23 17 54.16 & $-$00 09 35.50 & 0.66210 & 0.00049 & 3 & 22.53 & 1.25 & [OII]3727,CaII K \\
23 & 23 17 52.03 & $-$00 09 43.99 & 0.91214 & 0.00080 & 2 & 22.42 & 0.53 & [OII]3727 \\
24 & 23 17 55.54 & $-$00 09 55.19 & 0.38519 & 0.00004 & 1 & 22.50 & 0.13 & [OII]3727,H$_{\gamma}$,H$_{\beta}$,[OIII]4959,[OIII]5007 \\
25 & 23 17 54.92 & $-$00 10 06.64 & 0.60990 & 0.00052 & 2 & 21.51 & 0.51 & [OII]3727,CaII H,CaII K \\
26 & 23 18 02.61 & $-$00 05 01.25 & 0.56462 & 0.00014 & 2 & 21.31 & 0.54 & [OII]3727,CaII K \\
27 & 23 17 55.64 & $-$00 05 12.37 & 0.47181 & 0.00035 & 2 & 20.05 & 0.78 & [OII]3727,CaII H,H$_{\beta}$ \\
28 & 23 18 00.96 & $-$00 05 37.10 & 0.47310 & 0.00014 & 2 & 21.53 & 0.61 & [OII]3727,H$_{\beta}$ \\
29 & 23 17 59.80 & $-$00 06 02.66 & 0.40260 & 0.00006 & 2 & 20.63 & 0.74 & [OII]3727,H$_{\beta}$,[OIII]5007 \\
30 & 23 17 58.01 & $-$00 05 52.12 & 0.75690 & 0.00085 & 3 & 21.68 & 1.46 & CaII H,CaII K \\
31 & 23 18 01.95 & $-$00 08 07.40 & 0.46606 & 0.00021 & 3 & 22.01 & 0.67 & [OII]3727,H$_{\beta}$ \\
32 & 23 18 01.71 & $-$00 07 43.64 & 0.38053 & 0.00006 & 1 & 21.33 & 0.31 & [OII]3727,H$_{\beta}$,[OIII]5007 \\
33 & 23 17 59.78 & $-$00 06 49.28 & \nodata & \nodata & 0 & 21.64 & 1.63 & \nodata \\
34 & 23 17 59.62 & $-$00 07 54.73 & 0.16752 & 0.00023 & 1 & 20.52 & 0.37 & H$_{\beta}$,[OIII]4959,[OIII]5007 \\
35 & 23 17 59.54 & $-$00 07 20.42 & 0.60088 & 0.00015 & 1 & 20.59 & 1.17 & [OII]3727,CaII H,CaII K,H$_{\gamma}$ \\
36 & 23 17 59.27 & $-$00 07 08.94 & 0.59913 & 0.00020 & 1 & 20.61 & 0.78 & [OII]3727,CaII H,CaII K,H$_{\delta}$ \\
37 & 23 17 57.80 & $-$00 06 59.51 & 0.59569 & 0.00028 & 2 & 21.18 & 1.17 & [OII]3727,CaII K \\
38 & 23 17 56.68 & $-$00 07 33.67 & 0.64769 & 0.00023 & 2 & 20.56 & 1.22 & CaII H,CaII K,G band \\
39 & 23 17 52.49 & $-$00 06 28.44 & 0.40436 & 0.00012 & 1 & 22.33 & 0.23 & H$_{\gamma}$,[OIII]4959,[OIII]5007 \\
40 & 23 17 50.95 & $-$00 06 38.23 & 0.61876 & 0.00001 & 2 & 21.48 & 0.58 & [OII]3727,CaII H \\
41 & 23 18 06.78 & $-$00 08 54.24 & 0.28994 & 0.00012 & 2 & 20.94 & 0.38 & [OII]3727,H$_{\beta}$,[OIII]5007 \\
42 & 23 18 06.58 & $-$00 09 12.49 & 0.28360 & 0.00007 & 1 & 22.12 & 0.32 & H$_{\beta}$,[OIII]5007 \\
43 & 23 17 58.64 & $-$00 08 25.40 & 0.58640 & 0.00069 & 3 & 21.57 & 0.84 & [OII]3727,CaII H,CaII K \\
44 & 23 17 55.44 & $-$00 09 04.72 & 0.71340 & 0.00023 & 1 & 21.05 & 0.76 & [OII]3727,CaII H,CaII K \\
45 & 23 17 55.14 & $-$00 08 39.34 & 0.56384 & 0.00006 & 1 & 21.30 & 0.26 & [OII]3727,CaII H,CaII K \\
46 & 23 17 56.00 & $-$00 09 36.04 & \nodata & \nodata & 0 & 21.31 & 1.50 & \nodata \\
47 & 23 17 51.31 & $-$00 09 24.59 & 0.75810 & 0.00021 & 2 & 21.04 & 0.70 & [OII]3727,CaII H \\ 
48 & 23 17 49.98 & $-$00 09 46.66 & 0.32390 & 0.00012 & 1 & 21.18 & 0.30 & H$_{\beta}$,[OIII]4959,[OIII]5007 \\
49 & 23 18 06.66 & $-$00 10 02.93 & 0.59449 & 0.00080 & 3 & 21.75 & 0.49 & [OII]3727 \\
\enddata
\tablenotetext{a}{Redshift reliability classifier.}
\tablecomments{Whenever information could not be obtained for a specific target, a 
symbol '...' is used. Stars found in our data have zero redshift.}
\label{tab231759}
\end{deluxetable}

\newpage
\begin{figure}
\begin{center}
\includegraphics[height=14cm,width=15cm]{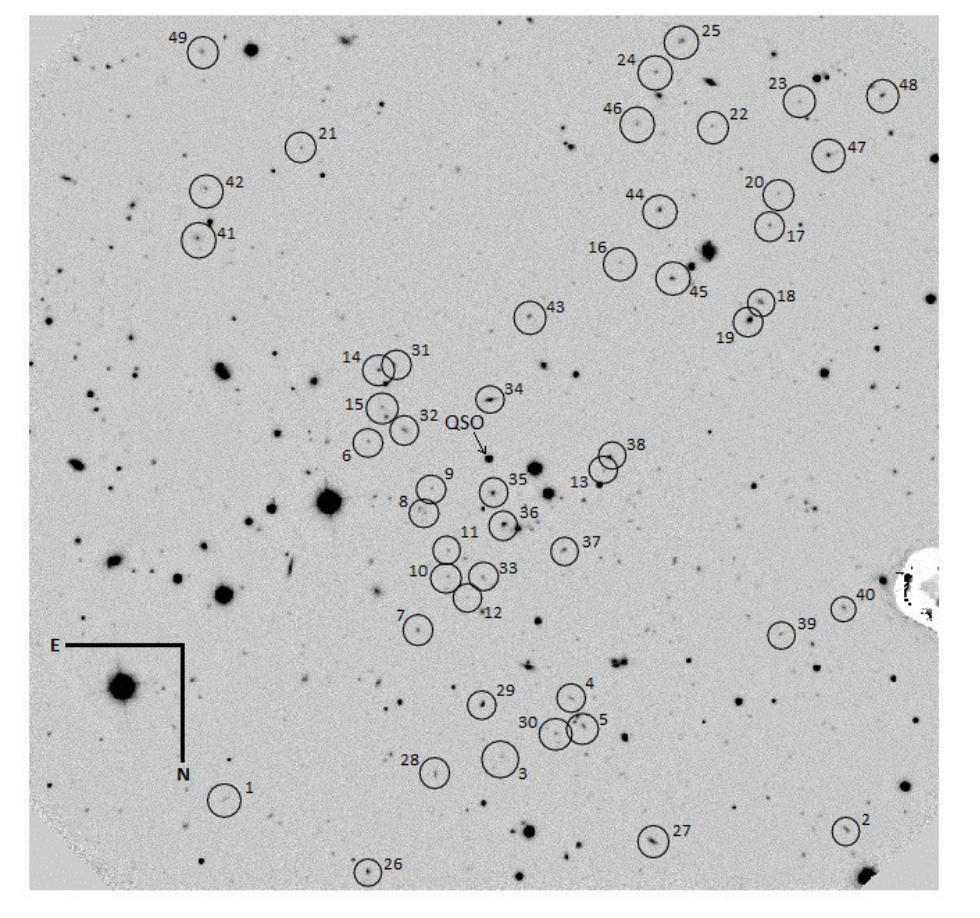}
\end{center}
\begin{flushright}
\includegraphics[height=5cm,width=7cm]{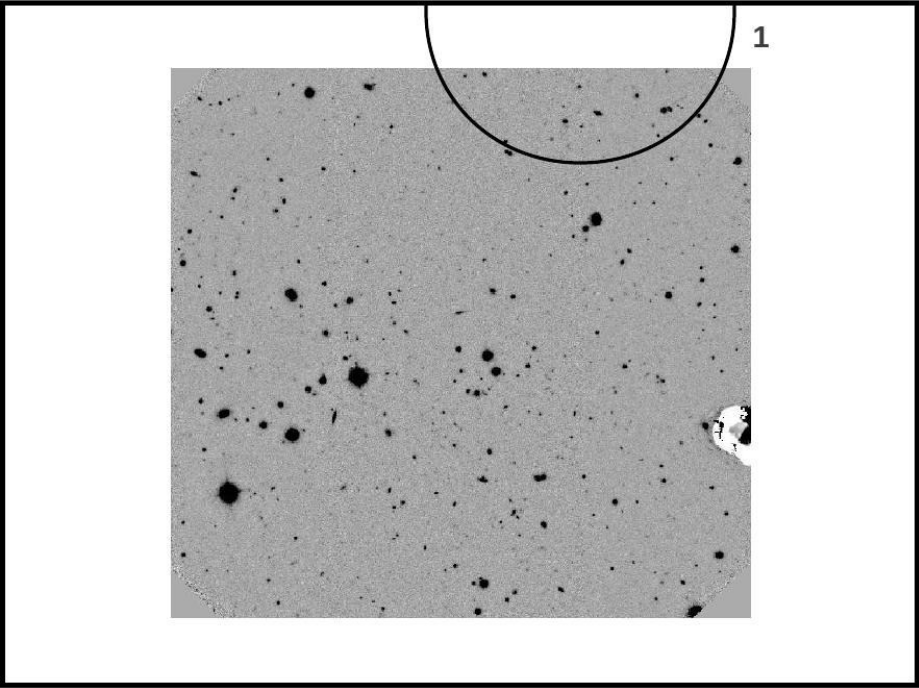} 
\end{flushright}
\begin{center}
\caption{{\it Top}: 5.5$\arcmin\times$5.5$\arcmin$ image of the field centered on the 
SDSS quasar 231759.63$-$000733.2. Galaxies are labeled according to the 
identification number given in Table \ref{tab231759}.  {\it Bottom}: A zoom-out of the image shown at the 
top. Center coordinates of each RCS1 cluster/group candidate are shown in circles. Each 
cluster is labeled according to their identification numbers 
given in the redshift histogram of Figure \ref{his231759}.}
\label{camp231759}
\end{center}
\end{figure}
\clearpage

\newpage
\begin{figure}
\begin{center}
\includegraphics[height=16cm,width=16cm]{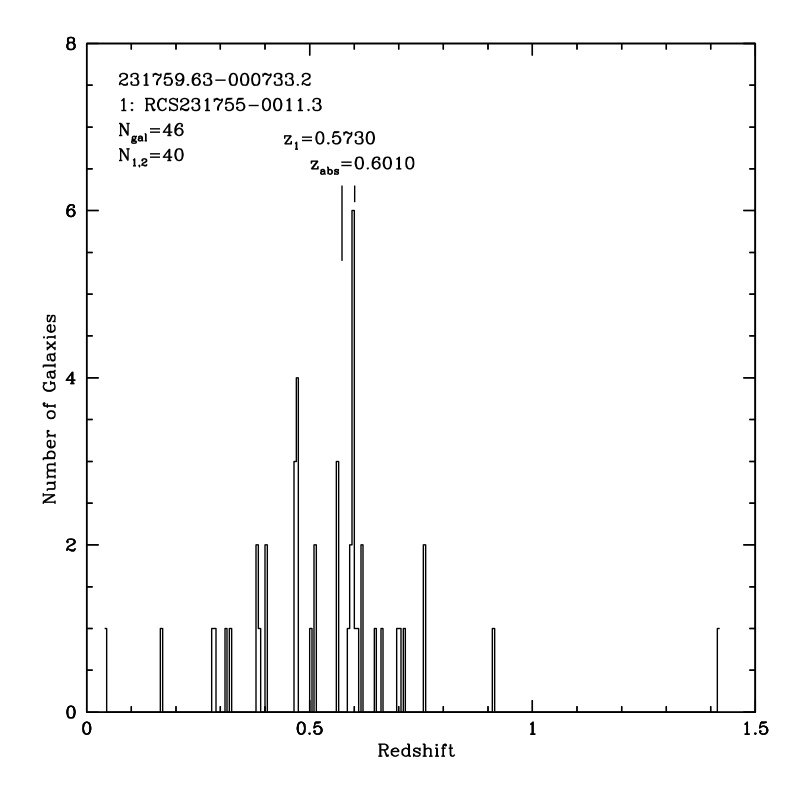}
\caption{Redshift histogram of the field centered on the SDSS quasar 
231759.63$-$000733.2. The bin size is of 0.005 in redshift space, which translates 
in $\Delta v \sim$ 1000 km/s at $z =$ 0.5. The total number of redshifts available 
for this field is given by N$_{gal}$, from which N$_{1,2}$ is the number of redshifts 
classified with reliability flag 1 or 2.}
\label{his231759}
\end{center}
\end{figure}
\clearpage

\newpage
\begin{deluxetable}{ccccccccp{3.2in}}
\tablewidth{0pc}
\tablecaption{Spectroscopic Targets of Field Centered on 231958.70$-$002449.3.}
\tabletypesize{\tiny}
\tablehead{ \colhead{No.} & \colhead{RA(J2000)} & \colhead{DEC(J2000)} & 
\colhead{$z_{gal}$} & \colhead{$\sigma_{z_{gal}}$} & \colhead{Flag\tablenotemark{a}} & 
\colhead{$z'$} & \colhead{$R_{c}$ - $z'$} & \colhead{Comments}}
\startdata
1 & 23 19 44.07 & $-$00 27 04.54 & 0.73498 & 0.00055 & 1 & 21.08 & 1.31 & CaII H,CaII K \\
2 & 23 19 47.51 & $-$00 26 55.72 & 0.55650 & 0.00080 & 3 & 21.37 & 0.67 & [OII]3727 \\
3 & 23 19 48.68 & $-$00 26 47.76 & 0.57801 & 0.00003 & 1 & 21.18 & 0.54 & [OII]3727,CaII H,CaII K \\
4 & 23 19 49.31 & $-$00 27 21.82 & 0.56133 & 0.00008 & 1 & 21.64 & 0.43 & [OII]3727,CaII H,CaII K,H$_{\delta}$,H$_{\gamma}$ \\
5 & 23 19 56.85 & $-$00 26 09.20 & 0.72041 & 0.00060 & 3 & 20.87 & 1.29 & CaII H,CaII K \\
6 & 23 19 59.07 & $-$00 26 25.01 & 0.35052 & 0.00032 & 3 & 20.25 & 0.58 & G band,H$_{\beta}$,[OIII]5007 \\
7 & 23 19 42.02 & $-$00 24 49.00 & 0.80880 & 0.00080 & 3 & 21.97 & 0.79 & [OII]3727 \\
8 & 23 19 43.77 & $-$00 25 46.02 & 0.29361 & 0.00007 & 1 & 19.36 & 0.48 & [OII]3727,CaII H,CaII K \\
9 & 23 19 59.00 & $-$00 25 10.81 & 0.71750 & 0.00023 & 2 & 20.44 & 1.32 & [OII]3727,CaII H,CaII K \\
10 & 23 19 59.97 & $-$00 24 21.60 & 0.84995 & 0.00080 & 3 & 22.94 & 0.28 & [OII]3727 \\
11 & 23 19 59.83 & $-$00 25 34.03 & 0.84727 & 0.00080 & 2 & 22.10 & 0.58 & [OII]3727 \\
12 & 23 19 58.14 & $-$00 24 57.49 & \nodata & \nodata & 0 & 23.30 & 0.64 & \nodata \\
13 & 23 20 00.28 & $-$00 24 41.29 & 0.10786 & 0.00003 & 2 & 25.82 & -0.84 & H$_{\alpha}$,[SII]6716,[SII]6730 \\
14 & 23 19 59.27 & $-$00 24 30.78 & 0.56299 & 0.00005 & 1 & 22.96 & 0.14 & [OII]3727,[NeIII]3868,HeI 3888,[NeIII]3967,H$_{\delta}$,H$_{\gamma}$,H$_{\beta}$ \\
15 & 23 19 59.26 & $-$00 25 18.70 & 0.72096 & 0.00010 & 2 & 21.27 & 1.28 & CaII H,CaII K \\
16 & 23 19 43.45 & $-$00 24 03.74 & 0.80923 & 0.00080 & 2 & 21.73 & 0.56 & [OII]3727 \\
17 & 23 19 55.00 & $-$00 22 52.10 & 0.44211 & 0.00006 & 1 & 21.53 & 0.33 & [OII]3727,H$_{\gamma}$,H$_{\beta}$,[OIII]4959,[OIII]5007 \\
18 & 23 19 55.53 & $-$00 22 38.24 & \nodata & \nodata & 0 & 20.11 & 0.47 & \nodata \\
19 & 23 19 59.91 & $-$00 23 51.22 & \nodata & \nodata & 0 & 21.24 & 0.79 & \nodata \\
20 & 23 19 58.61 & $-$00 23 32.96 & \nodata & \nodata & 0 & 22.37 & 0.79 & \nodata \\
21 & 23 19 58.30 & $-$00 23 06.43 & 0.59366 & 0.00017 & 1 & 21.76 & 0.75 & [OII]3727,CaII H,CaII K,G band \\
22 & 23 20 00.58 & $-$00 22 26.94 & \nodata & \nodata & 0 & 20.43 & 1.60 & \nodata \\
23 & 23 19 57.89 & $-$00 23 14.75 & \nodata & \nodata & 0 & 21.21 & 1.31 & \nodata \\
24 & 23 19 45.31 & $-$00 27 10.40 & \nodata & \nodata & 0 & 19.86 & 1.50 & \nodata \\
25 & 23 19 50.46 & $-$00 27 19.58 & 0.27483 & 0.00004 & 1 & 20.48 & 0.31 & [OII]3727,H$_{\gamma}$,H$_{\beta}$,[OIII]4959,[OIII]5007 \\
26 & 23 19 42.12 & $-$00 26 06.29 & 0.42579 & 0.00010 & 1 & 20.96 & 0.36 & [OII]3727,H$_{\beta}$,[OIII]4959,[OIII]5007 \\
27 & 23 19 42.64 & $-$00 26 56.15 & 0.39177 & 0.00007 & 1 & 21.17 & 0.74 & [OII]3727,H$_{\beta}$ \\
28 & 23 19 43.76 & $-$00 26 39.95 & \nodata & \nodata & 0 & 21.50 & 1.41 & \nodata \\
29 & 23 19 44.64 & $-$00 25 52.28 & 0.25271 & 0.00005 & 1 & 18.56 & 0.62 & CaII H,CaII K,H$_{\beta}$,[OIII]5007 \\
30 & 23 19 44.89 & $-$00 26 20.04 & 0.18640 & 0.00014 & 2 & 19.44 & 0.57 & [OII]3727,H$_{\beta}$,H$_{\alpha}$ \\
31 & 23 19 55.02 & $-$00 26 30.37 & 0.38420 & 0.00005 & 1 & 20.99 & 0.31 & [OII]3727,H$_{\beta}$,[OIII]4959,[OIII]5007 \\
32 & 23 19 56.12 & $-$00 25 30.72 & 0.35157 & 0.00035 & 3 & 20.56 & 0.43 & CaII K,H$_{\beta}$ \\
33 & 23 19 58.13 & $-$00 25 21.00 & 0.49035 & 0.00006 & 1 & 20.19 & 0.70 & [OII]3727,CaII H,CaII K \\
34 & 23 19 58.97 & $-$00 25 39.29 & \nodata & \nodata & 0 & 20.82 & 0.70 & \nodata \\
35 & 23 19 50.11 & $-$00 25 10.74 & 0.15560 & 0.00002 & 1 & 19.50 & 0.30 & H$_{\gamma}$,H$_{\beta}$,[OIII]4959,[OIII]5007,H$_{\alpha}$ \\
36 & 23 19 51.67 & $-$00 24 20.30 & 0.45332 & 0.00037 & 2 & 20.28 & 0.73 & CaII H,CaII K,G band \\
37 & 23 19 53.37 & $-$00 24 30.82 & 0.63093 & 0.00010 & 1 & 21.46 & 0.12 & [OII]3727,H$_{\gamma}$,[OIII]5007 \\
38 & 23 20 01.50 & $-$00 24 42.37 & 0.08539 & 0.00008 & 1 & 19.35 & 0.49 & H$_{\alpha}$,[NII]6583,[SII]6716,[SII]6730 \\
39 & 23 19 55.72 & $-$00 23 59.71 & 0.18561 & 0.00003 & 1 & 20.40 & 0.21 & H$_{\beta}$,[OIII]4959,[OIII]5007,H$_{\alpha}$ \\
40 & 23 19 46.78 & $-$00 23 23.68 & 0.42541 & 0.00040 & 3 & 20.29 & 0.25 & H$_{\beta}$,[OIII]5007 \\
41 & 23 19 49.40 & $-$00 23 39.95 & 0.76568 & 0.00020 & 3 & 21.54 & 0.63 & [OII]3727,H$_{\gamma}$ \\
42 & 23 19 51.50 & $-$00 23 13.06 & 0.28258 & 0.00005 & 1 & 19.00 & 0.60 & CaII H,CaII K \\
43 & 23 19 50.03 & $-$00 22 39.90 & 0.42397 & 0.00015 & 1 & 19.37 & 0.50 & [OII]3727,CaII H,CaII K,H$_{\beta}$ \\
44 & 23 19 51.76 & $-$00 22 49.69 & 0.71920 & 0.00018 & 3 & 20.83 & 1.05 & [OII]3727,CaII H,CaII K \\
45 & 23 19 47.54 & $-$00 22 21.18 & 0.22817 & 0.00012 & 1 & 22.16 & -0.12 & [OII]3727,[OIII]4959,[OIII]5007 \\
46 & 23 19 47.60 & $-$00 22 29.03 & 0.44134 & 0.00023 & 1 & 19.16 & 0.69 & [OII]3727,CaII H,G band \\
\enddata
\tablenotetext{a}{Redshift reliability classifier.}
\tablecomments{Whenever information could not be obtained for a specific target, a 
symbol '...' is used. Stars found in our data have zero redshift.}
\label{tab231958}
\end{deluxetable}

\newpage 
\begin{figure}
\begin{center}
\includegraphics[height=14cm,width=15cm]{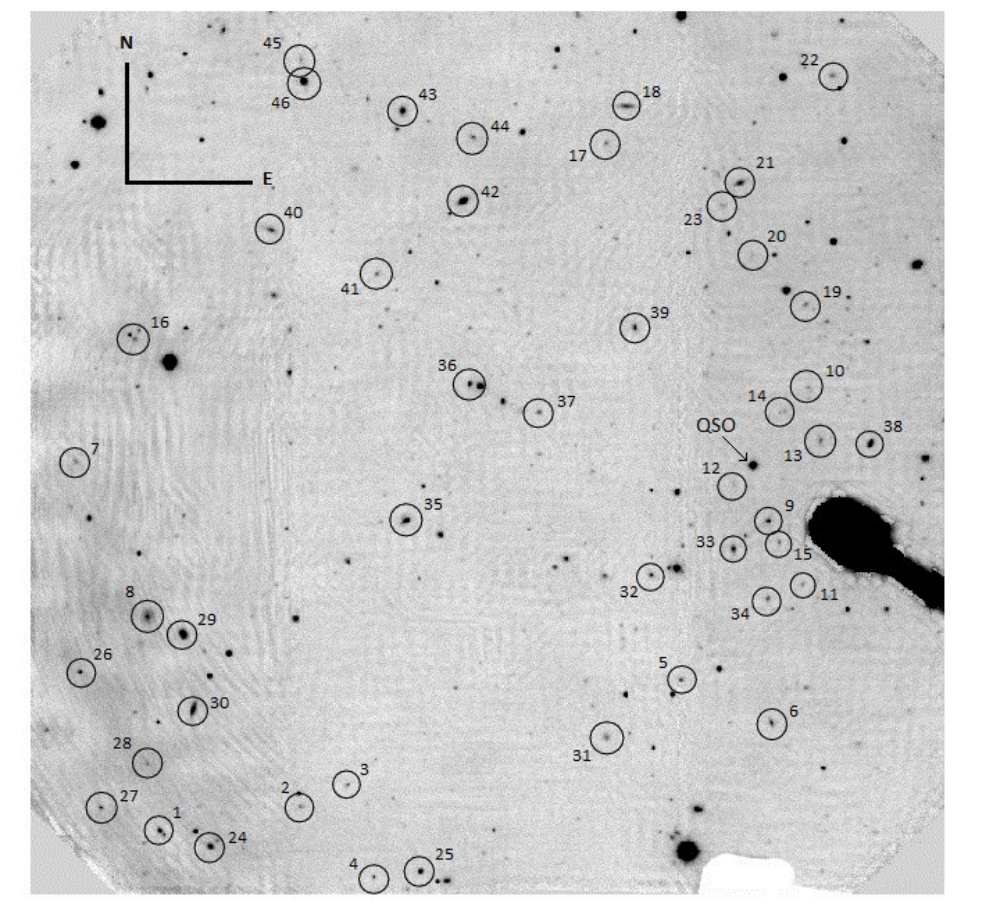}
\end{center}
\begin{flushright}
\includegraphics[height=5cm,width=7cm]{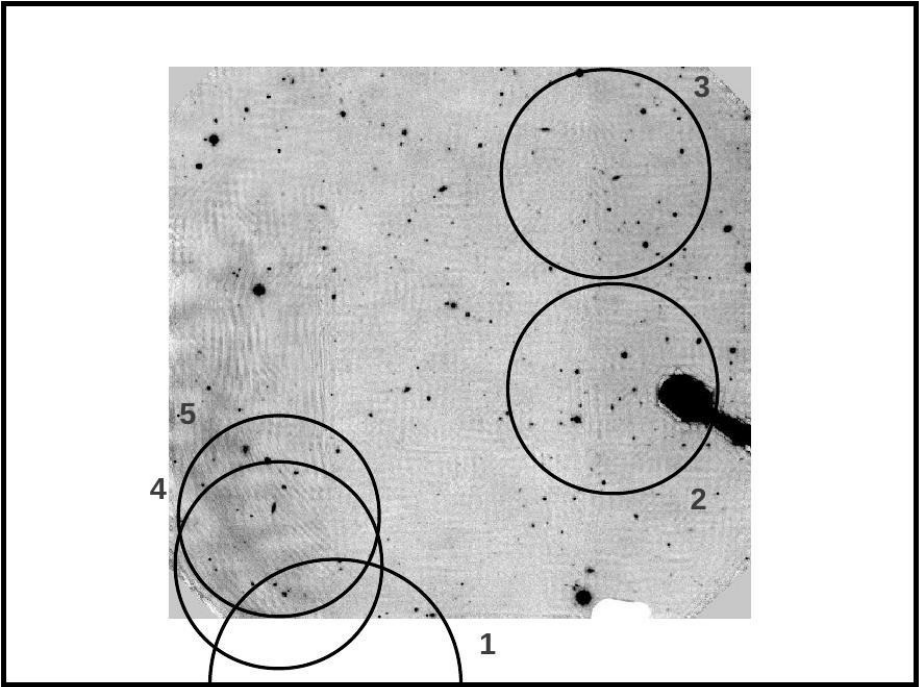} 
\end{flushright}
\begin{center}
\caption{{\it Top}: 5.5$\arcmin\times$5.5$\arcmin$ image of the field centered on the 
SDSS quasar 231958.70$-$002449.3. Galaxies are labeled according to the 
identification number given in Table \ref{tab231958}.  {\it Bottom}: A zoom-out of the image shown at the 
top. Center coordinates of each RCS1 cluster/group candidate are shown in circles. Each 
cluster is labeled according to their identification numbers 
given in the redshift histogram of Figure \ref{his231958}.}
\label{camp231958}
\end{center}
\end{figure}
\clearpage

\newpage
\begin{figure}
\begin{center}
\includegraphics[height=16cm,width=16cm]{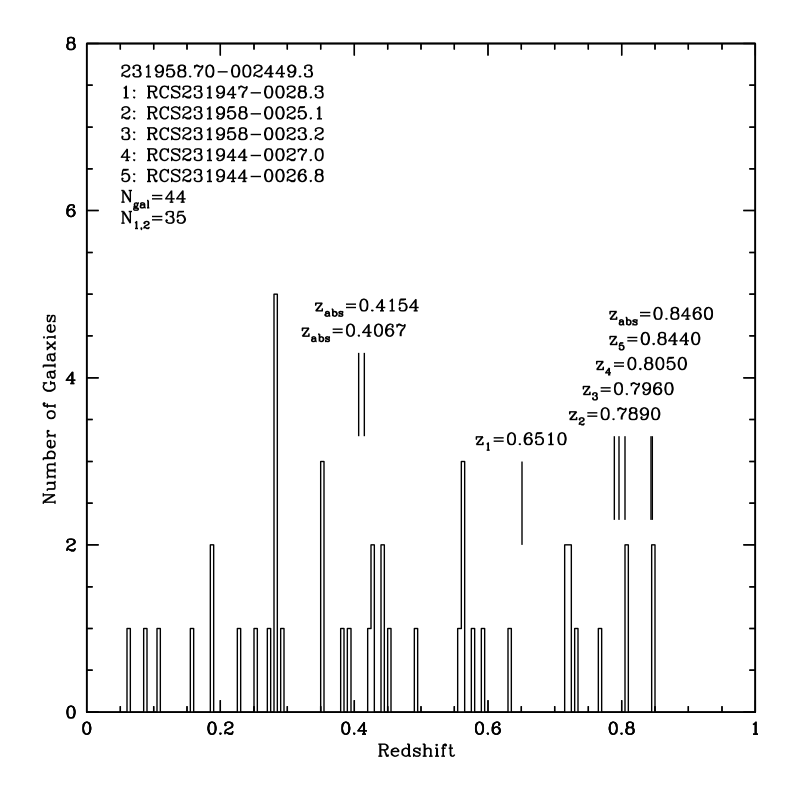}
\caption{Redshift histogram of the field centered on the SDSS quasar 
231958.70$-$002449.3. The bin size is of 0.005 in redshift space, which translates 
in $\Delta v \sim$ 1000 km/s at $z =$ 0.5. The total number of redshifts available 
for this field is given by N$_{gal}$, from which N$_{1,2}$ is the number of redshifts 
classified with reliability flag 1 or 2.}
\label{his231958}
\end{center}
\end{figure}
\clearpage

\end{document}